\pdfoutput=1
\documentclass[journal]{IEEEtran}
\usepackage{amsmath,amsfonts}
\usepackage{mathtools}
\usepackage{algorithm}
\usepackage{algpseudocode}
\usepackage{array}
\usepackage{textcomp}
\usepackage{stfloats}
\usepackage{url}
\usepackage{verbatim}
\usepackage{graphicx}

\usepackage{cite}
\usepackage{mathtools} 

\usepackage{listings} % For verbatim code blocks
\usepackage{graphicx}
\usepackage{booktabs} % For professional tables

\usepackage[utf8]{inputenc} % 必须添加
\usepackage[T1]{fontenc}    % 建议添加

\usepackage{listings}
\usepackage{float} % 允许使用 [H] 强制浮动体位置

\renewcommand{\thefigure}{\arabic{figure}}
\newtheorem{prob}{Problem}
\newtheorem{thm}{Theorem}
\newtheorem{lemma}{Lemma}
\newtheorem{remark}{Remark}

\usepackage{seqsplit}

\usepackage{graphicx}
\usepackage{algpseudocode}

\usepackage{svg}

\usepackage{mathabx}

\newboolean{showcomments}
\setboolean{showcomments}{true}

\ifthenelse{\boolean{showcomments}}
{ \newcommand{\mynote}[3]{
   \protect\fbox{\bfseries\sffamily\scriptsize#1}
   {\small$\blacktriangleright$\textsf{\emph{\color{#3}{#2}}}$\blacktriangleleft$}}}
{ \newcommand{\mynote}[3]{}}

\usepackage{subcaption}
\algdef{SE}[REQUIRE]{Require}{EndRequire}{\textbf{Input:}}{\endRequire}% If needed, but algpseudocode defines \Require
\algdef{SE}[ENSURE]{Ensure}{EndEnsure}{\textbf{Output:}}{\endEnsure}% Similarly for \Ensure

\usepackage[absolute]{textpos}
\begin{document}

\title{COMETS: Coordinated Multi-Destination Video Transmission with In-Network Rate Adaptation}

\author{Yulong~Zhang,~Ying~Cui,~Zili~Meng,~Abhishek~Kumar,~and~Dirk~Kutscher%
% \thanks{This work was supported in part by the Guangzhou Municipal Key Laboratory on Future Networked Systems (024A03J0623), the National Natural Science Foundation of China (62371412), the National Key Research and Development Program of China (2024YFE0200600), the Guangdong Basic and Applied Basic Research Natural Science Funding Scheme (2024A1515011184), and the Guangzhou-HKUST (GZ) Joint Funding Scheme (2024A03J0539). (Corresponding author: Dirk Kutscher)}%
% \thanks{Yulong Zhang, Ying Cui, and Dirk Kutscher are with the Internet of Things Thrust, The Hong Kong University of Science and Technology, Guangzhou, China (e-mail: Y.~Zhang: yzhang893@connect.hkust-gz.edu.cn; Y.~Cui: yingcui@ust.hk; D.~Kutscher: dku@dkutscher.net).}%
% \thanks{Zili Meng is with the Department of Electronic and Computer Engineering, The Hong Kong University of Science and Technology, Hong Kong, China (e-mail: zilim@ieee.org).}%
% \thanks{Abhishek Kumar is with the Department of Computer Science, University of Helsinki, Helsinki, Finland (e-mail: abhishek.kumar@helsinki.fi).}%
\thanks{Yulong Zhang, Ying Cui, and Dirk Kutscher are with the Internet of Things Thrust, The Hong Kong University of Science and Technology, Guangzhou, China (e-mail: Y.~Zhang: yzhang893@connect.hkust-gz.edu.cn; Y.~Cui: yingcui@ust.hk; D.~Kutscher: dku@dkutscher.net). Zili Meng is with the Department of Electronic and Computer Engineering, The Hong Kong University of Science and Technology, Hong Kong, China (e-mail: zilim@ieee.org). Abhishek Kumar is with the Faculty of Information Technology, University of Jyväskylä, Finland (e-mail: abhishek.k.kumar@jyu.fi).}%
}

\markboth{Accepted to IEEE Transactions on Multimedia (TMM), 2026}{Zhang \MakeLowercase{\textit{et al.}}: COMETS}

% \IEEEpubid{0000--0000/00\$00.00~\copyright~2021 IEEE}
% Remember, if you use this you must call \IEEEpubidadjcol in the second
% column for its text to clear the IEEEpubid mark.

\maketitle

% === 版权声明 (放在首页底部) ===
% 坐标 (1.5, 26.5) 可能需要根据双栏排版微调
% 如果文字挡住了正文，可以调整垂直坐标 26.5
\begin{textblock}{18}(1.5,26.5)
    \noindent \footnotesize \copyright 2026 IEEE. Personal use of this material is permitted. Permission from IEEE must be obtained for all other uses, in any current or future media, including reprinting/republishing this material for advertising or promotional purposes, creating new collective works, for resale or redistribution to servers or lists, or reuse of any copyrighted component of this work in other works.
\end{textblock}

\begin{abstract}
Large-scale video streaming events attract millions of simultaneous viewers, stressing existing delivery infrastructures. Client-driven adaptation reacts slowly to shared congestion, while server-based coordination introduces scalability bottlenecks and single points of failure. We present COMETS, a coordinated multi-destination video transmission framework that leverages information-centric networking principles such as request aggregation and in-network state awareness to enable scalable, fair, and adaptive rate control. COMETS introduces a novel range-interest protocol and distributed in-network decision process that aligns video quality across receiver groups while minimizing redundant transmissions. To achieve this, we develop a lightweight distributed optimization framework that guides per-hop quality adaptation without centralized control. Extensive emulation shows that COMETS consistently improves bandwidth utilization, fairness, and user-perceived quality of experience over DASH, MoQ, and ICN baselines, particularly under high concurrency. The results highlight COMETS as a practical, deployable approach for next-generation scalable video delivery.

\end{abstract}

% \begin{IEEEkeywords}
% Adaptive Video Streaming, Data-Oriented Networking, Distributed Optimization, and Multi-user Coordination
% % ,  Quality of Experience (QoE)
% \end{IEEEkeywords}

\section{Introduction}
\label{intro}

Nowadays, large streaming events typically attract millions of viewers, and the demand for concurrent video consumption is also expanding dramatically. For example, the number of monthly sports streaming viewers have grown from 57 million in 2021 to more than 90 million in 2025 \cite{pwc2024}, with more than 17\% users participating in multiple streams simultaneously \cite{rossvideo2025,wang2025vifusion}. This explosive growth exposes fundamental limitations in existing video delivery architectures: how to maintain consistent, fair Quality of Experience (QoE) when thousands of users compete for shared bottleneck resources.

% These patterns pose challenges for existing video delivery infrastructure in maintaining consistent QoE at scale, as e
Existing infrastructures are not designed for effective coordination and resource sharing among large numbers of simultaneous viewers, resulting in inefficient management of concurrent requests for the same content segments and insufficient coordination of network resource allocation among users of the shared infrastructure \cite{zhang2024networked}.
These inefficiencies lead to redundant data transmission and suboptimal bandwidth utilization, ultimately impairing user QoE by increasing network congestion, unstable bitrates, and higher incidences of buffering, especially during peak usage scenarios.
To address these challenges, an ideal video delivery system must possess coordinated, scalable, and adaptive capabilities to maximize bandwidth utilization while ensuring a fair, high-quality experience for all users. Such a system should aggregate requests for the same content to eliminate redundancy, make intelligent in-network decisions and distribute computational load to avoid bottlenecks.
% This architecture should quickly converge to the optimal bitrate distribution even under dynamic conditions, transforming shared network paths into efficient, resilient delivery pipelines.

% \dirk{here, you could give a first intuition of the COMETS idea. I.e., "what should be done, ideally.}

\begin{figure}[ht]
\centering
\begin{subfigure}[b]{0.235\textwidth}
    \centering
    \includegraphics[width=\textwidth,height=0.75\textwidth,keepaspectratio]{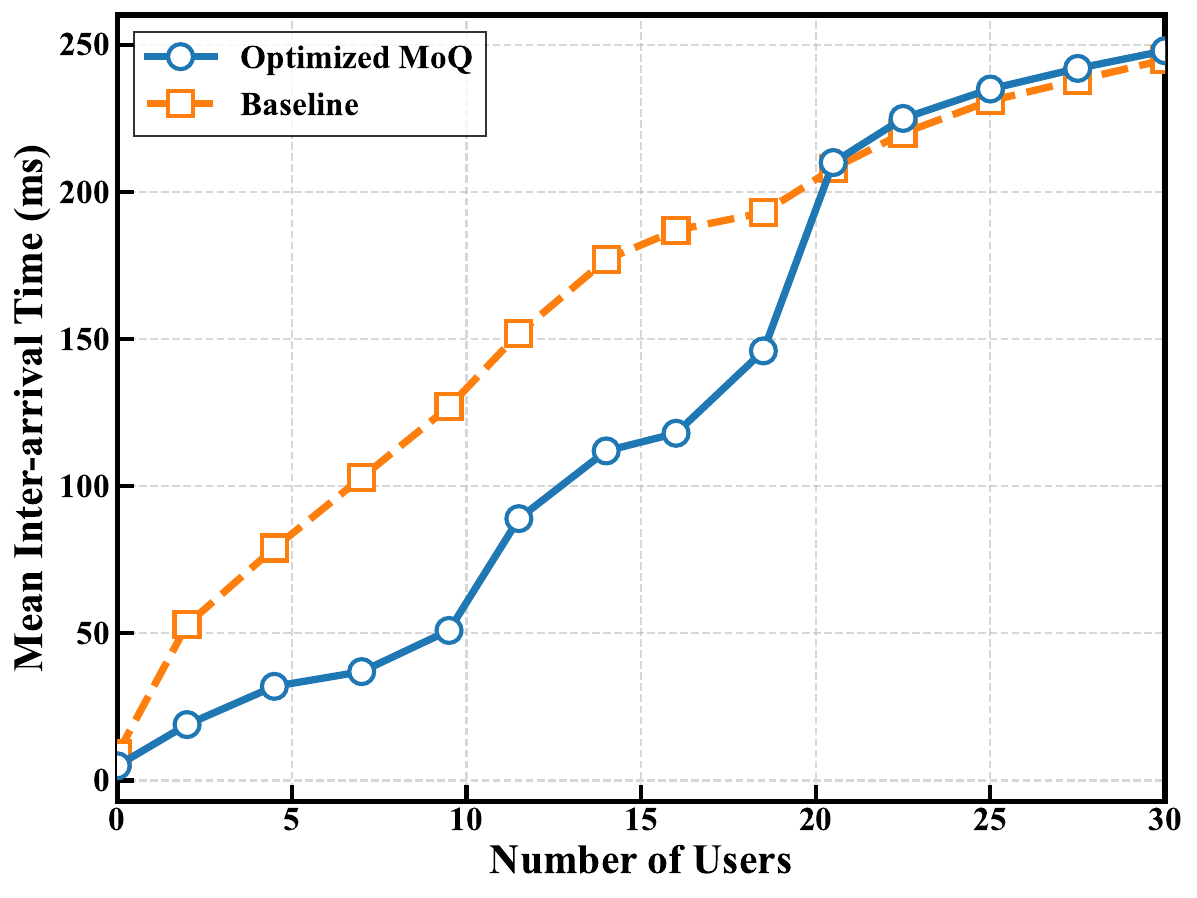}
    \caption{Latency vs. User Load}
    \label{fig:latency_vs_load}
\end{subfigure}
\hfill 
\begin{subfigure}[b]{0.235\textwidth}
    \centering
    \includegraphics[width=\textwidth,height=0.75\textwidth,keepaspectratio]{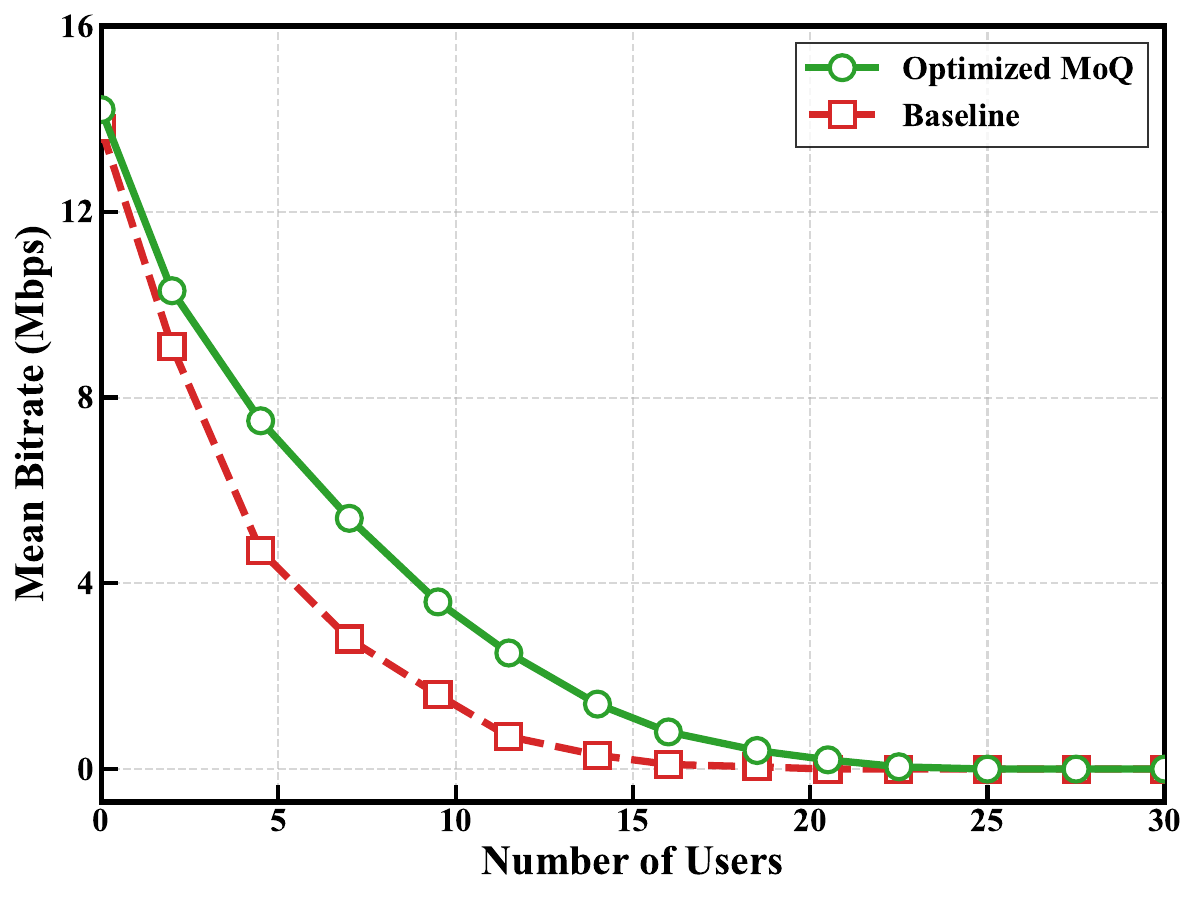}
    \caption{Mean Bitrate vs. User Load} 
    \label{fig:bitrate_vs_load}
\end{subfigure}
\caption{Performance Comparison between baseline MoQ and server-optimized MoQ under increasing user load.}
\label{network_performance}
\end{figure}

Current solutions exhibit fundamental limitations with respect to coordination and scalability. Client-adaptive approaches like Dynamic Adaptive Streaming over HTTP (DASH) enable individual clients to select video representations independently \cite{li2014probe, zhou2016mdash, spiteri2019theory}. However, their uncoordinated decisions, based on delayed and localized network views, lag behind the actual state of shared network bottlenecks, leading to bandwidth contention and bitrate oscillations.
Server-side approaches address these limitations by centralizing adaptation logic \cite{altamimi2020qoe, 10689417}, enabling optimal resource allocation through comprehensive network and user demand assessments. However, managing state and control interactions for numerous users introduces scalability challenges, and centralized decision architectures create single points of failure that compromise real-time performance.
Our experiments (Figure~\ref{network_performance}) demonstrate that even state-of-the-art server-optimized Media over QUIC (MoQ) ultimately encounters the same scalability barriers as baseline approaches under high concurrency. \footnote{Experiment setup: Mininet emulation with 10 Mbps link bandwidth, 40 Mbps backbone bandwidth, network delay 5 ms, testing on Intel Xeon Platinum 8358P CPU (2.60GHz, 64 cores) with 32GB RAM.}
% \begin{figure}[ht]
% \centering
% \includegraphics[width=0.45\textwidth]{figures/moq_performance_comparison.pdf}
% \caption{Performance Comparison between baseline MoQ and server-optimized MoQ under increasing user load.}
% \label{network_performance}
% \end{figure}

% Our experimental analysis reveals the severity of these scalability challenges. Figure \ref{network_performance} compares traditional MoQ implementation against state-of-the-art server-optimized MoQ \cite{10689417}. While server-side optimization demonstrates improvements during moderate loads (8-20 users), both approaches ultimately encounter the same scalability barriers, revealing that centralized optimization merely delays rather than resolves the fundamental architectural bottleneck.\footnote{Experiment setup: Mininet emulation with 10 Mbps link bandwidth, 40 Mbps backbone bandwidth, network delay 5 ms, testing on Intel Xeon Platinum 8358P CPU (2.60GHz, 64 cores) with 32GB RAM.}

\textbf{Key Insight.} We observe that effective multi-user video streaming requires two properties: I). \emph{aggregation-aware delivery}, where identical requests are merged to eliminate redundant transmissions, and II). \emph{distributed coordination}, where adaptation decisions are made at points of request convergence rather than at centralized endpoints.
This leads us to consider Information-Centric Networking (ICN) \cite{chai2025inds}. ICN provides inherent advantages for multi-user content distribution through in-network caching and request aggregation in systems like CCNx/NDN \cite{liu2016hop,7995137}. While these features reduce redundant transmissions by merging duplicate requests at forwarders, existing ICN-based solutions \cite{9195016,8007216,Alt} focus on hop-by-hop adaptation rather than coordinated multi-user rate adaptation, suffering from decision lag and failing to ensure efficient convergence toward stable, fair rate allocations (i.e., equitable QoE distribution).
To address these limitations, we present COMETS (Coordinated Multi-Destination Video Transmission with In-Network Rate Adaptation), a scalable, ICN-based multi-destination video streaming framework engineered to resolve challenges in large-scale video delivery: redundant data transmission, lack of scalable coordination, and inefficient system convergence.

\textbf{Design Philosophy.}
COMETS is based on three principles that distinguish it from prior work:
\emph{I). Group-aware rather than individual optimization.} Instead of each client independently selecting bitrates, COMETS groups receivers with similar capabilities and network conditions, then aligns video quality across each group. This transforms the combinatorial complexity of individual decisions into tractable group-level optimization.
\emph{II). Proactive rather than reactive adaptation.} Unlike existing ICN approaches that react to congestion signals, COMETS uses a distributed Lagrangian framework where forwarders exchange dual variables (price signals) to anticipate upstream constraints. This enables proactive coordination without centralized state collection.
\emph{III). Deployable overlay architecture.} COMETS requires no modifications to network infrastructure. To ensure deployability, COMETS is architecturally flexible and can be deployed as an application-layer overlay network over existing Internet protocols (e.g., HTTP/QUIC over UDP), similar to Content Delivery Networks (CDNs) like Akamai \cite{su2009drafting} or CloudFlare \cite{shobiri2023cdns}. It requires no infrastructure modifications and assumes trusted intermediate nodes under the same administrative domain \cite{pathan2007taxonomy}, enabling immediate integration into today's networks without network-layer changes.
While COMETS shares MoQ’s vision of moving intelligence into the network, it avoids central bottlenecks by enabling per-hop optimization via ICN primitives, and is deployable over MoQ-capable infrastructures as an overlay.

\textbf{Our Approach.} COMETS transforms video streaming from isolated endpoint control into coordinated in-network negotiation, with four key contributions:
\begin{itemize}
\item \textbf{Range-interest protocol for coordinated adaptation.} We introduce a novel protocol where clients express resolution \emph{ranges} rather than specific quality levels. This enables forwarders to aggregate requests and optimize resolution assignments across user groups, shifting adaptation logic from endpoints to the network fabric.
\item \textbf{Scalable architecture without central bottlenecks.} COMETS distributes adaptation logic across forwarders, combining request aggregation with per-hop decision-making. 
% The system is deployable as an application-layer overlay over existing Internet protocols (e.g., HTTP/QUIC), requiring no infrastructure modifications.
\item \textbf{Distributed optimization with closed-form solutions.} We formalize coordinated multi-destination video transmission as a unified Integer Linear Programming (ILP) problem and develop a two-stage distributed algorithm. Unlike prior ICN approaches that rely on heuristics or reactive congestion signals, our method derives \emph{analytical closed-form solutions} for per-hop quality decisions, enabling proactive, group-aware rate allocation with provable convergence guarantees.
\item \textbf{Implementation and Evaluation:} Through extensive emulation on Mini-NDN with up to 300 concurrent clients, we demonstrate that COMETS achieves consistent QoE scores above 0.7 across all tested scales, while baselines degrade below 0.5 at high concurrency. COMETS maintains near-perfect fairness (Jain's index $\geq$ 0.93) and achieves optimization convergence within 50ms—up to 3.7$\times$ faster than centralized approaches.
\end{itemize}

Table~\ref{tab:acronyms} summarizes the key acronyms used throughout this paper. The remaining sections of this paper are structured as follows: Section~\ref{Related_Works} examines related work in this field; Section~\ref{design} introduces a comprehensive system design; Sections~\ref{sec:model} and~\ref{Distributed Decision Algorithm} detail our implemented distributed decision algorithm; Section ~\ref{experiment} provides extensive experimental evaluation results; and Section~\ref{Conclusion} summarizes the significance of this paper and future research directions.

\begin{table}[h]
\centering
\caption{Key Acronyms}
\label{tab:acronyms}
\begin{tabular}{ll}
\toprule
\textbf{Acronym} & \textbf{Full Form} \\
\midrule
ABR & Adaptive Bitrate \\
AIMD & Additive Increase/Multiplicative Decrease \\
CDN & Content Delivery Network \\
DAG & Directed Acyclic Graph \\
DASH & Dynamic Adaptive Streaming over HTTP \\
ICN & Information-Centric Networking \\
ILP & Integer Linear Programming \\
MoQ & Media over QUIC \\
NDN & Named Data Networking \\
PIT & Pending Interest Table \\
QoE & Quality of Experience \\
VMAF & Video Multi-Method Assessment Fusion \\
\bottomrule
\end{tabular}
\end{table}

\section{Related Work}
\label{Related_Works}

% \dirk{The separation into "connection-oriented" and "data-oriented" may not be ideal. Are the discussed systems and their features dependent on these categories?}

% Existing multi-destination video delivery approaches primarily comprise \textbf{connection-oriented} approaches based on traditional network architectures and \textbf{data-oriented} approaches leveraging content-centric networking principles.

% Client-adaptive approaches enable individual clients to select video representations independently.

\subsection*{Client-based \& server-based adaptation}

Early DASH research focused on single-user adaptation using client-side information. For example, Li et al. \cite{li2014probe} developed a probe-and-adapt scheme analogous to TCP's additive increase multiplicative decrease (AIMD) mechanism. Zhou et al. \cite{zhou2016mdash} employed Markov decision processes optimizing for quality, switching, and buffer status. 
While effective individually, these approaches lack awareness of broader network conditions and cannot coordinate when users compete for shared resources.
In contrast, COMETS leverages ICN's in-network aggregation to enable coordinated, group-aware adaptations, avoiding the oscillations and inefficiencies inherent in isolated client decisions.
For multi-user extensions, approaches like Tan et al.'s game theory model \cite{tan2021game} use payoff functions for fairness while network-assisted methods like SAND \cite{thomas2016applications} provide QoS hints. Recent reinforcement learning approaches \cite{altamimi2020qoe,bentaleb2022bob} enhance adaptation decisions but adds overhead, often prioritizing individual over global gains. QUIC-based solutions, such as Gurel et al.'s I-frame bandwidth estimation \cite{gurel2023media} and Ravuri et al.'s  hybrid reliable/unreliable transmission \cite{ravuri2023adaptive}, improve individual decisions but cannot prevent oscillations when users compete for bandwidth.
Altamimi et al. \cite{altamimi2020qoe} address coordination by centralizing adaptation logic, but face computational bottlenecks that worsen with scale. 
COMETS differs by distributing optimization across forwarders, eliminating central bottlenecks and enabling scalable, resilient coordination without global state management.
\subsection*{ICN-based streaming and aggregation}

ICN-based approaches offer inherent content aggregation capabilities through Pending Interest Tables (PITs) \cite{geng2023sok}, which store Interest state at each forwarder to perform stateful forwarding operations, including Internet aggregation.
% \dirk{"which store Interest state at each forwarder to perform stateful forwarding operations, including Internet aggregation."}
For single-user adaptation, Liu et al. \cite{liu2016hop} developed HAVS-CCN with selective layer dropping during congestion.
% while Awiphan et al. \cite{awiphan2018proactive} added interest adaptation and proactive caching. 
Saltarin et al. \cite{8007216} made changes to the data structures of ICN, exploiting network coding to use the multiple paths connecting the clients to the sources. For multi-user scenarios, 
% the GB algorithm \cite{tan2021game} models streaming as a Bayesian game with Nash Equilibrium-based decisions. 
Wu et al. \cite{9195016} proposed NDN-MMRA for multi-stage multicast rate adaptation in wireless networks, adjusting rates through multi-stage transmission to balance high- and low-speed consumers. Alt et al. \cite{Alt} proposed a Contextual Quality Adaptation (CBA) algorithm, formulating adaptation as a contextual multi-armed bandit problem.
% without fully leveraging ICN's aggregation capabilities. 
Despite their merits, these approaches primarily rely on localized reactions (i.e., congestion signals) without sophisticated mechanisms for anticipating upstream constraints or coordinating across user groups. This leads to decision lag and instability when facing high user concurrency and fluctuating network conditions. COMETS advances beyond reactive local adaptations by introducing a distributed optimization framework with analytical closed-form solutions, enabling proactive group-aware rate allocation that fully exploits ICN's aggregation benefits while ensuring rapid convergence and fairness.

% \vspace{-5pt}
\subsection*{Overlay/multicast systems}

COMETS also differs fundamentally from traditional hierarchical overlay multicast streaming systems.
These systems primarily focus on constructing efficient overlay topologies (trees, meshes, or hybrid structures) and maintaining them under node dynamics.
Representative systems include BAHMO (Bandwidth Adapted Hierarchical Multicast Overlay) \cite{bagheri2010bandwidth}, which constructs hierarchical multicast trees by grouping peers into clusters with servers in each cluster; OMTP (Overlay Multicast Tree Protocol) \cite{kwan2005overlay}, which focuses on improving joining and maintenance procedures of overlay multicast trees. These systems invest computational resources in optimizing tree structures based on factors like delay, bandwidth availability, and node stability. In contrast, COMETS addresses the orthogonal problem of coordinated multi-user rate adaptation within any distribution network. While overlay multicast optimizes `where data flows', COMETS optimizes `what quality variant flows to each user group to maximize aggregate QoE'. In essence, COMETS \textit{complements} overlay multicast; it can operate over any underlying topology, including an overlay multicast tree, to add a layer of adaptive intelligence that these systems traditionally lack.

\section{COMETS System Design}
\label{design}
Given the shortcomings in both end-host adaptation and purely centralized coordination, COMETS distributes adaptation logic into the network itself. We present the COMETS system design, including architecture, protocol design, and security aspects.
Fundamentally, COMETS reimagines video streaming as a distributed negotiation within a network fabric, where forwarders (acting as intelligent intermediaries) recursively aggregate, optimize, and adjust content flows in a hierarchical tree.
% This section elucidates the design of COMETS, modularized into architecture, protocol, and security facets to illuminate its core innovations.
% \vspace{-5pt}
\subsection{System Architecture}

COMETS employs a distributed hierarchical architecture where network forwarders actively participate in content delivery optimization. At a high level, forwarders serve as decentralized decision points, aggregating requests from downstream clients, monitoring local network conditions (e.g., bandwidth, congestion), and computing optimal video resolutions for user groups to maximize collective QoE while minimizing redundancy. This occurs recursively throughout the distribution tree: leaf forwarders handle direct clients, propagating aggregated data upstream, where parent forwarders optimize for larger subtrees, ensuring scalability without global synchronization.

We assume forwarders are deployed by the video provider or CDN operator (e.g., edge PoPs), under a single administrative domain. This enables control over trust anchors and supports efficient cache placement and Range Interest handling.

\begin{figure}[ht]
\centering
\includegraphics[width=0.5\textwidth]{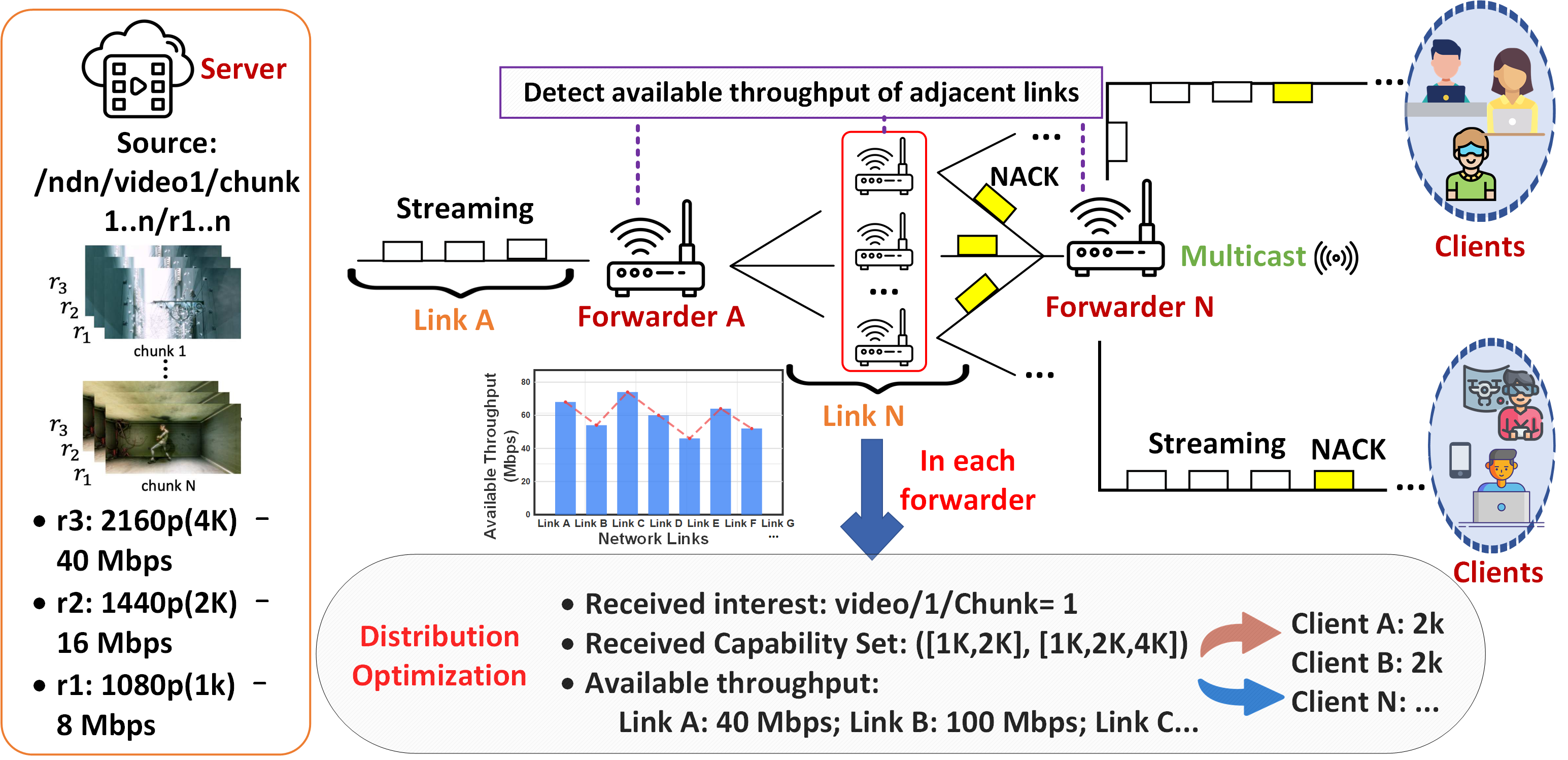}
\caption{System Architecture. Forwarders optimize quality decisions based on client capabilities and available bandwidth.}
\label{fig:system_architecture}
\end{figure}

\begin{figure}[ht]
\centering
\includegraphics[width=0.5\textwidth]{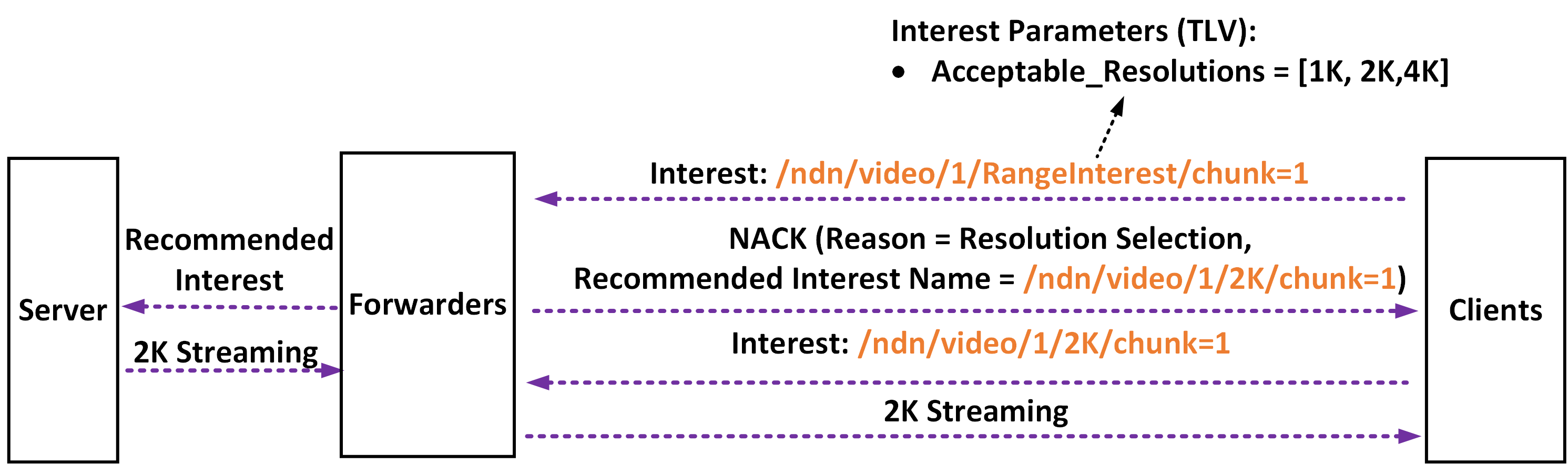}
\caption{Client-forwarder-server communication via interests.}
\label{fig:Protocol}
\end{figure}

Figure \ref{fig:system_architecture} illustrates the system architecture and Figure \ref{fig:Protocol} shows the communication flow between clients, forwarders, and servers. When requesting video content, clients send Range-based Interests that include their supported resolution capabilities (e.g., [2K, 4K, 8K]) instead of requesting specific resolutions directly. Forwarders process these to compute optimal configurations—specifically, determining the best resolution assignments that balance client capabilities, upstream bandwidth availability, and fairness metrics (e.g., minimizing QoE variance)—based on aggregated downstream needs and local constraints.
We initially developed a centralized ILP approach as our theoretical baseline (Problem~\ref{prob_qoe}), which solves for global optimality but faces deployment limits in data collection and scalability. To overcome this, we devised a distributed Lagrangian algorithm (Problem~\ref{master_prob}) that enables forwarders to solve localized subproblems, approaching centralized performance with only neighborhood information—fundamentally, this decomposes the combinatorial challenge of resolution allocation into analytically tractable parts, enabling real-time decisions that scale with network size.

% \vspace{-10pt}

% \vspace{-5pt}

\subsection{Protocol Design}

COMETS operation relies on structured message exchanges between clients, forwarders, and coordination nodes, as illustrated in Figure \ref{fig:Protocol}. Clients initiate the process by periodically (every $T=4$ seconds, aligned with standard adaptive streaming practices) sending ICN Range Interest messages following the naming convention \textit{/ndn/video/TitleID/RangeInterest/chunk=seq[/nonce=val]}. These messages convey the client's identifier, acceptable resolution levels $L_u$, and QoE weights $w_u$ (defined in Section \ref{Distributed Decision Algorithm}). Clients then await resolution assignments and use them to request video data via standard ICN Interests.
Forwarders serve as primary computational and interaction entities, aggregating downstream client capabilities and monitoring network conditions. Depending on deployment configuration, forwarders implement:
% one of two coordination approaches:

\begin{enumerate}
    \item \textbf{Centralized Coordination Mode:} Forwarders report aggregated information to a designated Optimizer using ICN Interests \(I_{\text{report}}\) (naming pattern: \textit{/ndn/opt/report/forwarder=FwdID/...}). The Optimizer computes decisions and distributes signed configuration data, which forwarders retrieve through versioned Interests \(I_{\text{config}}\) (e.g., \textit{/ndn/opt/config/forwarder=FwdID/v=version}). Outdated version requests trigger NACKs indicating the latest version, prompting proper requests. Upon verification, forwarders apply these decisions. The scalability is limited by the central node.

    \item \textbf{Distributed Optimization Mode:} Forwarders exchange state variables and Lagrangian multipliers with peers via versioned, signed ICN Interest/Data pairs (e.g., \textit{\(I_{\text{state\_req}}(\text{NodeID, Var, Level, } t)\)} following \textit{/ndn/comets/state/node/NodeID/Var/Level/v=}\(t\)). Each forwarder solves local subproblems using analytical solutions from Section \ref{Distributed Decision Algorithm}, updates its dual variables per equations, and publishes results as signed, versioned ICN data (\(D_{\text{state\_pub}}(\text{NodeID, Var, Level, } t+1)\)). This fetch-compute-update-publish cycle iterates within the 4-second optimization window.
\end{enumerate}

Upon determining the optimal resolution \(Res_u^*\) for client \(u\), the forwarder proactively fetches the video segment by issuing a standard ICN Interest for the data \(N_{video}(TitleID, Res_u^*, seq)\). Subsequently, the forwarder notifies the client of this optimal resolution via a signed NACK with a Recommended Name field specifying the pattern \textit{/ndn/video//\(Res_u^*\)/chunk=}, which serves as a verified instruction for the client. Following this recommendation, the client issues its own standard ICN Interest for the segment. This preemptive fetching strategy by the forwarder is designed to ensure the required data segment is potentially already cached locally or in transit when the client's request arrives.

Range Interests resemble \textit{Remote Procedure Calls (RPCs)} in their request-response pattern, but provide different semantics in that they do not assume a specific "server" as the RPC target.
% differ fundamentally in their distributed nature. While RPCs emphasize procedure calls to specific servers with concentrated decision-making authority,
Range Interests function as a distributed negotiation protocol: they are sent into the network without targeting specific servers, allowing dedicated trusted forwarders to intercept, aggregate, and respond collaboratively based on local and propagated state. The key distinction is that COMETS embeds rate negotiation semantics into the Interest–Data exchange itself, enabling decentralized optimization decisions without relying on repeated feedback loops or per-client control paths.

\subsection{Reliability, Congestion Control, and Security}
\label{Reliability}

To ensure robust video delivery over varying network conditions, COMETS integrates a consumer-driven transport mechanism with proactive in-network congestion control and a rigorous trust model. This design allows the system to handle packet loss, node failures, and security threats without relying on heavy connection-oriented protocols.

\textbf{Transport Mechanism and Loss Recovery.} COMETS relies on ICN’s stateful forwarding plane for reliability. We adopt a receiver-driven recovery strategy to handle packet loss effectively. Clients maintain a retransmission timer for each expressed Interest. A timeout indicates potential packet loss or link failure. Upon detecting a timeout, the client initiates a limited retransmission (e.g., up to 2 attempts) to recover from transient random losses such as wireless interference. Since Data packets are cached at intermediate forwarders, re-expressed Interests can often be satisfied by a nearby upstream node via local repair, significantly reducing recovery latency compared to end-to-end retransmissions. However, if retransmissions fail or if multiple consecutive timeouts occur, the client interprets this as network congestion or path failure rather than random loss. In this scenario, the client suppresses further retransmissions for the current segment and triggers the rate adaptation logic to request a lower quality resolution in subsequent intervals. This aligns with the optimization framework’s objective to maximize aggregate QoE under constrained capacity. 

\textbf{Congestion Control and Dynamics.} COMETS employs a dual-loop congestion control mechanism that handles dynamic bandwidth fluctuations and user mobility. On the client side, an AIMD-based window control is utilized for Interest pipelining. The window size adjusts based on Data arrival rates, ensuring the request rate matches the bottleneck bandwidth. To address the "decision lag" typical in purely client-side approaches, COMETS incorporates in-network backpressure. When a forwarder detects that its downstream link capacity is saturated or its queue exceeds a threshold, it proactively issues a NACK packet carrying a specific reason code and a recommendation (e.g., pointing to a lower resolution representation). This allows the client to immediately switch to a sustainable bitrate without waiting for a timeout. This mechanism also supports seamless mobility; as users switch access points, the stateless nature of Interest packets allows the new serving forwarder to immediately process requests or issue NACKs based on the new path’s state. 
% \footnote{A detailed technical discussion of these mechanisms, including comparison with remote procedure calls (RPCs), failure assumptions, reliability mechanisms, and congestion control feedback loops, is provided in Appendix K.}}

\textbf{Security.} COMETS ensures content authenticity and coordination integrity through a CDN-like trust model within a single administrative domain (e.g., similar to Akamai \cite{su2009drafting} or CloudFlare \cite{shobiri2023cdns} infrastructures). The content producer serves as the root of trust, issuing certificates to authorized entities including forwarders and optimizers. Critical control messages—such as state variables, optimizer configurations, and proactive NACKs—are signed by the issuer. Forwarders verify these signatures against the trust hierarchy to prevent unauthorized state injection or spoofing attacks. Furthermore, content chunks are signed by the producer or an authorized transcoder. Clients verify the signature of every Data packet upon receipt. This decouples security from the transmission channel, allowing content to be safely retrieved from any cache or peer without risking data tampering. 
Signing every segment with public-key cryptography can be expensive in large deployments \cite{yu2024secure}. 
For efficiency, COMETS can adopt manifest-based approaches where multiple data objects are linked through hash chains, requiring only one signature per manifest rather than per object \cite{irtf-icnrg-flic-06}. This reduces cryptographic overhead while maintaining authenticity.

Following sections formalize these coordination mechanisms through a distributed optimization framework that mathematically ensures optimal QoE allocation across all users.

% \vspace{-0.7 em}

\section{System Model}\label{sec:model}
The network is modeled as a directed acyclic graph (DAG) $\mathcal{G} = (\mathcal{V}, \mathcal{E})$, where $\mathcal{V}$ is the set of nodes, and $\mathcal{E}$ is the set of edges. The node set is partitioned into three disjoint subsets: the set of servers  $\mathcal{S}$ (content sources), the set of forwarders  $\mathcal{F}$ (intermediate nodes), and the set of users $\mathcal{U}$ (end nodes). Note that $\mathcal{V} = \mathcal{S} \cup \mathcal{F} \cup \mathcal{U}$, $\mathcal{S} \cap \mathcal{F} = \emptyset$, $\mathcal{F} \cap \mathcal{U} = \emptyset$, and $\mathcal{S} \cap \mathcal{U} = \emptyset$.
For each node $i \in \mathcal{V}$, we define the upstream node set and the downstream node set as  $\mathcal{I}_i = \{j \mid (j,i) \in \mathcal{E}\}$ and  $\mathcal{O}_i = \{j \mid (i,j) \in \mathcal{E}\}$, respectively. The system provides a set of $L$ resolution levels $\mathcal{L} = \{1, 2, \ldots, L\}$, where each level $l \in \mathcal{L}$ is associated with a bandwidth requirement $B_l$ and a quality score $Q_l$. To capture the diminishing marginal benefit of increasing resolution on user experience,  we model $Q_l$ as a increasing, concave function of $l$, e.g.,  we can choose \cite{raake2017bitstream}:

\small
\begin{equation}\label{eq:quality}
Q_l = a + b \cdot \ln\left(\frac{R_l}{R_1}\right), \quad l \in \mathcal{L}.
\end{equation}
\normalsize
Here, $R_l$ denotes the resolution corresponding to level $l$, and $a$ and $b$ are constant coefficients. 
Each user $u\in \mathcal{U}$ has a weight $w_u > 0$ reflecting its priority.
Each network link $(i,j) \in \mathcal{E}$ has a bandwidth capacity limit $C_{i,j}>0$. 
% Each user $u \in \mathcal{U}$ is assigned a priority weight $w_u$ to reflect its relative importance in the system optimization. 

Let $\mathbf{x} \triangleq (x_{i,l})_{i \in \mathcal{V}, l \in \mathcal{L}} \in \{0,1\}^{|\mathcal{V}| \times |\mathcal{L}|}$ represent the node-resolution selection, where $x_{i,l} = 1$ indicates that node $i$ selects content of resolution $l$, for all $i \in \mathcal{V}$ and  $l \in \mathcal{L}$.  
\begin{itemize}
  \item For all $i \in \mathcal{U}$, $x_{i,l} = 1$ indicates that user $i$ requests content of resolution $l$.
  \item For all $i \in \mathcal{F}$, $x_{i,l} = 1$ indicates that forwarder $i$ retrieves and forwards content of resolution $l$.
  \item For all $i \in \mathcal{S}$, $x_{i,l} = 1$ indicates that server $i$ provides content of resolution $l$.
\end{itemize}
Let $\mathbf{y} \triangleq (y_{i,j,l})_{(i,j) \in \mathcal{E}, l \in \mathcal{L}} \in \{0,1\}^{|\mathcal{E}| \times |\mathcal{L}|}$ represent the edge-resolution transmission, where $y_{i,j,l} = 1$ indicates that edge $(i,j)$ transmits content of resolution $l$, for all $(i,j) \in \mathcal{E}$  and $ l \in \mathcal{L}$. Thus, $\mathbf{x}$ and $\mathbf{y}$ satisfy:
% \small
\begin{align}
& x_{s,l} = 1, \quad s \in \mathcal{S}, \; l \in \mathcal{L}, \label{c1}
\end{align}
\begin{align}
& x_{u,l} = 0, \quad u \in \mathcal{U}, \; l \in \mathcal{L} \setminus \mathcal{L}_u, \label{c2}
\end{align}
\begin{align}
& \sum_{l \in \mathcal{L}} x_{u,l} = 1, \quad u \in \mathcal{U}, \label{c3}
\end{align}
\begin{align}
& y_{i,j,l} \leq x_{i,l}, \quad (i,j) \in \mathcal{E}, \; l \in \mathcal{L}, \label{c4}
\end{align}
\begin{align}
& x_{i,l} \leq \sum_{j:(j,i) \in \mathcal{E}} y_{j,i,l}, \quad i \in \mathcal{U} \cup \mathcal{F}, \; l \in \mathcal{L}, \label{c5}
\end{align}
\begin{align}
& \sum_{l \in \mathcal{L}} B_l y_{i,j,l} \leq C_{i,j}, \quad (i,j) \in \mathcal{E}, \label{c6}
\end{align}
\begin{align}
& x_{i,l}, y_{i,j,l} \in \{0, 1\}, \quad i \in \mathcal{V}, \; (i,j) \in \mathcal{E}, \; l \in \mathcal{L}. \label{c7}
\end{align}
% \normalsize

Specifically, \eqref{c1} indicates that each server provides content at all resolution levels; \eqref{c2} indicates that each user can only select resolution levels supported by their hardware or network conditions; \eqref{c3} indicates that each user selects exactly one resolution level; \eqref{c4} indicates that each node can only transmit content of resolution levels they have acquired; \eqref{c5} indicates that each non-server node can only acquire content of resolution levels delivered by the upstream edges; \eqref{c6} indicates that each edge's  transmission bandwidth must not exceed its bandwidth capacity limit; and \eqref{c7} indicates that all decision variables are binary. The network performance metric is set as the weighted sum QoE of all users:
% \footnote{Key notation is summarized in Table I of Appendix A.}
% \small
$$Z(\mathbf{x}, \mathbf{y}) \triangleq \sum_{u \in \mathcal{U}} w_u \sum_{l \in \mathcal{L}} Q_l x_{u,l}.$$ 
% \normalsize

\section{Multi-user QoE Optimization}
\label{Distributed Decision Algorithm}
We now formalize the resolution selection problem introduced in Section \ref{design}. Table~\ref{tab:notation} provides a comprehensive list of notations used throughout the paper.

\begin{table}[h]
\centering
\caption{LIST OF NOTATIONS}
\label{tab:notation}
\resizebox{0.47\textwidth}{!}{%
\begin{tabular}{ll}
\hline
\textbf{Symbol} & \textbf{Description} \\
\hline
\multicolumn{2}{l}{\textit{Network Structure}} \\
$\mathcal{G} = (\mathcal{V}, \mathcal{E})$ & Directed acyclic graph representing the network \\
$\mathcal{V}$ & Set of all nodes \\
$\mathcal{S}$ & Set of server nodes \\
$\mathcal{F}$ & Set of forwarder nodes \\
$\mathcal{U}$ & Set of user nodes \\
$\mathcal{I}_i$ & Upstream nodes of node $i$ (incoming neighbors) \\
$\mathcal{O}_i$ & Downstream nodes of node $i$ (outgoing neighbors) \\
\hline
\multicolumn{2}{l}{\textit{Video Parameters}} \\
$\mathcal{L} = \{1, 2, \dots, L\}$ & Set of resolution levels \\
$\mathcal{L}_u \subseteq \mathcal{L}$ & Resolutions supported by user $u$ \\
$B_l > 0$ & Bandwidth requirement for resolution $l$ \\
& (increasing in $l$) \\
$Q_l = a + b \cdot \ln(l / l_0)$ & Quality score for resolution $l$ \\
& (logarithmic model) \\
$w_u > 0$ & Priority weight for user $u$ \\
$C_{i,j}$ & Capacity of link $(i,j) \in \mathcal{E}$ \\
\hline
\multicolumn{2}{l}{\textit{Decision Variables}} \\
$x_{i,l} \in \{0,1\}$ & 1 if node $i$ selects resolution $l$ \\
$y_{i,j,l} \in \{0,1\}$ & 1 if resolution $l$ is transmitted on link $(i,j)$ \\
$\mathbf{x} = (x_{i,l})_{i \in \mathcal{V}, l \in \mathcal{L}}$ & Matrix of all selection variables \\
$\mathbf{y} = (y_{i,j,l})_{(i,j) \in \mathcal{E}, l \in \mathcal{L}}$ & Matrix of all transmission variables \\
\hline
\multicolumn{2}{l}{\textit{Dual Variables}} \\
$\boldsymbol{\lambda}_1 = \{\lambda_{1,i,l} \ge 0\}$ & Multipliers for incoming transmission \\
& constraints (\ref{c5}) \\
$\boldsymbol{\lambda}_2 = \{\lambda_{2,f,k,l} \ge 0\}$ & Multipliers for outgoing transmission \\
& constraints (\ref{c4}) \\
\hline
\multicolumn{2}{l}{\textit{Algorithm Parameters}} \\
$\alpha^{(t)}, \beta^{(t)}$ & Step sizes at iteration $t$ (diminishing) \\
$\alpha_0, \beta_0, \gamma$ & Initial step size parameters and decay rate \\
$T_{\max}$ & Maximum number of iterations \\
$\epsilon$ & Convergence tolerance threshold \\
$\sigma_i$ & Permutation ordering resolutions by density $\lambda_{1,i,l}/B_l$ \\
$\theta_i$ &
Cutoff index for capacity-constrained knapsack filling \\
\hline
\multicolumn{2}{l}{\textit{Functions and Objectives}} \\
$Z(\mathbf{x}, \mathbf{y})$ & Weighted average QoE objective \\
$= \sum_{u \in \mathcal{U}} w_u \sum_{l \in \mathcal{L}} Q_l x_{u,l}$ & \\
$\mathcal{L}(\mathbf{x}, \mathbf{y}, \boldsymbol{\lambda}_1, \boldsymbol{\lambda}_2)$ & Lagrangian function of the relaxed problem \\
$g(\boldsymbol{\lambda}_1, \boldsymbol{\lambda}_2)$ & Dual function (maximization over Lagrangian) \\
\hline
\end{tabular}}
\end{table}

\subsection{Problem Formulation and Transformation}\label{sec:formulation}

We aim to optimize the resolution allocation $\mathbf{x}$ and $\mathbf{y}$ to maximize the weighted sum QoE $Z(\mathbf{x}, \mathbf{y})$ subject to the feasibility constraints in (\ref{c1})-(\ref{c7}). Specifically, we formulate the following optimization problem, which is supposed feasible.
\begin{prob}[QoE Maximization]\label{prob_qoe}
% \small
\begin{align}
\max_{\mathbf{x},\mathbf{y}} \quad&\sum_{u \in \mathcal{U}} w_u \sum_{l \in \mathcal{L}} Q_l x_{u,l} \notag\\
\text{s.t.} \quad&(\ref{c1}),(\ref{c2}),(\ref{c3}),(\ref{c4}),(\ref{c5}),(\ref{c6}),(\ref{c7}).\notag
\end{align}
\end{prob}
% \normalsize

The objective function and the constraints in (\ref{c1})-(\ref{c6}) are linear in $\mathbf{x}$, while the constraints in (\ref{c7}) enforce discreteness. Thus, Problem~\ref{prob_qoe} is a challenging ILP problem, generally NP-hard~\cite{boyd2004convex}. To solve Problem~\ref{prob_qoe} efficiently, we first consider its continuous relaxation, resulting in linear programming (LP).

\begin{prob}[Continuous Relaxation of Problem~\ref{prob_qoe}]\label{prob_relax}
\begin{align}
\max_{\mathbf{x},\mathbf{y}} \quad&\sum_{u \in \mathcal{U}} w_u \sum_{l \in \mathcal{L}} Q_l x_{u,l} \notag\\
\text{s.t.} \quad& (\ref{c1}),(\ref{c2}),(\ref{c3}),(\ref{c4}),(\ref{c5}),(\ref{c6}), \notag\\
& x_{i,l}, y_{i,j,l} \in [0,1], \quad i \in \mathcal{V}, \; (i,j) \in \mathcal{E}, \; l \in \mathcal{L}. \label{c9}
\end{align}
\end{prob}
% \normalsize

Problem~\ref{prob_relax} (with the continuous constraints in (\ref{c9}) can be viewed as the continuous relaxation of Problem~\ref{prob_qoe} (with the discrete constraints in (\ref{c7})). Problem~\ref{prob_relax}'s objective function and constraint functions are all linear in $\mathbf{x},\mathbf{y}$. Thus, Problem~\ref{prob_relax} is an LP~\cite[pp. 146]{boyd2004convex} and can be solved in polynomial time using the simplex method or interior point method. However, for large $L$, the simplex method (per-iteration complexity $\mathcal{O}(L^2)$) and the interior point method (overall complexity $\mathcal{O}(L^3)$) are very time consuming. Besides, the optimal solution of the LP in Problem~\ref{prob_relax} may not be integers and hence not feasible for the original ILP in Problem~\ref{prob_qoe}. 

\subsection{Scalable Method}

We develop a scalable method to obtain a near-optimal solution of Problem~\ref{prob_qoe}. \footnote{A detailed analysis comparing centralized ILP solvers, AI-based methods, and our proposed method is provided in Appendix A.} Specifically, we first propose a dual decomposition algorithm to obtain an optimal solution of the LP in Problem~\ref{prob_relax} using the Dual Decomposition Algorithm \cite{boyd2007notes}. Then, based on the optimal solution of the LP in Problem~\ref{prob_relax}, we propose a feasibility reconstruction algorithm to obtain a feasible point of Problem~\ref{prob_qoe} that is near optimal. 
% \footnote{All theoretical proofs are provided in the Appendix.}

\subsubsection{Dual Decomposition Algorithm}\label{sec:dualalg}
% In this section, we propose an efficient algorithm to obtain an optimal point of Problem~\ref{prob_relax} using the partial Dual Decomposition Algorithm~\cite{boyd2004convex}, 

Given that Problem~\ref{prob_relax} is convex and satisfies strong duality, and the variables are coupled through the constraints in \eqref{c4} and \eqref{c5}, we form the Lagrangian function $\mathcal{L}(\mathbf{x}, \mathbf{y}, \boldsymbol{\lambda}_1, \boldsymbol{\lambda}_2)$ in \eqref{eq:lagrangian_function}, as shown at the top of this page, by augmenting the objective function with a weighted sum of the constraints (\ref{c4}) and (\ref{c5}).
Here, $\boldsymbol{\lambda}_1 \triangleq (\lambda_{1,i,l})_{i \in \mathcal{U} \cup \mathcal{F}, l \in \mathcal{L}}\succeq \mathbf 0$ and $\boldsymbol{\lambda}_2 \triangleq (\lambda_{2,f,k,l}) _{f \in \mathcal{F}, k \in \mathcal{O}_f, l \in \mathcal{L}} \succeq \mathbf 0$ are the Lagrange multipliers associated with the constraints in \eqref{c5} and \eqref{c4}, respectively.

% \small
\begin{figure*}[t]
\begin{equation}\label{eq:lagrangian_function}
\begin{aligned}
\mathcal{L}(\mathbf{x}, \mathbf{y}, \boldsymbol{\lambda}_1, \boldsymbol{\lambda}_2) &= \sum_{u \in \mathcal{U}} w_u \sum_{l \in \mathcal{L}} Q_l x_{u,l} + \sum_{i \in \mathcal{U} \cup \mathcal{F}} \sum_{l \in \mathcal{L}} \lambda_{1,i,l} \left( x_{i,l} - \sum_{j:(j,i) \in \mathcal{E}} y_{j,i,l} \right) 
+ \sum_{f \in \mathcal{F}} \sum_{k \in \mathcal{O}_f} \sum_{l \in \mathcal{L}} \lambda_{2,f,k,l} ( y_{f,k,l} - x_{f,l})
\end{aligned}
\end{equation}
\hrulefill
\end{figure*}
% \normalsize

The Lagrangian dual function is obtained by maximizing the Lagrangian function $\mathcal{L}(\mathbf{x}, \mathbf{y}, \boldsymbol{\lambda}_1, \boldsymbol{\lambda}_2)$ over the primal variables $(\mathbf{x},\mathbf{y})$ subject to the remaining  constraints:

% \small
\begin{equation}
\begin{aligned} \label{dual_function}
g(\boldsymbol{\lambda}_1,\boldsymbol{\lambda}_2) \triangleq&  \max_{\mathbf{x},\mathbf{y}} \mathcal{L}(\mathbf{x},\mathbf{y},\boldsymbol{\lambda}_1,\boldsymbol{\lambda}_2) \\ 
\text{s.t.} \quad& (\ref{c1}),(\ref{c2}),(\ref{c3}),  (\ref{c6}), (\ref{c9}) 
\end{aligned}
\end{equation}
% \normalsize
% The Lagrangian dual function is obtained by maximizing the Lagrangian function $\mathcal{L}(\mathbf{x}, \mathbf{y}, \boldsymbol{\lambda}_1, \boldsymbol{\lambda}_2)$ over the primal variables $(\mathbf{x},\mathbf{y})$ subject to the remaining  constraints:
% \begin{equation}
% \begin{aligned} \label{dual_function}
% g(\boldsymbol{\lambda}_1,\boldsymbol{\lambda}_2) \triangleq&  \max_{\mathbf{x},\mathbf{y}} \mathcal{L}(\mathbf{x},\mathbf{y},\boldsymbol{\lambda}_1,\boldsymbol{\lambda}_2) \\
% =& \sum_{s \in \mathcal{S}} \sum_{i \in \mathcal{O}_s} \max_{\mathbf{y}_{s,i}} \sum_{l \in \mathcal{L}} \lambda_{1,i,l} y_{s,i,l} \\
% &+ \sum_{u \in \mathcal{U}} \max_{\mathbf{x}_u} \sum_{l \in \mathcal{L}} (w_u Q_l - \lambda_{1,u,l}) x_{u,l} \\
% &+ \sum_{f \in \mathcal{F}} \max_{\mathbf{x}_f} \sum_{l \in \mathcal{L}} \left( \sum_{k \in \mathcal{O}_f} \lambda_{2,f,k,l} - \lambda_{1,f,l} \right) x_{f,l} \\
% &+ \sum_{f \in \mathcal{F}} \sum_{k \in \mathcal{O}_f} \max_{\mathbf{y}_{f,k}} \sum_{l \in \mathcal{L}} (\lambda_{1,k,l} - \lambda_{2,f,k,l}) y_{f,k,l} \\ 
% \text{s.t.} \quad& (\ref{c1}),(\ref{c2}),(\ref{c3}),  (\ref{c6}), (\ref{c9}) 
% \end{aligned}
% \end{equation} 
The dual problem is to find the optimal dual variables (Lagrange multipliers) that minimize the dual function:
\begin{equation}
\begin{aligned}\label{dual_problem}
\min_{\boldsymbol{\lambda}_1 \succeq 0, \boldsymbol{\lambda}_2 \succeq 0} \quad& g(\boldsymbol{\lambda}_1, \boldsymbol{\lambda}_2).
\end{aligned}
\end{equation}
Let \(\boldsymbol{\lambda}^* \triangleq (\boldsymbol{\lambda}_1^*, \boldsymbol{\lambda}_2^*)\) denote the optimal Lagrange multipliers.

Since $\mathcal{L}(\mathbf{x}, \mathbf{y}, \boldsymbol{\lambda}_1, \boldsymbol{\lambda}_2)$ in (\ref{eq:lagrangian_function}) is \textbf{separable} with respect to 
$\mathbf{y}_{s,i} \triangleq (y_{s,i,l})_{l \in \mathcal{L}}$, $s \in \mathcal{S}$, $i \in \mathcal{O}_s$, 
$\mathbf{x}_u \triangleq (x_{u,l})_{l \in \mathcal{L}}$, $u \in \mathcal{U}$, 
$\mathbf{x}_f \triangleq (x_{f,l})_{l \in \mathcal{L}}$, and $\mathbf{y}_{f,k} \triangleq (y_{f,k,l})_{l \in \mathcal{L}}$, $f \in \mathcal{F}$, $k \in \mathcal{O}_f$, \eqref{dual_problem} can be separated into four subproblems coordinated by a master problem. %Specifically, the dual function decomposes as:
\begin{prob}[Master Dual Problem]\label{master_prob}
% \small
% {\fontsize{7.8pt}{10pt}\selectfont
\begin{align}
\min_{\boldsymbol{\lambda}_1 \succeq 0, \boldsymbol{\lambda}_2 \succeq 0}  \sum_{s \in \mathcal{S}, i \in \mathcal{O}_s} \sum_{l \in \mathcal{L}} \lambda_{1,i,l} y_{s,i,l}^*+  \sum_{u \in \mathcal{U}} \sum_{l \in \mathcal{L}} (w_u Q_l - \lambda_{1,u,l}) x_{u,l}^* \\ 
+ \sum_{f \in \mathcal{F}} \sum_{l \in \mathcal{L}} \left( \sum_{k \in \mathcal{O}_f} \lambda_{2,f,k,l} - \lambda_{1,f,l} \right) x_{f,l}^* \notag \\ + \sum_{f \in \mathcal{F}, k \in \mathcal{O}_f} \sum_{l \in \mathcal{L}} (\lambda_{1,k,l} - \lambda_{2,f,k,l}) y_{f,k,l}^*
 \notag 
\end{align}
% }
\end{prob}
% \normalsize
\textbf{where $y_{s,i,l}^*$, $x_{u,l}^*$, $x_{f,l}^*$, and $y_{f,k,l}^*$ are the optimal solutions of Problems~\ref{subprob_server}, \ref{subprob_user}, \ref{subprob_forwarder_x}, and \ref{subprob_forwarder_y}, respectively, given the dual variables $(\boldsymbol{\lambda}_1, \boldsymbol{\lambda}_2)$.}

\begin{prob}[Server Subproblem w.r.t. $\mathbf{y}_{s,i}$]\label{subprob_server}
For all $\lambda_{1,i,l} \ge 0$, $s \in \mathcal{S}$, $i \in \mathcal{O}_s$, we have: 
\begin{align}
\max_{\mathbf{y}_{s,i}} \quad& \sum_{l \in \mathcal{L}} \lambda_{1,i,l} y_{s,i,l} \notag\\
\text{s.t.} \quad& (\ref{c6}), (\ref{c9}) \notag
\end{align}
% The optimal point is denoted by $\mathbf{y}_{s,i}^*(\boldsymbol{\lambda}_1)$.
\end{prob}

\begin{prob}[User Subproblem w.r.t. $\mathbf{x}_u$]\label{subprob_user}
For all $\lambda_{1,u,l} \ge 0$, $u \in \mathcal{U}$, we have:
% \small
\begin{align}
\max_{\mathbf{x}_u} \quad& \sum_{l \in \mathcal{L}} (w_u Q_l - \lambda_{1,u,l}) x_{u,l} \notag\\
\text{s.t.} \quad& (\ref{c2}), (\ref{c3}), (\ref{c9}). \notag
\end{align}
% The optimal point is denoted by $\mathbf{x}_u^*(\boldsymbol{\lambda}_1)$.
\end{prob}
% \normalsize
\begin{prob}[Forwarder Subproblem w.r.t. $\mathbf{x}_f$]\label{subprob_forwarder_x}
For all $\lambda_{1,f,l}, \lambda_{2,f,k,l} \ge 0$, $f \in \mathcal{F}$, we have: 
% \small
\begin{align}
\max_{\mathbf{x}_f} \quad& \sum_{l \in \mathcal{L}} \left( \sum_{k \in \mathcal{O}_f} \lambda_{2,f,k,l} - \lambda_{1,f,l} \right) x_{f,l} \notag\\
\text{s.t.} \quad&  x_{f,l} \in [0,1], \quad l \in \mathcal{L}. \notag
\end{align}
% \normalsizes
% The optimal point is denoted by $\mathbf{x}_f^*(\boldsymbol{\lambda}_1, \boldsymbol{\lambda}_2)$.
\end{prob}

\begin{prob}[Forwarder Subproblem w.r.t. $\mathbf{y}_{f,k}$]\label{subprob_forwarder_y}
For all $\lambda_{1,f,l}, \lambda_{2,f,k,l} \ge 0$, $f \in \mathcal{F}$, $k \in \mathcal{O}_f$, we have:
\begin{align}
\max_{\mathbf{y}_{f,k}} \quad& \sum_{l \in \mathcal{L}} (\lambda_{1,k,l} - \lambda_{2,f,k,l}) y_{f,k,l} \notag\\
\text{s.t.} \quad&  (\ref{c6}), (\ref{c9}). \notag
\end{align}
% The optimal point is denoted by $\mathbf{y}_{f,k}^*(\boldsymbol{\lambda}_1, \boldsymbol{\lambda}_2)$.
\end{prob}
% the nodes. This allows the inner maximization problem in (\ref{dual}) to be \textbf{decomposed} into independent subproblems, one for each node (or edge originating from a server). These subproblems can be solved in parallel at each node, using only local information and the current values of the dual variables communicated from neighbors.

\begin{lemma}[Decomposition of Dual Problem]\label{lemma_decomposition}
The dual problem in \eqref{dual_problem} with the dual function given by \eqref{dual_function} is equivalent to the master dual problem in Problem~\ref{master_prob} together with the subproblems in Problems~\ref{subprob_server}, \ref{subprob_user}, \ref{subprob_forwarder_x}, and \ref{subprob_forwarder_y}.
% decomposes into the sum of optimal values from Problems~\ref{subprob_server}, \ref{subprob_user}, \ref{subprob_forwarder_x}, and \ref{subprob_forwarder_y}, coordinated by the master problem in Problem~\ref{master_prob}. Problem~\ref{master_prob} is equivalent to the dual problem \eqref{dual_problem}.
\end{lemma}
\begin{IEEEproof}
See Appendix B for detailed proof.
\end{IEEEproof}

% ---
% Problems~\ref{subprob_server}--\ref{subprob_forwarder_y} are linear programs (LPs) or continuous knapsack problems, solved using standard techniques from optimization textbooks~\cite[Ch. 3]{bertsimas1997introduction}.

% For Problem~\ref{subprob_server}, this is a continuous knapsack problem~\cite{bertsekas2003convex}. 

In the following, we first solve the subproblems for any given $\boldsymbol{\lambda}_1 \succeq 0, \boldsymbol{\lambda}_2 \succeq 0$ and then  solve the master problem to obtain  $(\boldsymbol{\lambda}_1^*, \boldsymbol{\lambda}_2^*)$.

% 排序唯一性（profit/weight）的条件：iff所有物品的价值密度严格不同（strictly distinct）。 在我们的背包问题中，只能保证B_l严格递增，不能保证 λ_{1,i,l}/B_l 一定不同，因此存在ties。

% 背包问题的解唯一性的条件：iff 在贪婪填充后，部分物品（fractional item）是唯一的。

% 我们的背包问题可以保证最大唯一，因为贪婪填充最大化价值密度，且ties不改变总价值。

% 设想的保证解唯一性的方法：在实现时添加稳定排序键：按密度降序，tie时按B_l升序（e.g., 优先小重量的）。
% 或者对密度添加 ε 扰动以实现严格排序

Problem~\ref{subprob_server} is a continuous knapsack problem~\cite{bertsekas2003convex}. The optimal solution is obtained by sorting resolutions in descending order of the value density $\lambda_{1,i,l}/B_l, l \in \mathcal{L}$. For all $s \in \mathcal{S}$ and $i \in \mathcal{O}_s$, 
let $\sigma_i: \mathcal{L} \to \mathcal{L}$ be the permutation such that
% let $\sigma_{i}$ be the permutation such that 
$\lambda_{1,i,\sigma_i(1)}/B_{\sigma_i(1)} \geq \cdots \geq \lambda_{1,i,\sigma_i(L)}/B_{\sigma_i(L)}$. 
Define the cutoff index $\theta_i$ as

% \small
\begin{align}\notag
\theta_i \triangleq
\begin{cases}
L+1, & \hspace*{-7em}\text{if } \sum_{k=1}^L B_{\sigma_i(k)} \leq C_{s,i}, \\
\min\left\{ p \in \mathcal{L} \;\middle|\; \sum_{k=1}^p B_{\sigma_i(k)} > C_{s,i} \right\}, & \text{otherwise}. 
\end{cases}
\end{align}
% \normalsize
Then, the optimal solution of 
Problem~\ref{subprob_server} is given by:
\small
\begin{equation}\label{eq:server_solution}
\begin{aligned}
y_{s,i,\sigma_{i}(k)}^*(\lambda_{1,i,\sigma_{i}(k)}) =
\begin{cases}
1, & k < \theta_i, \\
\frac{C_{s,i} - \sum_{m=1}^{\theta_i-1} B_{\sigma_{i}(m)}}{B_{\sigma_{i}(\theta_i)}}, & k = \theta_i, \\ 
0, & \text{otherwise}, 
\end{cases}
\quad  k \in \mathcal{L}.
\end{aligned}
\end{equation}
\normalsize

\begin{IEEEproof}
Problem~\ref{subprob_server} is a continuous knapsack problem with objective $\sum_l \lambda_{1,i,l} y_{s,i,l}$ and capacity $C_{s,i}$, where items have values $\lambda_{1,i,l} \geq 0$ and weights $B_l > 0$. The greedy algorithm sorts items by density $\rho_l = \lambda_{1,i,l} / B_l$ in descending order, assigns full units to the highest-density items until the capacity is nearly filled, and assigns a fraction to the marginal item.
Formally, let $\sigma_i$ be the permutation such that $\rho_{\sigma_i(1)} \geq \rho_{\sigma_i(2)} \geq \cdots \geq \rho_{\sigma_i(L)}$. Find the smallest $\theta_i$ where the cumulative weight exceeds capacity. The solution assigns 1 to the first $\theta_i - 1$ items and the exact fraction to the $\theta_i$-th, maximizing the objective under monotonicity. If all densities are zero, the solution is trivial (all zero). This interprets $\lambda_{1,i,l}$ as the "delivery value" per unit bandwidth.
\end{IEEEproof}
% \textcolor{red}{Denote \(\mathbf{y}_1^*(\boldsymbol{\lambda}_{1}) \triangleq (y_{s,i,\sigma_{i}(k)}^*(\lambda_{1,i,\sigma_{i}(k)}))_{s \in \mathcal{S}, i \in \mathcal{O}_s, k \in \mathcal{L}}\).}
% Define the cutoff index $\theta_i$ as 
% $\theta_i \triangleq \min\left\{p \in \{1,\dots,L+1\} \;\middle|\; p = L+1 \;\text{or}\; \sum_{k=1}^p B_{\sigma_i(k)} > C_{s,i},\; p \leq L \right\}.$
% p \in \mathbb{Z}^+

% Problem~\ref{subprob_server} is a continuous knapsack problem~\cite{bertsekas2003convex}. The optimal solution is obtained by sorting resolutions in descending order of the value density $\lambda_{1,i,l}/B_l$. Let $\sigma_{i}$ be the permutation such that $\lambda_{1,i,\sigma_i(1)}/B_{\sigma_i(1)} \geq \cdots \geq \lambda_{1,i,\sigma_i(L)}/B_{\sigma_i(L)}$. Define the cutoff index $\theta_i = \min\{p \mid \sum_{k=1}^p B_{\sigma_i(k)} > C_{s,i}, p \leq L\}$ or $\theta_i = L+1$ if $C_{s,i} \geq \sum_{k=1}^L B_{\sigma_i(k)}$. If $C_{s,i} = 0$, set $\theta_i = 1$. For each $k = 1, \dots, L$,

Problem~\ref{subprob_user} is an LP with a closed-form solution.  Choose $l^* \in \arg\max_{l' \in \mathcal{L}_u} (w_u Q_{l'} - \lambda_{1,u,l'})$.  The optimal solution of Problem~\ref{subprob_user} is given by:
\begin{equation}\label{eq:user_solution}
x_{u,l}^*(\lambda_{1,u,l}) =
\begin{cases}
1, & \text{if } l = l^*, \\
0, & \text{otherwise}.
\end{cases}
\end{equation}
% This indicates that the optimal solution selects the single resolution maximizing the net utility.
  % \textcolor{red}{Denote \(\mathbf{x}_1^*(\boldsymbol{\lambda}_{1}) \triangleq (x_{u,l}^*(\lambda_{1,u,l}))_{u \in \mathcal{U}, l \in \mathcal{L}_u}\).}

\begin{IEEEproof}
Problem~\ref{subprob_user} is a linear program over a simplex: maximize $\sum_{l \in \mathcal{L}_u} (w_u Q_l - \lambda_{1,u,l}) x_{u,l}$ subject to $\sum_{l \in \mathcal{L}_u} x_{u,l} = 1$, $x_{u,l} \in [0,1]$, and $x_{u,l} = 0$ for $l \notin \mathcal{L}_u$. The optimum lies at a vertex of the feasible polytope, which are the standard basis vectors (one-hot). Thus, select $l^* = \arg\max_{l' \in \mathcal{L}_u} (w_u Q_{l'} - \lambda_{1,u,l'})$, set $x_{u,l^*}^* = 1$, and others to 0.
\end{IEEEproof}
  
Problem~\ref{subprob_forwarder_x} is an LP with 
the optimal solution  given by:
% \small
\begin{equation}\label{eq:forwarder_x_solution}
x_{f,l}^*(\lambda_{1,f,l}, \boldsymbol{\lambda}_{2}) =
\begin{cases}
1, & \sum_{k \in \mathcal{O}_f} \lambda_{2,f,k,l} - \lambda_{1,f,l} > 0, \\
0, & \text{otherwise},
\end{cases}
\end{equation}
% \normalsize

\begin{IEEEproof}
Problem~\ref{subprob_forwarder_x} has a simple analytical solution. For each resolution $l$, if the coefficient $\sum_{k \in \mathcal{O}_f} \lambda_{2,f,k,l} - \lambda_{1,f,l} > 0$, set $x_{f,l}^* = 1$; otherwise, set $x_{f,l}^* = 0$. This maximizes the linear objective over the box constraints.
\end{IEEEproof}

% where $\boldsymbol{\lambda}_{2,f,l} \triangleq (\lambda_{2,f,k,l})_{k \in \mathcal{O}_f}$. 
% Denote \textcolor{red}{\(\mathbf{x}_2^*(\boldsymbol{\lambda}_{1}, \boldsymbol{\lambda}_{2}) \triangleq (x_{f,l}^*(\lambda_{1,f,l}, \boldsymbol{\lambda}_{2,f,l}))_{f \in \mathcal{F}, l \in \mathcal{L}}\).}

% $\mathbf{x}^*(\boldsymbol{\lambda}_{1}, \boldsymbol{\lambda}_{2}) = \left[ (x_{u,l}^*(\lambda_{1,u,l}))_{u \in \mathcal{U}, l \in \mathcal{L}_u}^T (x_{f,l}^*(\lambda_{1,f,l}, \boldsymbol{\lambda}_{2,f,l}))_{f \in \mathcal{F}, l \in \mathcal{L}}^T \right]^T$

% This indicates that the optimal solution thresholds on non-negative net revenue per resolution.

Problem~\ref{subprob_forwarder_y} is also a continuous knapsack problem~\cite{bertsekas2003convex} and can be solved in a similar way to Problem~\ref{subprob_server}. For all $f \in \mathcal{F}$ and $k \in \mathcal{O}_f$, let $\sigma_{f,k}: \mathcal{L} \to \mathcal{L}$ be the permutation such that $(\lambda_{1,k,\sigma_{f,k}(1)} - \lambda_{2,f,k,\sigma_{f,k}(1)})/B_{\sigma_{f,k}(1)} \geq \cdots \geq (\lambda_{1,k,\sigma_{f,k}(L)} - \lambda_{2,f,k,\sigma_{f,k}(L)})/B_{\sigma_{f,k}(L)}$. 
Define the cutoff index
% \small
\begin{equation*}\notag
\theta_{f,k} \triangleq
\begin{cases}
L+1, & \hspace*{-7em} \text{if } \sum_{m=1}^L B_{\sigma_{f,k}(m)} \leq C_{f,k}, \\
\min\left\{p \in \mathcal{L} \;\middle|\; \sum_{m=1}^p B_{\sigma_{f,k}(m)} > C_{f,k} \right\}, & \text{otherwise}.
\end{cases}
\end{equation*}
% \normalsize
Then, the optimal solution of Problem~\ref{subprob_forwarder_y} is given by: 
% \small
\begin{equation} \label{eq:forwarder_y_solution}
\begin{aligned}
&y_{f,k,\sigma_{f,k}(m)}^*(\lambda_{1,k,\sigma_{f,k}(m)}, \lambda_{2,f,k,\sigma_{f,k}(m)}) \\
=&\begin{cases}
1, & m < \theta_{f,k}, \\
\frac{C_{f,k} - \sum_{j=1}^{\theta_{f,k}-1} B_{\sigma_{f,k}(j)}}{B_{\sigma_{f,k}(\theta_{f,k})}}, & m = \theta_{f,k}, \\
0, & \text{otherwise},
\end{cases} \quad  m \in \mathcal{L}.
\end{aligned}
\end{equation}
% \normalsize

\begin{IEEEproof}
Problem~\ref{subprob_forwarder_y} is a continuous knapsack problem for each downstream node $k \in \mathcal{O}_f$. The objective is to maximize $\sum_l (\lambda_{1,k,l} - \lambda_{2,f,k,l}) y_{f,k,l}$ subject to $\sum_l B_l y_{f,k,l} \leq C_{f,k}$ and $y_{f,k,l} \in [0,1]$. Sort resolutions by density $(\lambda_{1,k,l} - \lambda_{2,f,k,l})/B_l$ in descending order, excluding any with negative coefficients. Apply the greedy algorithm: assign full allocation to high-density resolutions until capacity, with fractional allocation to the marginal resolution.
\end{IEEEproof}
% \textcolor{red}{Denote $\mathbf{y}_2^*(\boldsymbol{\lambda}_{1}, \boldsymbol{\lambda}_{2}) \triangleq (y_{f,k,\sigma_{f,k}(m)}^*(\lambda_{1,k,\sigma_{f,k}(m)},$ $ \lambda_{2,f,k,\sigma_{f,k}(m)}))_{f \in \mathcal{F}, k \in \mathcal{O}_f, m \in \mathcal{L}}$.}

Denote
% \small
\begin{multline} \notag
\mathbf{x}^*(\boldsymbol{\lambda}_1, \boldsymbol{\lambda}_2) \triangleq \\
\Big( (x_{u,l}^*(\lambda_{1,u,l}))_{u \in \mathcal{U}, l \in \mathcal{L}}, 
(x_{f,l}^*(\lambda_{1,f,l}, \boldsymbol{\lambda}_{2}))_{f \in \mathcal{F}, l \in \mathcal{L}} \Big),
\end{multline}
\begin{multline} \notag
\mathbf{y}^*(\boldsymbol{\lambda}_1, \boldsymbol{\lambda}_2) \triangleq \\ \Big( (y_{s,i,l}^*(\lambda_{1,i,l}))_{s \in \mathcal{S}, i \in \mathcal{O}_s, l \in \mathcal{L}}, (y_{f,k,l}^*(\lambda_{1,k,l}, \boldsymbol{\lambda}_{2}))_{f \in \mathcal{F}, k \in \mathcal{O}_f, l \in \mathcal{L}} \Big).
\end{multline}
% \normalsize

\begin{remark}[Properties of Subproblem Solutions]\label{rem1}
All coordinates of the optimal solution of Problem~\ref{subprob_user} (Problem~\ref{subprob_forwarder_x}) are binary. For the optimal solution of Problem~\ref{subprob_server} (Problem~\ref{subprob_forwarder_y}), at most one coordinate is not binary, and all the other coordinates are binary.
The optimal solutions of Problem~\ref{subprob_server} (Problem~\ref{subprob_forwarder_y}), Problem~\ref{subprob_user}, and Problem~\ref{subprob_forwarder_x} are unique iff the corresponding permutation is unique, $\arg\max_{l' \in \mathcal{L}u} (w_u Q{l'} - \lambda_{1,u,l'})$ is a singleton, and $\sum_{k \in \mathcal{O}_f} \lambda_{2,f,k,l} - \lambda_{1,f,l} \neq 0$, respectively. The conditions of uniqueness hold in most practical cases.
% According to \eqref{eq:user_solution}, \textcolor{red}{$x_{u,l}^*(\lambda_{1,u,l})$ } are binary.
% % (exactly one 1, others 0)
% According to \eqref{eq:forwarder_x_solution}, \textcolor{red}{$x_{f,l}^*(\lambda_{1,f,l}, \boldsymbol{\lambda}_{2,f,l})$} are binary.
% % (rare, can be set to 0 or 1). 
% \textcolor{red}{According to \eqref{eq:server_solution}, only when $k = \theta_{i}$, 
% $y_{s,i,l}^*(\lambda_{1,i,l})$ is not binary}; similarly for \eqref{}. 

% \begin{remark}[Integer Properties of Subproblem Solutions]
% According to \eqref{eq:user_solution}, for each user $u$, exactly one element of 
% $\mathbf{x}_u^*$ equals 1 and all others equal 0. According to \eqref{eq:forwarder_x_solution}, 
% all elements of $\mathbf{x}_f^*$ are binary. \textcolor{red}{According to \eqref{eq:server_solution}, only one of} among all 
% $l \in \mathcal{L}$, at most one $y_{s,i,l}^*$ can be fractional (specifically at index $\theta_i$).
% \end{remark}

% Thus, fractional variables are sparse ($\leq |\mathcal{E}|$).
\end{remark}

Problem~\ref{master_prob} is a convex problem with a non-smooth objective function. We solve Problem~\ref{master_prob} by the projected subgradient method [30, Sec. 8.2].
First, we derive the partial subgradients of $g(\boldsymbol{\lambda}_1, \boldsymbol{\lambda}_2)$ w.r.t. $\boldsymbol{\lambda}_1$ and $\boldsymbol{\lambda}_2$ 
% ~\cite[pp. 667]{bertsekas2016nonlinear}:
% \yulong{$x_{u,l}^*(\lambda_{1,u,l})$ --> $x_{u,l}^*$ ?}
% \textcolor{red}{
% \small
% \begin{align}
% &\partial_{\lambda_{1,i,l}}g(\boldsymbol{\lambda}_1, \boldsymbol{\lambda}_2) = \begin{cases}
% \sum_{j \in \mathcal{I}_u, j \in \mathcal{S}} y_{j,u,l}^*(\lambda_{1,u,l}) - x_{u,l}^*(\lambda_{1,u,l}), \\
% \sum_{j \in \mathcal{I}_u, j \in \mathcal{F}} y_{j,u,l}^*(\lambda_{1,u,l}, \boldsymbol{\lambda}_{2,f,l}) - x_{u,l}^*(\lambda_{1,u,l}), \\
% \sum_{j \in \mathcal{I}_f,j \in \mathcal{S}}, y_{j,f,l}^*(\lambda_{1,f,l}) - x_{f,l}^*(\lambda_{1,f,l}, \boldsymbol{\lambda}_{2,f,l}), \\
% \sum_{j \in \mathcal{I}_f,j \in \mathcal{F}}, y_{j,f,l}^*(\lambda_{1,f,l}, \boldsymbol{\lambda}_{2,f,l}) - x_{f,l}^*(\lambda_{1,f,l}, \boldsymbol{\lambda}_{2,f,l}), 
% \end{cases} \label{eq:subgrad_lambda1_corrected}\\
% &\partial_{\lambda_{2,f,k,l}}g(\boldsymbol{\lambda}_1, \boldsymbol{\lambda}_2) = x_{f,l}^*(\lambda_{1,f,l}, \boldsymbol{\lambda}_{2,f,l}) - y_{f,k,l}^*(\lambda_{1,k,l}, \boldsymbol{\lambda}_{2,f,l}). \label{eq:subgrad_lambda2_corrected}
% \end{align}}
% \normalsize

% {\fontsize{7.4pt}{10pt}\selectfont
% \small
\begin{equation}\label{eq:subgrad_lambda1}
\begin{aligned}
&\partial_{\lambda_{1,i,l}} g(\boldsymbol{\lambda}_1, \boldsymbol{\lambda}_2) = \\
&\begin{cases} 
\displaystyle\sum_{j \in \mathcal{I}_i \cap \mathcal{S}} y_{j,i,l}^*(\lambda_{1,i,l}) + \sum_{j \in \mathcal{I}_i \cap \mathcal{F}} y_{j,i,l}^*(\lambda_{1,i,l}, \boldsymbol{\lambda}_{2}) - x_{i,l}^*(\lambda_{1,i,l}), \\ \quad  i \in \mathcal{U}, \\
\displaystyle\sum_{j \in \mathcal{I}_i \cap \mathcal{S}} y_{j,i,l}^*(\lambda_{1,i,l}) + \sum_{j \in \mathcal{I}_i \cap \mathcal{F}} y_{j,i,l}^*(\lambda_{1,i,l}, \boldsymbol{\lambda}_{2}) \\ \quad - x_{i,l}^*(\lambda_{1,i,l}, \boldsymbol{\lambda}_{2}), i \in \mathcal{F}.
\end{cases}
\end{aligned}
\end{equation}
% }
\vspace{-0.7 em}
% \begin{align}
% \partial_{\lambda_{1,i,l}} g(\boldsymbol{\lambda}_1, \boldsymbol{\lambda}_2) &= \sum_{j \in \mathcal{I}_i} \hat{y_{j,i,l}} - \hat{x_{i,l}}, \\
% &\text{where } \hat{y_{j,i,l}} \triangleq 
% \begin{cases}
% y_{j,i,l}^*(\lambda_{1,i,l}), & j \in \mathcal{S}, \\
% y_{j,i,l}^*(\lambda_{1,i,l}, \boldsymbol{\lambda}_{2,i,l}), & j \in \mathcal{F},
% \end{cases} \notag \\
% &\text{, and } \hat{ x_{i,l}} \triangleq
% \begin{cases}
% x_{i,l}^*(\lambda_{1,i,l}), &  i \in \mathcal{U}, \\
% x_{i,l}^*(\lambda_{1,i,l}, \boldsymbol{\lambda}_{2,i,l}), & i \in \mathcal{F} .
% \end{cases} \notag 
% \end{align}
% \begin{align}
% \small
\begin{align}\label{eq:subgrad_lambda2}
\partial_{\lambda_{2,f,k,l}} g(\boldsymbol{\lambda}_1, \boldsymbol{\lambda}_2) &= y_{f,k,l}^*(\lambda_{1,k,l}, \boldsymbol{\lambda}_{2}) - x_{f,l}^*(\lambda_{1,f,l}, \boldsymbol{\lambda}_{2}) . 
\end{align}
% \normalsize

% Then, we apply the projected subgradient method to solve Problem~\ref{master_prob}~\cite[Sec. 8.2]{bertsekas2003convex}. 
% % 哪有写函数不写input,你要写括号
% At iteration $t$, denote:
% \textcolor{red}{
% $x_{u}^{(t)} \triangleq x_{u}^*(\lambda_{1,u}^{(t)})_{u \in \mathcal{U}}$,
% $x_{f}^{(t)} \triangleq x_{f}^*(\lambda_{1,f}^{(t)}, (\lambda_{2,f,k}^{(t)})_{k \in \mathcal{O}_f})_{f \in \mathcal{F}}$,
% $y_{s,i}^{(t)} \triangleq y_{s,i}^*(\lambda_{1,i}^{(t)})_{s \in \mathcal{S}, i \in \mathcal{O}_s}$,
% $y_{f,k}^{(t)} \triangleq y_{f,k}^*(\lambda_{1,k}^{(t)}, \lambda_{2,f,k}^{(t)})\}_{f \in \mathcal{F}, k \in \mathcal{O}_f}$.}

% $\mathbf{x}_f^{(t)}(\boldsymbol{\lambda}_{1,f}, \boldsymbol{\lambda}_{2,f,l}) \triangleq (x_{f,l}(\lambda_{1,f,l}^{(t)}, \boldsymbol{\lambda}_{2,f,l}^{(t)}))_{f \in \mathcal{F}, l \in \mathcal{L}}$

Then, we present the projected subgradient method. 
% At iteration $t$, we compute in parallel the optimal solutions \textcolor{red}{$x_{u,l}^*(\lambda_{1,u,l}^{(t)})$, $x_{f,l}^*(\lambda_{1,f,l}^{(t)}, \boldsymbol{\lambda}_{2,f,l}^{(t)}),$ $y_{s,i,l}^*(\lambda_{1,i,l}^{(t)})$, and $y_{f,k,l}^*(\lambda_{1,k,l}^{(t)}, \boldsymbol{\lambda}_{2,f,l}^{(t)})$}
At iteration $t$, we compute in parallel the optimal solutions given in \eqref{eq:server_solution}, \eqref{eq:user_solution}, \eqref{eq:forwarder_x_solution}, and \eqref{eq:forwarder_y_solution}, respectively, and then update Lagrange multipliers $\boldsymbol{\lambda}_1^{(t+1)}$ and $\boldsymbol{\lambda}_2^{(t+1)}$ according to:

% \small
% {\fontsize{8pt}{10pt}\selectfont
\begin{align}
\lambda_{1,i,l}^{(t+1)} &= \left[ \lambda_{1,i,l}^{(t)} - \alpha^{(t)} \left( \partial_{\lambda_{1,i,l}} g(\boldsymbol{\lambda}_1^{(t)}, \boldsymbol{\lambda}_2^{(t)}) \right) \right]^+ , \\& \quad i \in \mathcal{F} \cup \mathcal{U}, l \in \mathcal{L} \notag \end{align}
% \small
% {\fontsize{7.4pt}{10pt}\selectfont
\begin{align}
\lambda_{2,f,k,l}^{(t+1)} &= \left[ \lambda_{2,f,k,l}^{(t)} - \beta^{(t)} \left( \partial_{\lambda_{2,f,k,l}} g(\boldsymbol{\lambda}_1^{(t)}, \boldsymbol{\lambda}_2^{(t)}) \right) \right]^+, \\& \quad f \in \mathcal{F}, k \in \mathcal{O}_f, l \in \mathcal{L} \notag 
\end{align}
% }
% \small
% \begin{align}
% \lambda_{1,i,l}^{(t+1)} &= \left[ \lambda_{1,i,l}^{(t)} - \alpha^{(t)} \left( x_{i}^*(\boldsymbol{\lambda}_1^{(t)}, \boldsymbol{\lambda}_2^{(t)}) - \sum_{j:(j,i) \in \mathcal{E}} y_{j,i}^*(\boldsymbol{\lambda}_1^{(t)}, \boldsymbol{\lambda}_2^{(t)}) \right) \right]^+, \\
% &\quad  i \in \mathcal{V} \notag
% \end{align}
% \begin{align}
% \lambda_{2,f,k,l}^{(t+1)} &= \left[ \lambda_{2,f,k,l}^{(t)} - \beta^{(t)} \left( y_{f,k,l}^*(\boldsymbol{\lambda}_1^{(t)}, \boldsymbol{\lambda}_2^{(t)}) - x_{f,l}^*(\boldsymbol{\lambda}_1^{(t)}, \boldsymbol{\lambda}_2^{(t)}) \right) \right]^+,\notag \\
% &\quad  (f,k) \in \mathcal{E}, \, l \in \mathcal{L}
% \end{align}
\normalsize
where $\alpha^{(t)}>0$ and $\beta^{(t)}>0$ are step sizes satisfying the square summable but not summable condition:
\begin{align}
\sum_{t=0}^\infty \alpha_t = \infty,
\quad
\sum_{t=0}^\infty \alpha_t^2 < \infty,
\quad
\sum_{t=0}^\infty \beta_t = \infty,
\quad
\sum_{t=0}^\infty \beta_t^2 < \infty.\notag
\end{align}

% $\mathbf{x}^{(t)} = \mathbf{x}^(\boldsymbol{\lambda}_1^{(t)}, \boldsymbol{\lambda}_2^{(t)})$ and $\mathbf{y}^{(t)} = \mathbf{y}^(\boldsymbol{\lambda}_1^{(t)}, \boldsymbol{\lambda}2^{(t)})$

\begin{thm}[Convergence Analysis of Algorithm~\ref{alg:distributed_lagrangian}]\label{lemma_converge}
\begin{enumerate}
    \item Sequence $\{(\boldsymbol{\lambda}_1^{(t)}, \boldsymbol{\lambda}_2^{(t)})\}_{t \in \mathbb{N}}$ generated by Algorithm~\ref{alg:distributed_lagrangian} converges to an optimal solution  $(\boldsymbol{\lambda}_1^*, \boldsymbol{\lambda}_2^*)$ of Problem~\ref{master_prob}.
    \item If $\mathbf{x}^*(\boldsymbol{\lambda}_1, \boldsymbol{\lambda}_2)$ and $\mathbf{y}^*(\boldsymbol{\lambda}_1, \boldsymbol{\lambda}_2)$ are unique, then they are also optimal for Problem 2.
    % $(x_{u,l}^*(\lambda_{1,u,l}^*))_{u \in \mathcal{U}, l \in \mathcal{L}_u}$, $(x_{f,l}^*(\lambda_{1,f,l}^*, \boldsymbol{\lambda}_{2,f,l}^*))_{f \in \mathcal{F}, l \in \mathcal{L}},$ $(y_{s,i,l}^*(\lambda_{1,i,l}^*))_{s \in \mathcal{S}, i \in \mathcal{O}_s, l \in \mathcal{L}}$, and $(y_{f,k,l}^*(\lambda_{1,k,l}^*, \boldsymbol{\lambda}_{2,f,l}^*))_{f \in \mathcal{F}, k \in \mathcal{O}_f}$ } 
    \item If $\mathbf{x}^*(\boldsymbol{\lambda}_1, \boldsymbol{\lambda}_2)$ and $\mathbf{y}^*(\boldsymbol{\lambda}_1, \boldsymbol{\lambda}_2)$ are unique and integer-valued, then they are also optimal for Problem~\ref{prob_qoe}.
\end{enumerate}
\end{thm}

\begin{IEEEproof}
Since \(\partial_{\lambda_{1,i,l}} g (\boldsymbol{\lambda}_1, \boldsymbol{\lambda}_2)\) and \(\partial_{\lambda_{2,f,k,l}} g (\boldsymbol{\lambda}_1, \boldsymbol{\lambda}_2)\) are bounded, the convergence of the projected subgradient algorithm with square-summable but nonsummable step sizes follows immediately from the standard convergence results \cite[Prop. 8.2.6]{bertsekas2003convex}. The sequence \(\{(\boldsymbol{\lambda}_1^{(t)}, \boldsymbol{\lambda}_2^{(t)})\}_{t \in \mathbb{N}}\) converges to an optimal solution of Problem~\ref{master_prob}.
% The dual function $g(\boldsymbol{\lambda}_1, \boldsymbol{\lambda}_2)$ is convex (as the pointwise supremum of linear functions in the multipliers). The projected subgradient method with appropriate step sizes converges as follows:
% \begin{itemize}
%     \item  \textit{Constant step sizes}: Converges to within a bounded distance of the optimum, with suboptimality $O(\alpha + \beta)$.
%     \item  \textit{Nonsummable diminishing}: Guarantees convergence to the exact optimum, as steps allow infinite travel but finite variance.
%     \item  \textit{Square summable but not summable or Polyak}: Ensures convergence to optimum with rate guarantees, e.g., $O(1/\sqrt{t})$ for suboptimality.
% \end{itemize}
\end{IEEEproof}

Denote $\mathbf{x}^{(t)}\triangleq \mathbf{x}^*
(\boldsymbol{\lambda}_1^{(t)}, \boldsymbol{\lambda}_2^{(t)}) $ and 
$\mathbf{y}^{(t)}\triangleq \mathbf{y}^*
(\boldsymbol{\lambda}_1^{(t)}, \boldsymbol{\lambda}_2^{(t)})$. 
The details are summarized in Algorithm~\ref{alg:distributed_lagrangian}. 
Step 3-Step 5 compute the optimal solutions of all subproblems. Step 6-Step 7 update the Lagrange multipliers. 
% In each iteration of Algorithm~\ref{alg:distributed_lagrangian}, the computational complexities of Step 3, Step 4, Step 5, Step 6, Step 7, and Step 8 are $\mathcal{O}(S L \log L)$, $\mathcal{O}(U L)$, $\mathcal{O}(F L \log L)$, $\mathcal{O}(E L)$, $\mathcal{O}(E L)$, and $\mathcal{O}(1)$, respectively, as $L \to \infty$. Thus, 
The per-iteration computational complexity 
% of Algorithm~\ref{alg:distributed_lagrangian} 
is $\mathcal{O}(V L \log L + E L)$, as $L \to \infty$.

\begin{algorithm}\small
\caption{Partial Dual Decomposition Algorithm}
\label{alg:distributed_lagrangian}
\begin{algorithmic}[1]
\Statex \textbf{Input:} $\mathcal{U}$, $\mathcal{S}$, $\mathcal{F}$, $B_l$, $Q_l$, $w_u$ and $C_{i,j}$, $(i,j) \in \mathcal{E}$.
\Statex \textbf{Output:} $\mathbf{x}^{(t)}$ and $\mathbf{y}^{(t)}$.
% \Statex \textbf{Output:} \textcolor{red}{$\mathbf{x}^*(\boldsymbol{\lambda}_1^{(t)}, \boldsymbol{\lambda}_2^{(t)})$ and $\mathbf{y}^*(\boldsymbol{\lambda}^{(t)}_1, \boldsymbol{\lambda}^{(t)}_2) $.}
% \State \textbf{Output:} $\hat{\mathbf{x}}$,$\hat{\mathbf{y}}$, $\boldsymbol{\lambda}_1$, and $\boldsymbol{\lambda}_2$.
\State Initialize $\boldsymbol{\lambda}_1^{(0)}$, $\boldsymbol{\lambda}_2^{(0)}$, and $t \gets 1$.
\Repeat
\State Compute $\mathbf{x}^{(t)}$ and $\mathbf{y}^{(t)}$ according to \eqref{eq:server_solution}, \eqref{eq:user_solution}, \eqref{eq:forwarder_x_solution}, and \eqref{eq:forwarder_y_solution}, all in parallel.
    % \State \textcolor{red}{Compute $\mathbf{x}^*(\boldsymbol{\lambda}_1^{(t)}, \boldsymbol{\lambda}_2^{(t)})$ and $\mathbf{y}^*(\boldsymbol{\lambda}_1^{(t)}, \boldsymbol{\lambda}_2^{(t)})$ according to \eqref{eq:server_solution}, \eqref{eq:user_solution}, \eqref{eq:forwarder_x_solution}, and \eqref{eq:forwarder_y_solution}, all in parallel.}
    \State Update $\boldsymbol{\lambda}_1^{(t)}$ and $\boldsymbol{\lambda}_2^{(t)}$ according to \eqref{eq:subgrad_lambda1} and \eqref{eq:subgrad_lambda2}, respectively, all in parallel. 
    \State $t \gets t + 1$.
\Until{stopping criterion is satisfied.}
\end{algorithmic}
\end{algorithm}

% \hspace{-0.5em} 

If the update stops at a finite iteration $t$ or converges but with non unique optimal solutions, the solutions may not be integer or feasible. Therefore, we propose a two-phase greedy method to obtain a feasible point, denoted by $\tilde{\mathbf{x}} \triangleq (\tilde{x}_{i,l})_{i \in \mathcal{U} \cup \mathcal{F}, l \in \mathcal{L}}$ and 
$\tilde{\mathbf{y}} \triangleq (\tilde{y}_{i,j,l})_{(i,j) \in \mathcal{E}, l \in \mathcal{L}}$, based on $\mathbf{x}^{(t)}$ and $ \mathbf{y}^{(t)}$ obtained by Algorithm~\ref{alg:distributed_lagrangian}. The details are summarized in Algorithm~\ref{alg:recovery}.
Specifically, we first round down the fractional values of $\mathbf{y}^{(t)}$ to integers $\tilde{\mathbf{y}}$ to ensure the capacity constraints given in \eqref{c6} (see Step 2).
Then, we adjust the values of $\tilde{\mathbf{x}}$ and $\tilde{\mathbf{y}}$ in non-decreasing order of depth to satisfy the flow conservation constraints given in \eqref{c4} and \eqref{c5} (see Steps 3-12). 
%Specifically,\textcolor{red}{Step 1} computes a depth-based ordering using the longest path algorithm for DAGs \cite{cormen2022introduction}, where each node's depth equals the maximum path length from any server. \textcolor{red}{Step 2} rounds down all fractional transmission variables to ensure capacity constraints. \textcolor{red}{Steps 3-14} process nodes in depth order: forwarders update $\tilde{x}_{f,l}$ based on incoming flows and adjust outgoing transmissions, while users are assigned their highest-quality feasible resolution.
The overall computational complexity is $\mathcal{O}((V + E)L)$. 
% \footnote{Details on Depth-Based Ordering are provided in the Appendix I.}
The depth-based ordering in Algorithm \ref{alg:recovery} serves two critical purposes:
\begin{itemize}
    \item \textbf{Dependency Resolution:} The depth-based ordering ensures that nodes with multiple paths are processed only after all their dependencies are resolved. For instance, consider a node that can be reached from servers via both a 2-hop path and a 3-hop path. This node is assigned depth 3 (the maximum path length), ensuring that all intermediate nodes on both paths are fully processed before this node is considered. This guarantees that when we determine $\tilde{x}_{i,l}$ for any node $i$, all incoming transmissions $\tilde{y}_{j,i,l}$ from upstream nodes have been finalized.
    \item \textbf{Deterministic Execution:} The ordering ensures deterministic and consistent execution of the algorithm. Nodes at depth $d$ are processed only after all nodes at depth $d-1$ are finalized. This prevents any ambiguity in the execution order and ensures reproducible results across different implementations. Without this ordering, the algorithm could produce different valid solutions depending on the arbitrary order in which nodes are processed, which would complicate analysis and debugging.
\end{itemize}

\textbf{Method Properties.} 
The proposed optimization framework is \textbf{distributed}—each forwarder computes link prices $\lambda_l$ using only local states, ensuring the system scales independently of the global network size or total flow count. 
It is \textbf{lightweight}—the algorithm exchanges only Lagrange multipliers between neighboring forwarders. Each message contains: I). forwarder ID: 4 bytes; II). iteration number: 4 bytes; III). multiplier values: $L \times 8$ bytes.
For 12 resolution levels, each message is about 104 bytes.
This is $<$0.01\% of the video data volume transmitted in the same period (typically tens of MB), confirming that coordination overhead is negligible.
It is \textbf{intrinsically robust}—treating packet loss as stochastic gradient noise, which allows the system to self-correct using stale multipliers without requiring aggressive control-plane retransmissions (as detailed in Section \ref{Reliability}). 
It is \textbf{fast}—converging rapidly toward the global optimum regardless of the number of users ($n = 50, 100, 200, 300$), enabling the control loop to effectively track dynamic viewer and channel fluctuations within the time-frame of a single video GOP.

\begin{algorithm}\small
\caption{Feasible Integer Solution Construction}
\label{alg:recovery}
\begin{algorithmic}[1]
\Statex \textbf{Input:} $\mathbf{x}^{(t)}$ and $\mathbf{y}^{(t)}$.
\Statex \textbf{Output:}  $\tilde{\mathbf{x}}$, $\tilde{\mathbf{y}}$.
\State Compute depth-based ordering of nodes using the longest path algorithm for DAGs \cite{cormen2022introduction}.
% \State Initialize $\tilde{y}_{i,j,l} \gets \lfloor \mathbf{y}^{(t)} \rfloor$ for all $(i,j) \in \mathcal{E}$, $l \in \mathcal{L}$.
\State Initialize $\tilde{\mathbf{y}} \gets \lfloor \mathbf{y}^{(t)} \rfloor$
\For{each node $i$ in non-decreasing order of depth} 
  \If{$i \in \mathcal{F}$}
    \State  $\tilde{x}_{i,l} \gets \min\{1, \sum_{j \in \mathcal{I}_i} \tilde{y}_{j,i,l}\}$, $l \in \mathcal{L}$ in parallel.
    \State  $\tilde{y}_{i,k,l} \gets \min\{\tilde{y}_{i,k,l}, \tilde{x}_{i,l}\}$, $k \in \mathcal{O}_i$, $l \in \mathcal{L}$ in parallel.
  \ElsIf{$i \in \mathcal{U}$}
    \State $\tilde{\mathcal{L}}_{i} \gets \{l \in \mathcal{L}_i : \sum_{j \in \mathcal{I}_i} \tilde{y}_{j,i,l} > 0\}$.
    \If{$\tilde{\mathcal{L}}_{i} \neq \emptyset$}
        \State $\tilde{x}_{i,l} \gets 
        \begin{cases}
        1, & \text{if } l = \arg\max_{l' \in \tilde{\mathcal{L}}_{i}} Q_{l'},\\
        0, & \text{otherwise},
        \end{cases}$ $l \in \mathcal{L}_i$.
    \Else
        \State $\tilde{x}_{i,l} \gets 0$, $l \in \mathcal{L}_i$.
    \EndIf
  \EndIf

\EndFor
\end{algorithmic}
\end{algorithm}

\section{Experimental Evaluation}
\label{experiment}

% This section evaluates COMETS against state-of-the-art video streaming approaches through systematic assessment of user QoE, bandwidth efficiency, and scalability under conditions with numerous concurrent users.

This section evaluates COMETS against state-of-the-art video streaming approaches. Our evaluation focuses on three properties: I). scalability—whether QoE remains stable as concurrent users increase; II). fairness—whether the distributed optimization achieves equitable resource allocation across heterogeneous clients; and III). convergence effectiveness—whether the algorithm reaches optimal decisions faster than centralized approaches. 

\subsection{Experimental Setup}

While Section \ref{Distributed Decision Algorithm} establishes analytical guarantees for these properties—Theorem \ref{lemma_converge} proves convergence, and the per-iteration complexity ensures linear scaling—these results assume idealized conditions—instantaneous message passing and perfect synchronization. Validating that these properties hold with actual protocol implementations, real video content, and system dynamics requires an emulation platform. 
Mini-NDN \footnote{https://minindn.memphis.edu/}, an NDN emulator built on Mininet \footnote{https://mininet.org/}, is suitable for this purpose: it implements the complete NDN forwarding plane (PITs, in-network caching, stateful forwarding) needed to test Range Interest aggregation under realistic conditions, provides a controlled environment for fair cross-system comparison, and captures wall-clock timing that reflects true computational and communication overhead.
Mini-NDN allows us to systematically scale experiments by gradually increasing the number of concurrent users and infrastructure nodes, enabling direct observation of how QoE, fairness, and optimization time evolve with network size.

All emulations ran on a server with an Intel Xeon Platinum 8358P CPU (2.60GHz, 64 cores) and 32GB RAM under Ubuntu 22.04.5 LTS. 
This setup allows for running experiments with up to 300 concurrent clients without affecting the real-time behavior of the system.
While testing with a larger number of users (e.g., $\geq$ 1000) was considered, our claims of scalability are supported by two forms of evidence. First, the analytical complexity of our distributed algorithm guarantees linear growth and scaling. Second, our empirical results up to 300 users demonstrate consistent performance trends with no indication of an impending scalability bottleneck, suggesting the primary limitation was the capacity of our emulation server, not the COMETS architecture itself.
Connections from clients to initial forwarders are configured at 100 Mbps and backbone links at 400 Mbps, with inter-node latency below 5 ms (refer to \cite{8712644,yang2021high}). 
% In the MoQ system, clients connected to the server via relay nodes, while in COMETS, clients connected to content producers through intermediate forwarders.
% \dirk{what are "downstream links" exactly, and why is 100 Mbps a good number?} 
 
We compare COMETS against four baseline systems representing the key approaches discussed in the Introduction: I). \textit{DASH-BOLA} \cite{spiteri2020bola}: a standard client-side buffer-based adaptation algorithm (client-adaptive approach), II). \textit{MoQ-based Approach} \cite{10689417}: implementing server-centric optimization with global bandwidth knowledge (server-driven approach), III). \textit{GB \cite{tan2021game}}: a game-theoretic NDN-based approach using Bayesian Nash equilibrium for bandwidth allocation, and IV). \textit{NDN-MMRA \cite{9195016}}: an NDN-based multicast media rate adaptation algorithm (both ICN-based approaches). A comparison with these methods is shown in Table \ref{tab:comparison}.
% \textcolor{red}{These baselines represent the main paradigms from Section \ref{intro}: client-adaptive (DASH-BOLA), server-driven (MoQ), and existing ICN-based approaches (GB, NDN-MMRA).
% \dirk{explain why these baselines were chosen, linking them back to the limitations discussed in the Introduction.}
To ensure fair comparison, identical source video content was used across all systems. For DASH and the MoQ-based approach, FFmpeg streams was segmented and delivered via HTTP and pipe respectively; For ICN-based approaches, FFmpeg streams was segmented into chunks of 2 seconds duration.

\begin{table}[ht]
\centering
\caption{Comparison with Representative Approaches}
\label{tab:comparison}
\small
\begin{tabular}{lccccc}
\toprule
\textbf{Approach} & \textbf{Coord.} & \textbf{Optim.} & \textbf{Conv.} & \textbf{Scale} \\
\midrule
DASH-BOLA & None & Local & N/A & Linear \\
MoQ-Server & Central & Global ILP & Slow & Limited  \\
NDN-MMRA & Local & Heuristic & N/A & Sub-linear \\
GB & Game & BNE & Slow & Sub-linear \\
\textbf{COMETS} & Distrib. & Lagrangian & Fast & Sub-linear \\
\bottomrule
\end{tabular}
\vspace{1mm}
\footnotesize{Coord.: Coordination mechanism; Optim.: Optimization approach; Conv.: Convergence speed; BNE: Bayesian Nash Equilibrium.}
\end{table}

% \textit{GB (Game-Based) Algorithm} \cite{tan2021game}: a game-theoretic approach where clients solve Bayesian Nash Equilibrium for distributed bitrate decisions with implicit aggregation through NDN's PIT mechanism, and IV). \textit{NDN-MMRA} \cite{9195016}: a multi-stage multicast rate adaptation system leveraging PIT for global demand monitoring and dynamic rate adjustment in wireless environments.

\subsection{Evaluation Metrics} \label{sec:evaluation_metrics}

We evaluate COMETS against baseline systems using metrics that assess video quality, playback continuity, fairness, and overall QoE. To capture both average performance and consistency, we report mean values alongside 95th percentile statistics for key metrics, emphasizing on scalability and mitigation of poor tail-end user experiences.

%\begin{itemize}
%\item
\textbf{Video Quality:} We employ VMAF (Video Multi-Method Assessment Fusion) \cite{garcia2019practical} to quantify perceptual quality. For analysis, we establish baseline VMAF ranges by resolution: 480p (50-75), 1K (65-85), 2K (70-85), 4K (80-95), and 8K (95-98). Both mean and 95th percentile VMAF values are tracked to assess quality consistency.

%\item
\textbf{Jitter:} Playback smoothness is assessed through two metrics:\textit{Chunk Inter-arrival Time:} Temporal variance between consecutive chunk arrivals, adjusted for request frequency,  and \textit{RFC 3550 Jitter \cite{schulzrinne2003rfc3550}:} Calculated using $J(i) = J(i-1) + (|D(i,j)| - J(i-1))/16$, where D represents deviation between actual and expected intervals.

%\item
\textbf{Playback Buffer Time:} The client buffer is simulated to increase by the chunk duration (2.0 seconds) upon arrival and to be exhausted during playback, demonstrating that the system is able to maintain continuity.

%\item
\textbf{Startup Delay:} The initial waiting time before playback commences (when buffer reaches threshold), reflecting user experience influenced by initial resolution assignment, network latency, and optimization computation time.

%\item
\textbf{Jain's Fairness Index \cite{ware2019beyond}:} We compute Jain's Fairness Index based on average delivered resolution across clients to evaluate resource distribution equity. The formula $f(x) = (\sum x_i)^2 / (n \cdot \sum x_i^2)$ ranges from 1/n to 1, where \(x_i\) denotes the average delivered resolution for client \(i\), and \(n\) represents the total number of clients. 1 indicating perfect fairness.

%\item  Drawing inspirations from prior QoE research~\cite{yin2015control,dobrian2011understanding}, 
\textbf{Overall QoE Score:} We formulate a comprehensive score through weighted normalization: 
{\fontsize{7.8pt}{10pt}\selectfont
% \small
\begin{align}
\begin{split}
QoE = & \max(0, \alpha \cdot normalized\_VMAF + \beta \cdot normalized\_fairness \\
& + \gamma \cdot normalized\_buffer + \delta \cdot (1-normalized\_jitter) \\
& + \epsilon \cdot (1-normalized\_startup\_delay)) \nonumber
\end{split}
\end{align}
}
Weights $[\alpha, \beta, \gamma, \delta, \epsilon] = [0.4, 0.2, 0.15, 0.15, 0.1]$ prioritize video quality and fairness while penalizing rebuffering, jitter, and startup delay. \footnote{These weights are not arbitrary; they are based on insights from established QoE sensitivity analysis research \cite{yin2015control}, which have validated them as representative of user preferences in most common streaming scenarios.}
% dobrian2011understanding
% \footnote{These weights are not arbitrary; they are based on insights from established QoE sensitivity analysis research \cite{yin2015control,dobrian2011understanding}, which have validated them as representative of user preferences in most common streaming scenarios.}
% \footnote{While comprehensive sensitivity analysis under varying network conditions (RTT variations, packet loss patterns, QoE weight variations) is important for production deployment, this paper focuses on validating the fundamental design and demonstrating COMETS' core advantages. We prioritized establishing the benefits of network-orchestrated coordination over traditional approaches considering the scope and space constraints. Extensive robustness testing under diverse network conditions is planned for next work.}. 
%\end{itemize}

% 

\subsection{Comparative Performance Evaluation} \label{sec:comparative_performance}

%This section presents a comparative performance evaluation of COMETS against the baseline DASH-BOLA and MoQ-based systems across varying numbers of concurrent users to assess scalability and Quality of Experience (QoE). We analyze key metrics defined in Section \ref{sec:evaluation_metrics}, examining both mean performance and tail behaviour (95th percentile) to quantify the consistency of user experience under increasing network load.

This subsection compares COMETS with DASH-BOLA, MoQ, GB, and NDN-MMRA across increasing concurrent users, focusing on scalability and QoE metrics. 
% Both mean and 95th percentile values are analyzed to highlight performance consistency under network load.
We expect COMETS to excel under high concurrency, maintaining stable QoE where others degrade.

\subsubsection{Video Quality (VMAF) Analysis}
\begin{figure}[htbp]
    \centering
    % 上半部分
    \begin{subfigure}{0.45\textwidth}
        \centering
        \includegraphics[width=\linewidth]{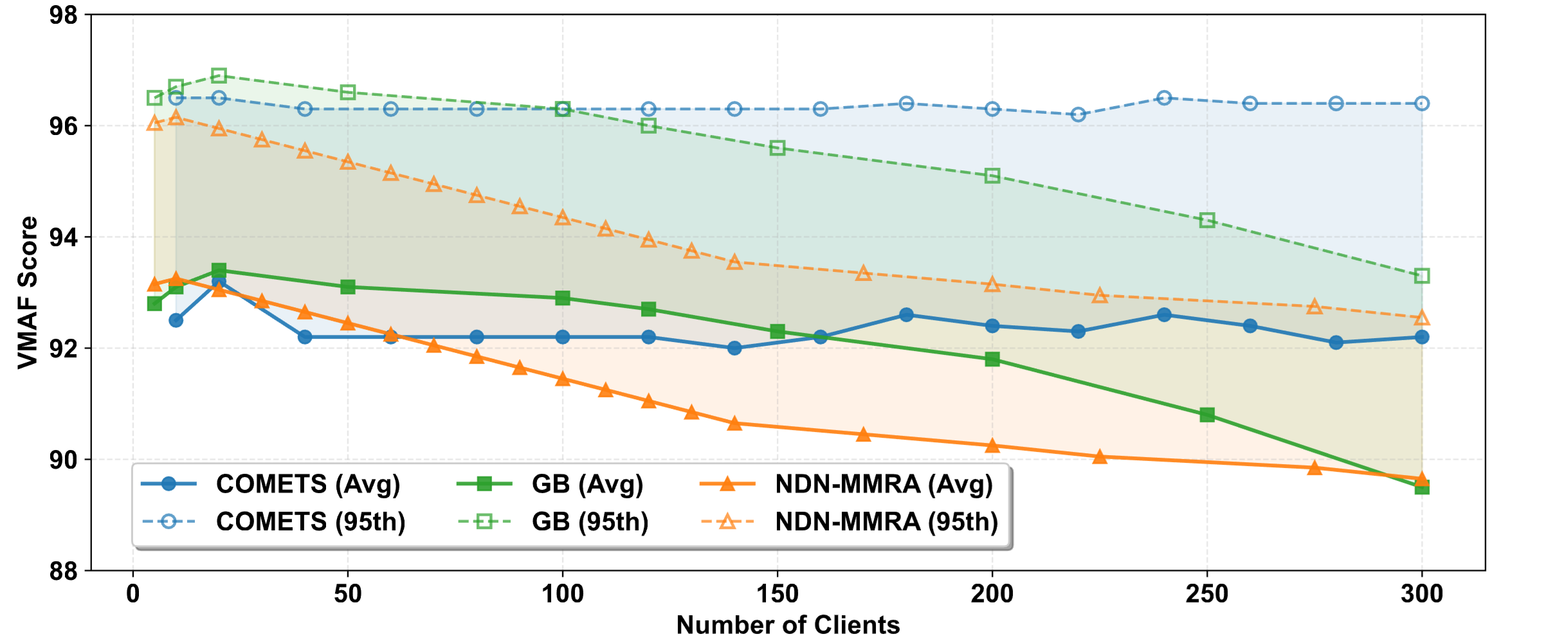}
        \caption{VMAF Scores of NDN-based System}
        \label{fig:ndn_vmaf}
    \end{subfigure}
    \vspace{0em} 
    \begin{subfigure}{0.45\textwidth}
        \centering
        \includegraphics[width=\linewidth]{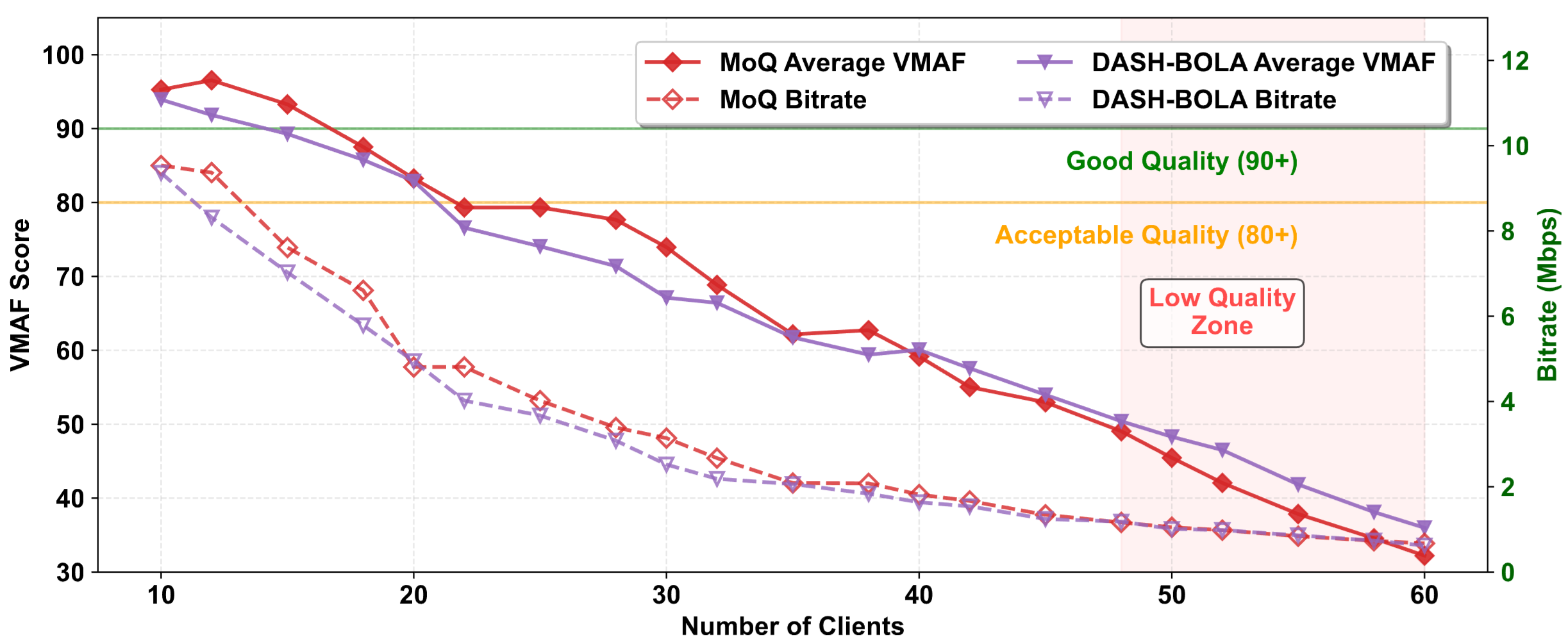}
        \caption{VMAF Scores of MoQ and DASH-BOLA}
        \label{fig:moq_dash_vmaf}
    \end{subfigure}
    \caption{Comparison of VMAF Scores}
    \label{fig:combined_vmaf}
\end{figure}

Figure~\ref{fig:combined_vmaf} demonstrates distinct performance patterns across systems. COMETS maintains exceptional stability with mean VMAF scores of 92.2 and 95th percentile values of 96.4 across the entire 10-300 client range, exhibiting minimal degradation even at maximum load. The GB algorithm shows moderate degradation from 93.4 to 89.5, attributed to information asymmetry in its Bayesian Nash Equilibrium (BNE) calculations—incomplete prior beliefs lead to suboptimal bitrate decisions under high concurrency. NDN-MMRA exhibits similar decline (93.25 to 89.65) as its PIT aggregation efficiency decreases with scale. Both DASH-BOLA and MoQ experience catastrophic quality collapse beyond 60 concurrent users, with VMAF dropping below 40—well beneath the acceptable quality threshold of 80; thus, testing beyond this point was unnecessary given their inability to maintain acceptable quality. 
% due to uncoordinated client decisions and centralized optimization bottlenecks.
COMETS' distributed resolution selection and interest aggregation enable this stability by optimizing bandwidth hierarchically, ensuring consistent quality despite scale.

\vspace{-0.3em}
\subsubsection{Jitter and Playback Stability}

\begin{figure*}[htbp]
\centering
\begin{subfigure}[b]{0.32\textwidth}
    \centering
    \includegraphics[width=\textwidth,height=0.75\textwidth,keepaspectratio]{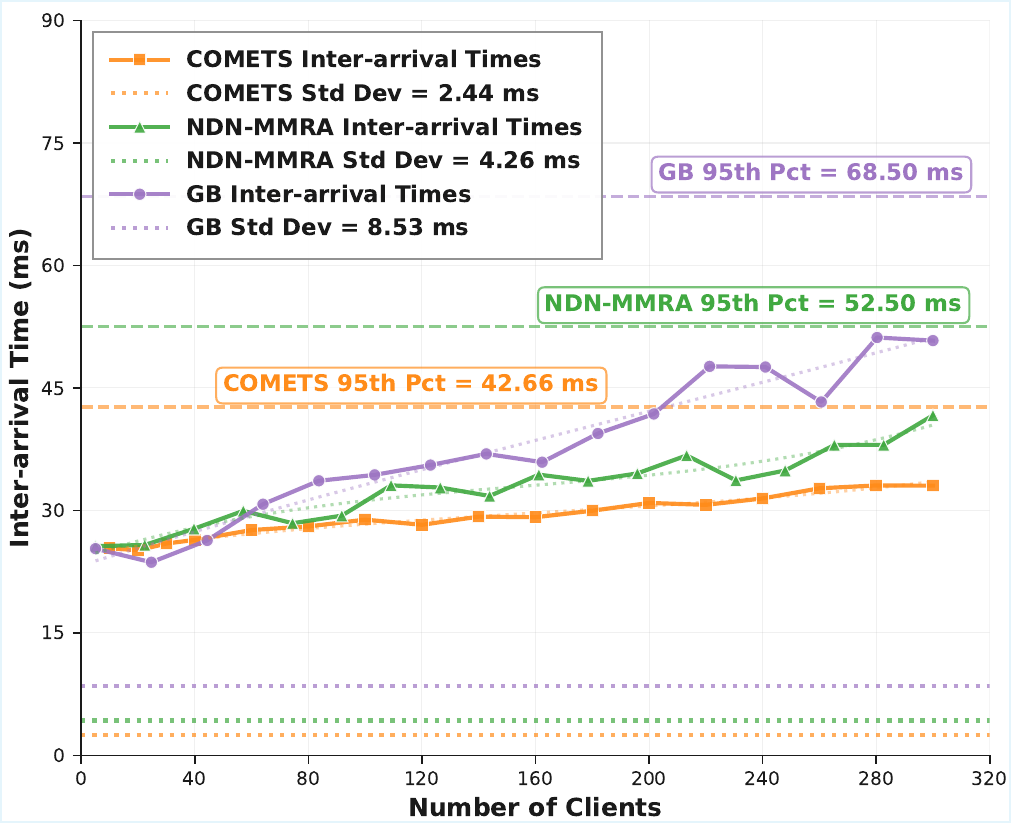}
    \caption{NDN-based system: inter-arrival time}
    \label{fig:jitter_inter}
\end{subfigure}
\hfill
\begin{subfigure}[b]{0.32\textwidth}
    \centering
    \includegraphics[width=\textwidth,height=0.75\textwidth,keepaspectratio]{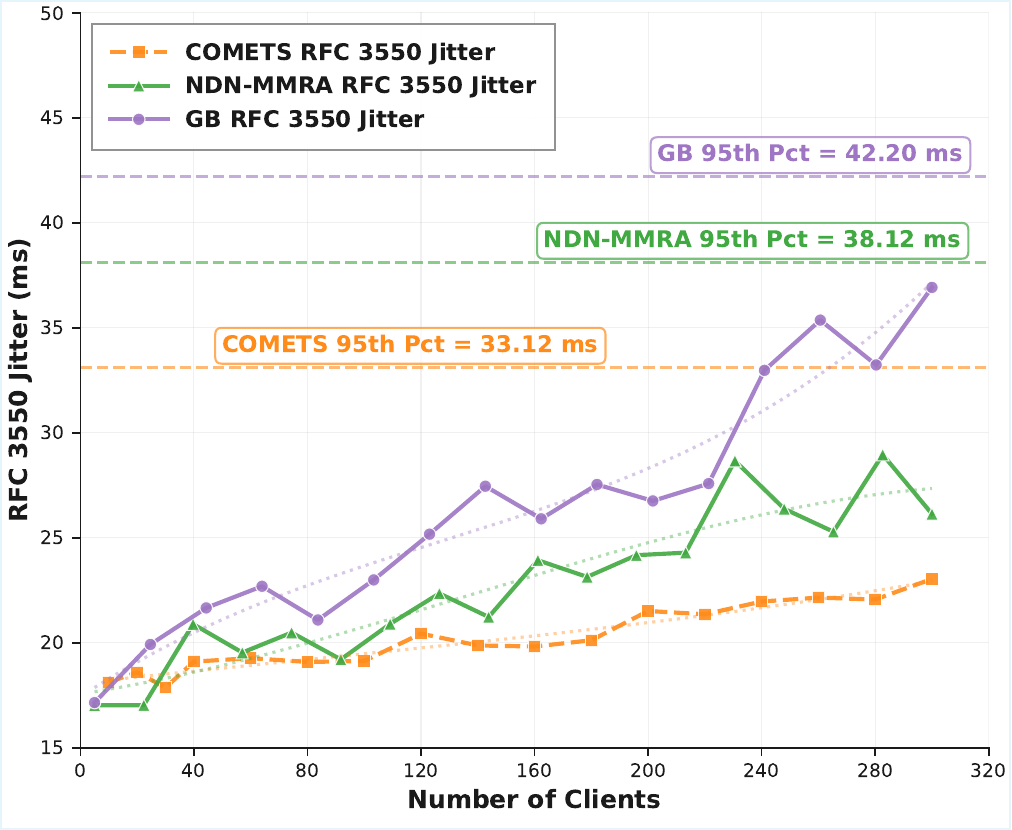}
    \caption{NDN-based system: RFC 3550 jitter}
    \label{fig:jitter_rfc}
\end{subfigure}
\hfill
\begin{subfigure}[b]{0.32\textwidth}
    \centering
    \includegraphics[width=\textwidth,height=0.75\textwidth,keepaspectratio]{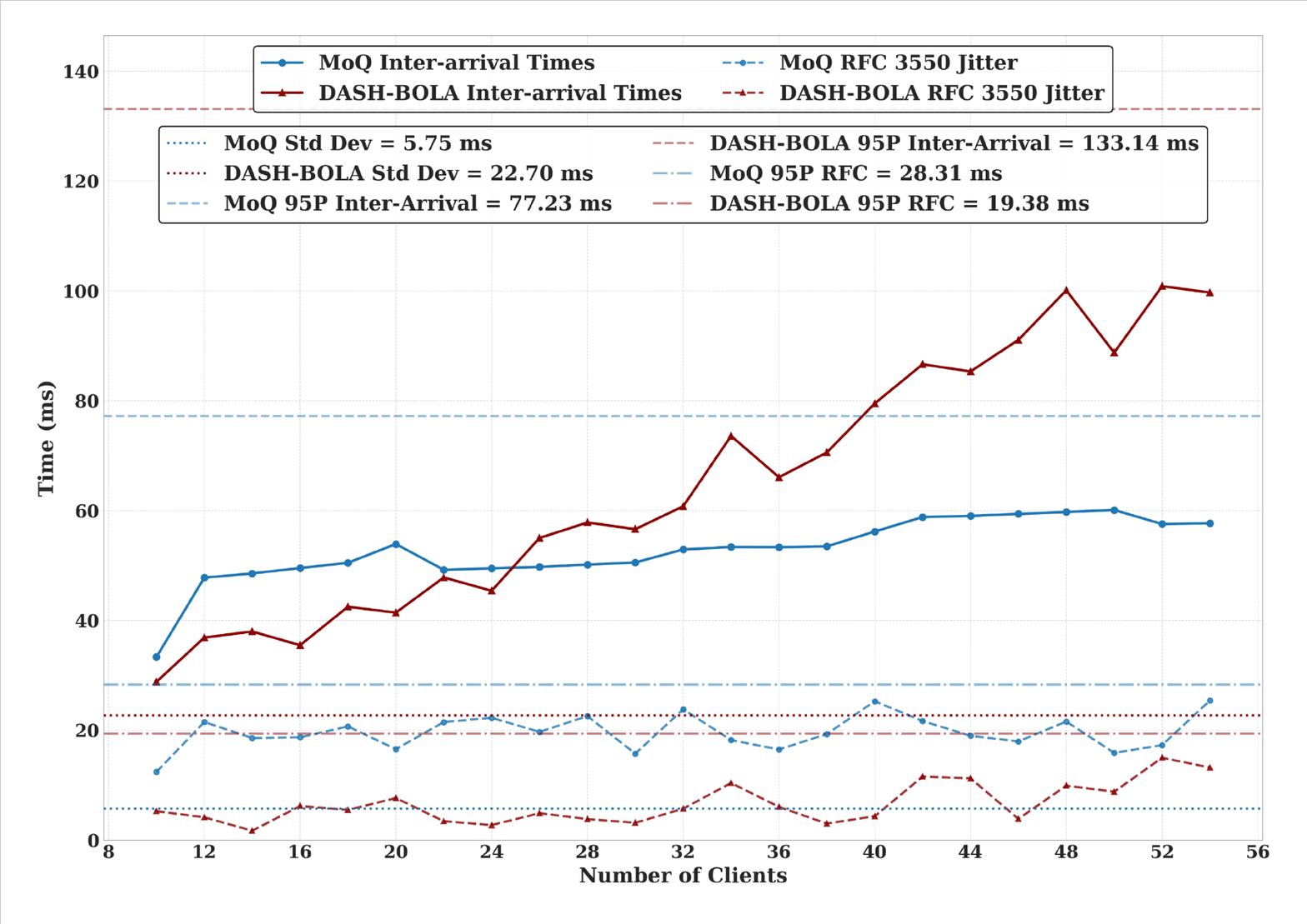}
    \caption{MoQ vs. DASH-BOLA jitter comparison}
    \label{fig:jitter_moq_dash}
\end{subfigure}

\caption{Comparison of Jitter Performance: (a) and (b) show two jitter metrics for an NDN-based system, while (c) compares jitter performance between MoQ and DASH-BOLA.}
\label{fig:jitter_comparison}
\end{figure*}

Figure~\ref{fig:jitter_inter} and \ref{fig:jitter_rfc} illustrates COMETS' low jitter, with inter-arrival times stabilizing at 30ms and RFC 3550 jitter below 23ms even at 300 clients.
% with mean inter-arrival times rising modestly from 15 ms to 23 ms and 95th percentile values from 32 ms to 43 ms, while RFC 3550 jitter remains below 13-19 ms across all loads. 
In comparison, GB exhibits increasing jitter variance (95th percentile: 68.50ms) due to periodic interest flooding when heterogeneous client priors disrupt PIT aggregation under suboptimal equilibria, while NDN-MMRA maintains moderate jitter (95th percentile: 52.50ms) through its multi-stage rate adaptation mechanism.
DASH-BOLA and MoQ exhibit higher jitter as client numbers increase from 5 to 56 (Figure~\ref{fig:jitter_moq_dash}).
DASH-BOLA's inter-arrival times escalate dramatically from 30ms to over 100ms as client competition intensifies, while maintains low RFC 3550 jitter (95th percentile: 19.38ms) due to its buffer-based algorithm's inherent smoothing. MoQ demonstrates more stable inter-arrival times (50-60ms) than DASH but higher RFC 3550 jitter (95th percentile: 28.31ms), reflecting the computational overhead of server-side optimization.
Buffer dynamics (Figure~\ref{fig:performance_comparisons}a) indicate all NDN-based systems maintain adequate playback buffers above the 8-second threshold. COMETS sustains 8.5-10.5 seconds throughout testing, while GB and NDN-MMRA show comparable patterns. The consistency across NDN variants demonstrates the effectiveness of in-network caching for buffer stability.

% These results underscore COMETS' advantage in reducing path contention through decentralized control, outperforming baselines that suffer from periodic flooding, feedback loops, or scaling inefficiencies.

% COMETS consistently achieves stable jitter performance, with inter-arrival times around 20ms and RFC 3550 jitter values between 10-18ms (Figure~\ref{fig:jitter_comets}).
% %demonstrates remarkable jitter stability (Figure~\ref{fig:jitter_comets}), with inter-arrival times consistently around 20ms and RFC 3550 jitter values between 10-18ms regardless of scale. 
% As shown in Figure~\ref{fig:jitter_comparison}, its 95th percentile inter-arrival time (42.66ms) is nearly one-third of DASH-BOLA's (133.14ms) and approximately half of MoQ's (77.23ms). This consistency results from COMETS' decentralized resolution control and interest aggregation that %distribute computational load while reducing
% reduce path contention.
% %competition. 
% Playback buffer dynamics (Figures~\ref{fig:buffer_comets} and~\ref{fig:buffer_comparison}) shows that all systems maintain stable average buffering times, ensuring smooth viewing experiences. %under most conditions.

% \begin{figure*}[htbp]
%     \centering
%     \begin{subfigure}[b]{0.24\textwidth} % Adjust width as needed
%         \centering
%         \includegraphics[width=\textwidth]{figures/jitter_comets.png}
%         \caption{Jitter Analysis for COMETS System}
%         \label{fig:jitter_comets}
%     \end{subfigure}
%     \hfill % Optional: Adds space between subfigures
%     \begin{subfigure}[b]{0.24\textwidth} % Adjust width as needed
%         \centering
%         \includegraphics[width=\textwidth]{figures/jitter_moq_dash.png}
%         \caption{Jitter Analysis between MoQ and DASH-BOLA}
%         \label{fig:jitter_comparison}
%     \end{subfigure}
%     \hfill % Optional: Adds space between subfigures
%     \begin{subfigure}[b]{0.24\textwidth} % Adjust width as needed
%         \centering
%         \includegraphics[width=\textwidth]{figures/buffer_comets.png}
%         \caption{Buffer Performance for COMETS System}
%         \label{fig:buffer_comets}
%     \end{subfigure}
%     \hfill % Optional: Adds space between subfigures
%     \begin{subfigure}[b]{0.24\textwidth} % Adjust width as needed
%         \centering
%         \includegraphics[width=\textwidth]{figures/buffer_moq_dash.png}
%         \caption{Buffer Performance between MoQ and DASH-BOLA}
%         \label{fig:buffer_comparison}
%     \end{subfigure}
%     \caption{Comparison of Jitter and Buffer Performance} % Add an overall caption
%     \label{fig:combined_jitter_buffer} % Add an overall label
% \end{figure*}

\subsubsection{Resource Fairness and Startup Delay}
\begin{figure*}[htbp]
    \centering
    \subfloat[Buffer Performance Comparison \label{fig:buffer_comparison}]{\includegraphics[width=0.32\textwidth]{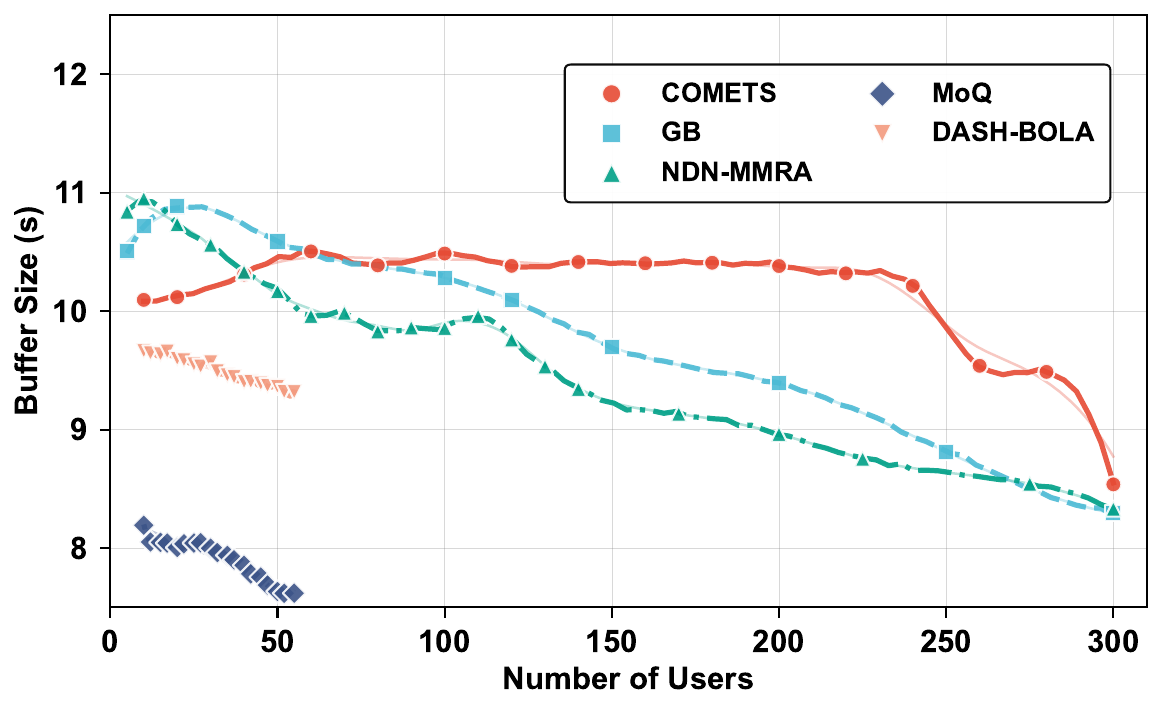}}
    \hfill
    \subfloat[Resource Fairness Comparison \label{fig:fairness_comparison}]{\includegraphics[width=0.32\textwidth]{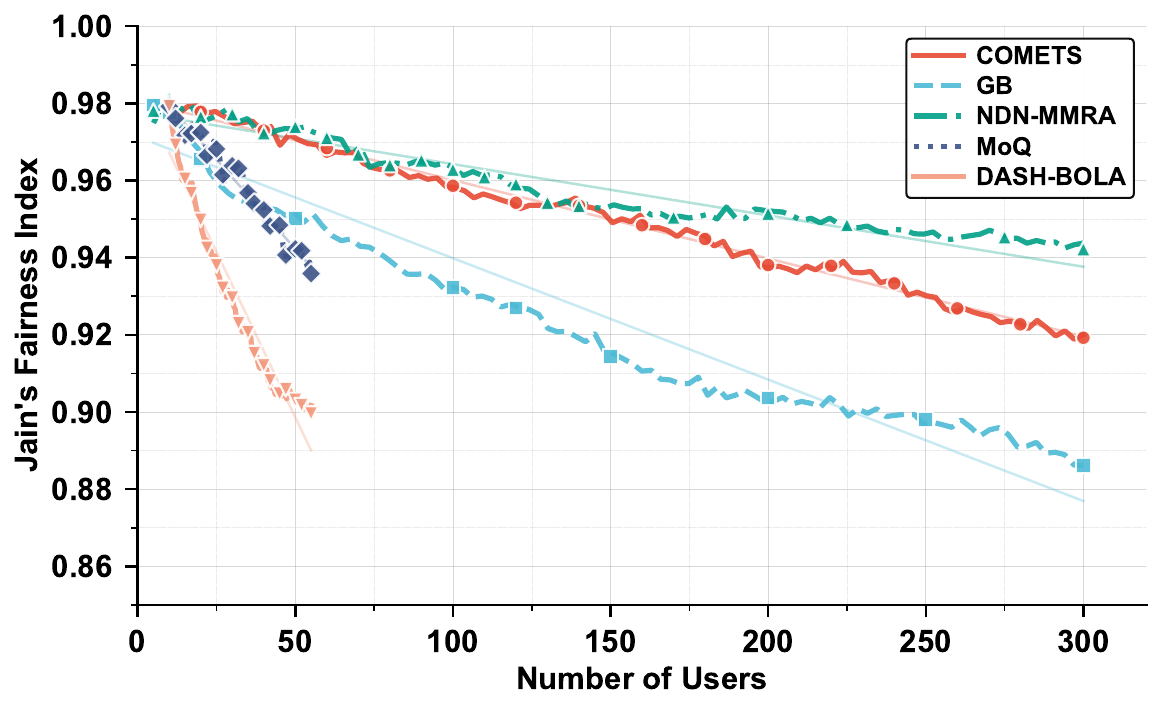}}
    \hfill
    \subfloat[Startup Delay Comparison \label{fig:startup_comparison}]{\includegraphics[width=0.32\textwidth]{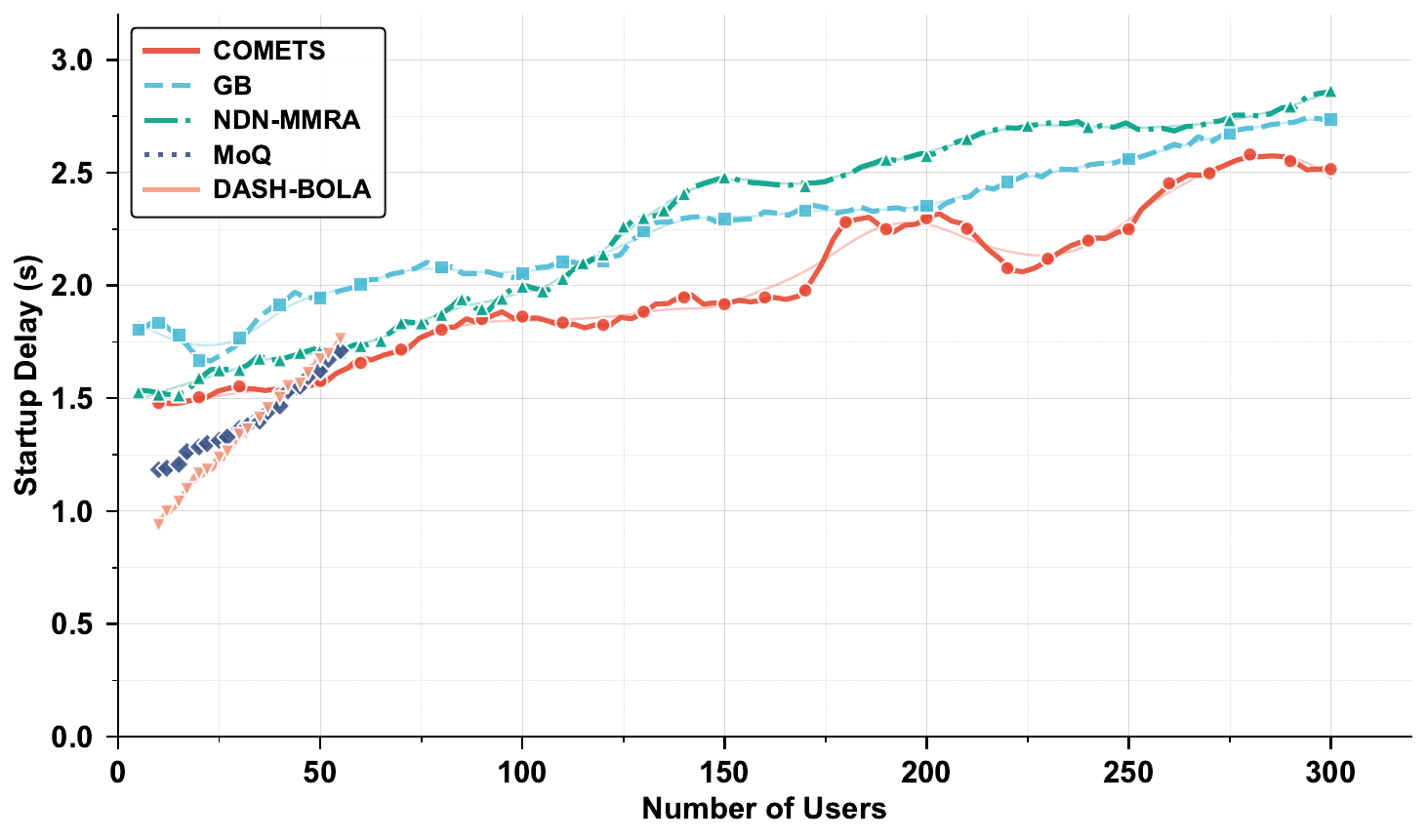}}
    \caption{Comparison of Buffer, Startup Delay and Fairness}
    \label{fig:performance_comparisons}
\end{figure*}

Figure~\ref{fig:performance_comparisons}b depicts COMETS' Jain's fairness index holding steady at 0.98-0.93, indicating equitable resolution distribution via network-wide coordination. NDN-MMRA maintains 0.98-0.94 through PIT-driven subset optimization that prioritizes weaker clients, whereas GB declines from 0.98 to 0.89 due to fairness penalties in its payoff function under information asymmetry. MoQ and DASH-BOLA drop from 0.98 to 0.87-0.90, highlighting inefficiencies in server-centric and client-driven resource allocation. All systems exceed 0.86, but COMETS' in-network aggregation ensures superior equity at high loads.
For startup delay (Figure~\ref{fig:performance_comparisons}(b)), COMETS stays below 2.3 s, with a gradual rise, owing to efficient initial assignments. NDN-MMRA and GB increase from 1.5-1.7 s to 2.6-2.8 s, influenced by initial probing and equilibrium solving overheads. MoQ and DASH-BOLA increase from 0.9-1.3 s to 1.8 s at scale due to client information and bandwidth probing. All systems can remain within the 3-second excellence threshold.

% \begin{figure*}[htbp]
%     \centering
%     \begin{subfigure}[b]{0.24\textwidth} % Adjust width as needed
%         \centering
%         \includegraphics[width=\textwidth]{figures/fairness_comets.png}
%         \caption{Resource Fairness for COMETS System}
%         \label{fig:fairness_comets}
%     \end{subfigure}
%     \hfill % Optional: Adds space between subfigures
%     \begin{subfigure}[b]{0.24\textwidth} % Adjust width as needed
%         \centering
%         \includegraphics[width=\textwidth]{figures/fairness_moq_dash.png}
%         \caption{Resource Fairness between MoQ and DASH-BOLA}
%         \label{fig:fairness_comparison}
%     \end{subfigure}
%     \hfill % Optional: Adds space between subfigures
%     \begin{subfigure}[b]{0.24\textwidth} % Adjust width as needed
%         \centering
%         \includegraphics[width=\textwidth]{figures/startup_delay_comets.png}
%         \caption{Startup Delay for COMETS System}
%         \label{fig:startup_comets}
%     \end{subfigure}
%     \hfill % Optional: Adds space between subfigures
%     \begin{subfigure}[b]{0.24\textwidth} % Adjust width as needed
%         \centering
%         \includegraphics[width=\textwidth]{figures/startup_delay_moq_dash.png}
%         \caption{Startup Delay between MoQ and DASH-BOLA}
%         \label{fig:startup_comparison}
%     \end{subfigure}
%     \caption{Comparison of Resource Fairness and Startup Delay} % Add an overall caption
%     \label{fig:combined_comparison} % Add an overall label
% \end{figure*}

\subsubsection{Algorithm Comparison}
\begin{figure}[htbp]
    \centering
    \includegraphics[width=0.45\textwidth]{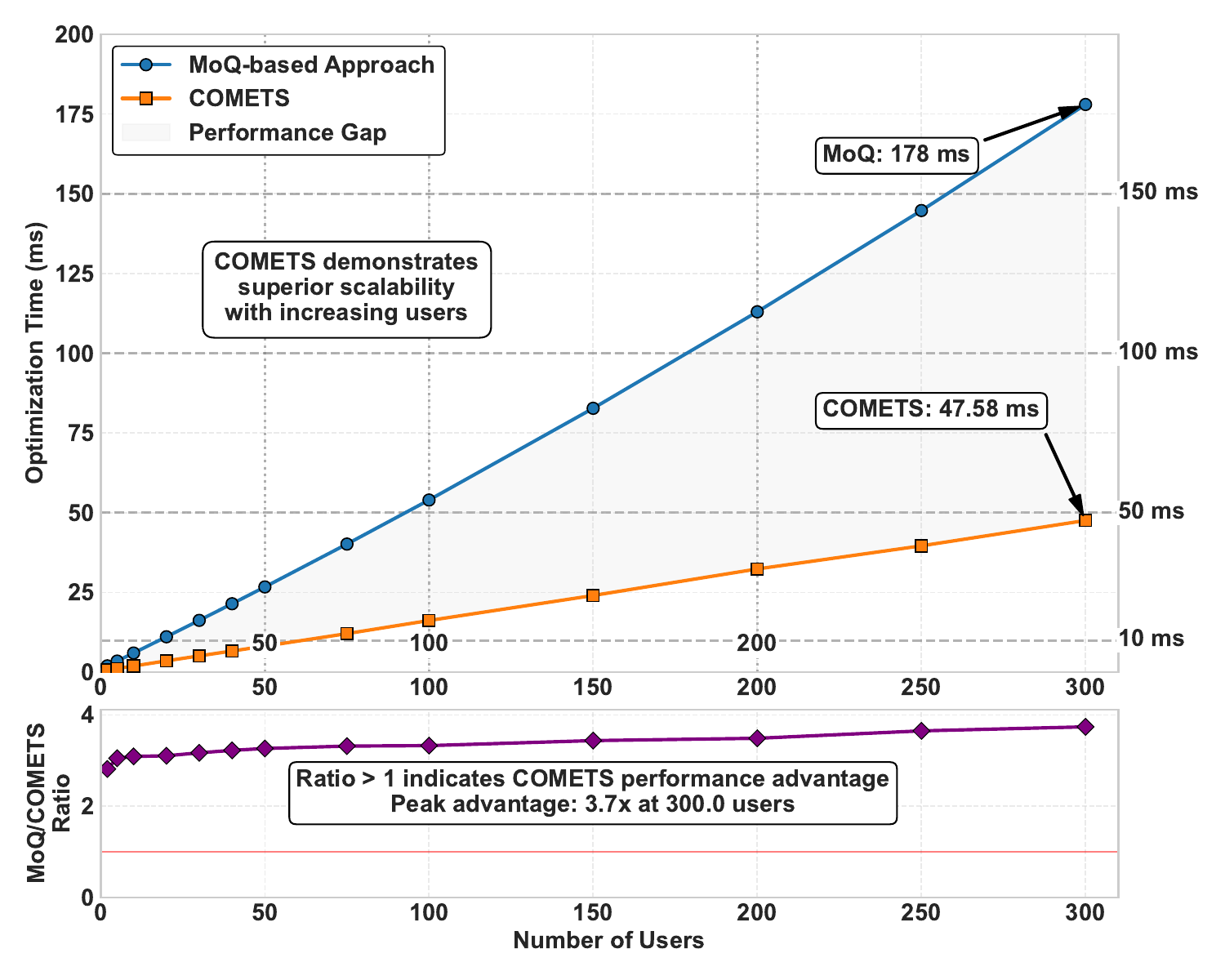}
    \caption{Optimization Time between MoQ and COMETS}
    \label{optimization}
\end{figure}

% As shown in Figure \ref{optimization}, the distributed optimization approach converges within 5–47ms for 10–300 concurrent users and achieves over 3.7× faster than a centralized solver \cite{10689417} — ensuring sub-50ms latency at scale.\footnote{Detailed optimization comparisons and convergence analyses are provided in the appendix H.}
Figure~\ref{optimization} demonstrates the fundamental computational advantage of distributed versus centralized optimization architectures. COMETS exhibits near-linear growth in optimization time from 5ms to 47.58ms across 10-300 concurrent users, maintaining sub-50ms latency even at maximum scale. In contrast, the MoQ-based approach shows steeper linear scaling from 10ms to 178ms, reflecting the inherent overhead of centralized global optimization.
The performance ratio (bottom panel) quantifies this efficiency gap, starting near parity at low client counts but diverging significantly as scale increases. COMETS maintains a consistent advantage throughout testing, reaching a 3.7× performance improvement at 300 users. This widening gap stems from architectural differences: while MoQ must solve increasingly complex global optimization problems with quadratic growth in decision variables, COMETS parallelizes resolution decisions across network nodes, constraining computational complexity to local subproblems.
The practical impact extends beyond raw computational efficiency. MoQ's 178ms optimization latency at 300 users approaches critical thresholds for real-time adaptation—any further scaling risks compromising responsiveness to network dynamics. Conversely, COMETS' 47.58ms latency provides substantial headroom for scale expansion while maintaining the sub-100ms response times essential for adaptive streaming. This computational efficiency directly enables the superior QoE stability observed in Section~\ref{sec:comparative_performance}, particularly the consistent startup delays and minimal jitter that characterize COMETS' performance at scale.

% \vspace{-5pt}
% \vspace{-5pt}

\subsubsection{Robustness Analysis}
\begin{figure}[ht]
\centering
\begin{subfigure}[b]{0.235\textwidth}
    \centering
    \includegraphics[width=\textwidth,height=0.75\textwidth,keepaspectratio]{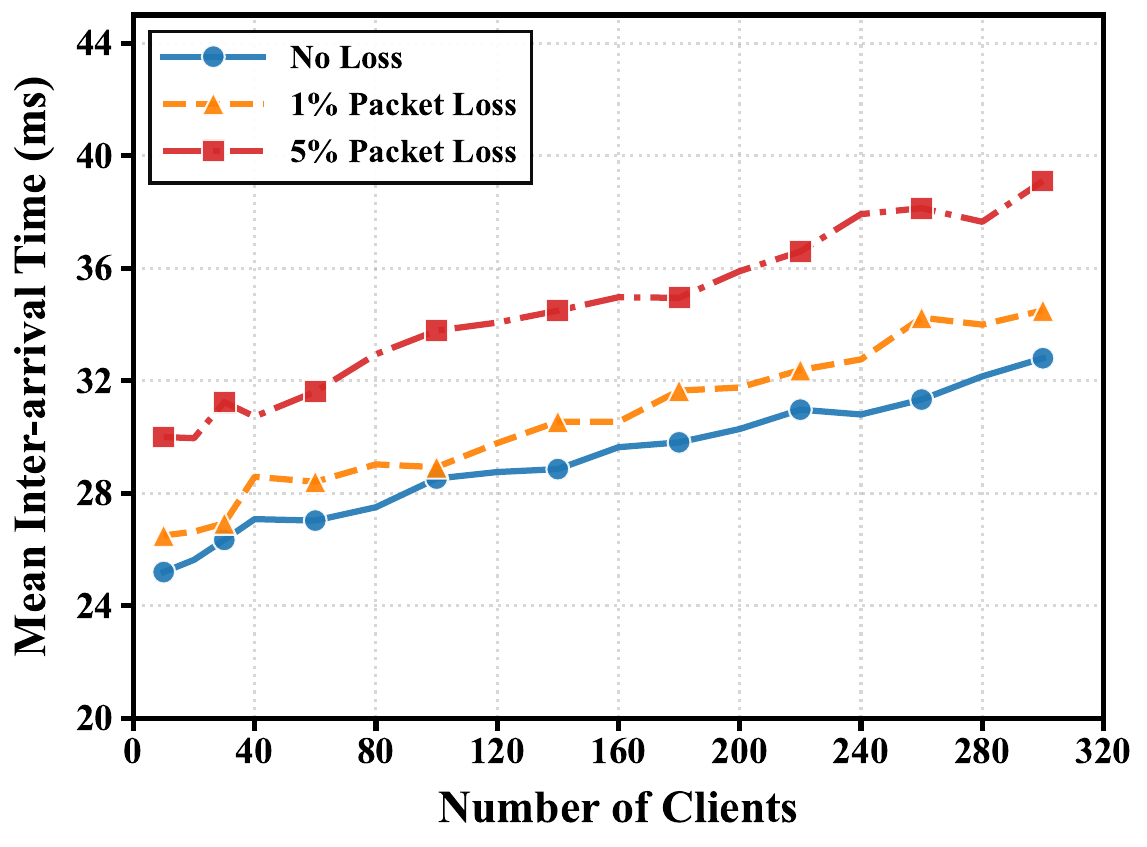}
    \caption{Packet Loss Behavior}
    \label{fig:packet_loss}
\end{subfigure}
\hfill 
\begin{subfigure}[b]{0.235\textwidth}
    \centering    \includegraphics[width=\textwidth,height=0.75\textwidth,keepaspectratio]{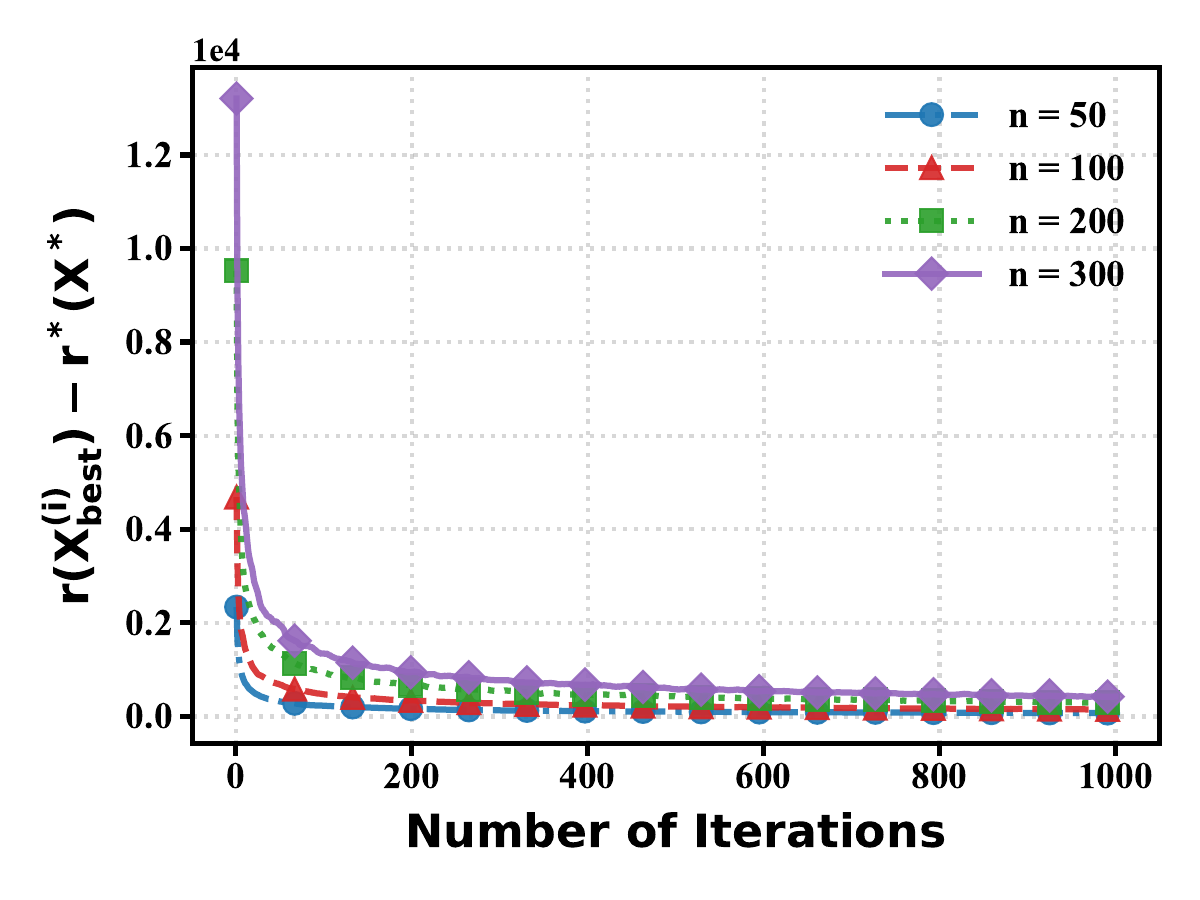}
    \caption{Convergence Behavior} 
    \label{fig:convergence}
\end{subfigure}
\caption{Optimization algorithm performance analysis.}
\label{fig:Optimization}
\end{figure}

To validate COMETS' resilience under realistic network impairments, we introduced random packet loss rates of 1\% and 5\% on bottleneck links. These rates represent typical congestion scenarios in wired networks and challenging conditions in wireless/mobile environments, respectively~\cite{cisco2024qos}.
As shown in Figure ~\ref{fig:packet_loss}, COMETS maintains stable playback performance under both loss scenarios. At 1\% packet loss, the mean inter-arrival time increases by approximately 5\% compared to the baseline (from 32.8\,ms to 34.5\,ms at 300 clients). At 5\% packet loss, inter-arrival times increase by approximately 19\% (to 39.1\,ms)---still well within acceptable streaming thresholds. The system exhibits graceful degradation rather than divergence, demonstrating inherent robustness across all tested user scales.
This resilience stems from COMETS' \textit{loss recovery} mechanism (Section \ref{Reliability}). Unlike end-to-end retransmission in TCP/QUIC where recovery latency scales with the round-trip time to the origin server, COMETS leverages hierarchical in-network caching. When a client detects packet loss via timeout, it re-expresses the Interest with limited retransmissions (up to 2 attempts). Due to the high redundancy of popular streams, these re-expressed Interests are frequently satisfied by nearby upstream forwarders rather than the distant origin.

To quantify this effect, we analyzed cache hit distributions across the topology (50 forwarders, 300 clients). Tier-3 nodes (edge forwarders closest to users) achieve effective hit ratios of approximately 75--80\%, absorbing the majority of retransmission requests locally. Tier-1 nodes (core forwarders closer to the source) sustain hit ratios around 50--60\%, filtering residual cross-regional traffic. This multi-level aggregation effectively masks physical-layer packet losses from the application layer, converting unicast request floods into efficient shared delivery trees.
Figure~\ref{fig:convergence} further confirms algorithmic stability under varying problem scales. The distributed optimization converges rapidly toward the global optimum regardless of the number of users ($n = 50, 100, 200, 300$). This fast convergence ensures that the control loop can track dynamic network fluctuations within the time-frame of a single video GOP, maintaining system stability even when underlying conditions change.

\subsubsection{Overall Quality of Experience}
\begin{figure}[htbp]
    \centering
    \includegraphics[width=0.45\textwidth]{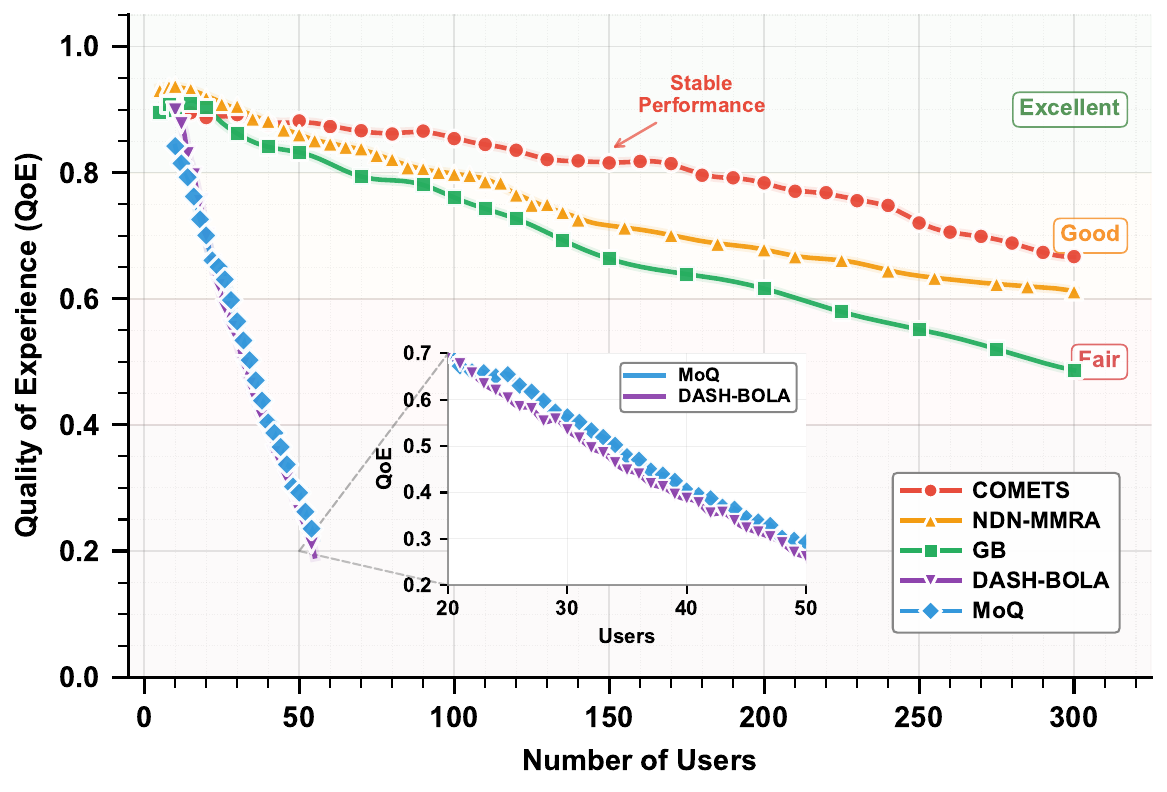}
    \caption{Quality of Experience Comparison}
    \label{fig:combined_qoe}
\end{figure}

The composite QoE metric (Figure~\ref{fig:combined_qoe}) provides evidence of COMETS' scalability advantages. COMETS maintains good QoE scores between 0.7-0.9 across the entire test range, balancing high VMAF, low jitter, and stable buffers through distributed optimization. NDN-MMRA degrades from 0.9 to 0.61 at 300 clients, limited by PIT-based adaptation delays and multicast staging overheads that accumulate under high client loads and dynamic network conditions. GB follows a steeper decline trajectory from 0.9 to 0.5. The game-theoretic approach's performance degradation reflects cumulative effects of information asymmetry—suboptimal BNE solutions compound as client heterogeneity increases.
Traditional approaches (MoQ and DASH-BOLA) become effectively unusable beyond 50 clients, with QoE plummeting below 0.3 due to centralized scaling issues. 
These results validate COMETS' architectural design: by combining ICN's intrinsic benefits (interest aggregation, in-network caching) with distributed intelligence, the system achieves robust QoE without sharp degradation at scale. 

% both bandwidth efficiency and fine-grained coordination 

\section{DISCUSSION}
%While COMETS demonstrates superior coordination, we acknowledge certain implementation boundaries:}

In addition to the analysis and experimental performance evaluation, we offer a  discussion of assumptions of the system, deployment aspects, and COMETS' technical novelty.

\textbf{Trust Model and Deployment Scope.}
COMETS assumes forwarders operate within a \emph{single administrative domain} under a common trust anchor. This model aligns with CDN deployments (e.g., Akamai, CloudFlare) where edge servers are operated by the content provider or a trusted partner. 
However, COMETS is not directly applicable to peer-to-peer (P2P) or federated scenarios where forwarders may be untrusted or have conflicting incentives. Extending the trust model to multi-domain settings with cryptographic verification or reputation systems is an interesting direction for future work.

\textbf{Scope of Experience Quality Assessment.} Our assessment uses objective metrics that have been validated for subjective quality in previous studies. A comprehensive subjective user study including diverse demographic characteristics could provide additional validation, which will be left to future work.

\textbf{Content Source Assumptions.}
The current design targets \emph{single-source video-on-demand (VoD) and near-live streaming}, where content originates from a known server and is pre-encoded at multiple quality levels.
For true live streaming with real-time encoding, additional challenges arise, including I). Encoding latency. The source must produce multiple quality variants in real-time, which may bottleneck at the encoder. II). Multi-source scenarios. User-generated content (e.g., Twitch) involves many concurrent sources, requiring per-stream optimization overhead.

% \textcolor{red}{\textbf{Limitations of Emulation-Based Evaluation. }
% While Mini-NDN provides a controlled environment for systematic evaluation, we acknowledge that it simplifies certain aspects of real networks, such as complex cross-traffic patterns and operational constraints in production deployments. 
% However, in real-world scenarios, the effects of competing TCP/QUIC flows primarily manifest as reduced available bandwidth and increased packet loss rates. Our distributed algorithm addresses these dynamics through Lagrange multiplier-based link pricing, which inherently captures changes in "effective bandwidth" as traffic conditions evolve. Furthermore, Section \ref{Reliability} details our mechanisms for ensuring robust video delivery under varying network conditions, including loss recovery and congestion control.
% Future work will pursue validation on testbeds with realistic Internet characteristics and integration with production CDN infrastructures to further validate these findings.}
% \dirk{see comment earlier.}

\textbf{Technical Novelty.}
\textbf{I). Novel problem formulation.} While prior work has applied optimization to video streaming, COMETS is the first to formulate request aggregation–enabled content multicasting with per-hop quality adaptation as a unified ILP problem, and tackles its NP-hardness with a two-stage distributed optimization: a dual-decomposition–based distributed algorithm with parallel closed-form updates from local information to solve the continuous relaxation exactly, followed by a low-complexity greedy algorithm that yields near-optimal integer-feasible solutions. Previous algorithms mostly used solvers or heuristic algorithms, which had many problems such as high complexity and long running time.
\textbf{II). Range-interest protocol.} The protocol transforms the client-forwarder interaction from ``request-response'' to ``capability announcement,'' enabling forwarders to make informed group-level decisions. This is fundamentally different from prior ICN interest semantics.
\textbf{III). Architectural integration.} While ICN caching and request aggregation are known techniques, their integration with distributed optimization for coordinated rate adaptation is novel. Prior ICN systems use aggregation for \emph{efficiency} but not for \emph{coordination}—they still make independent, reactive decisions at each hop.

% \textbf{II). Analytical closed-form solutions.} A key novelty is our derivation of closed-form solutions for the dual subproblems. Unlike iterative numerical solvers used in centralized approaches, our solutions enable constant-time per-iteration computation at each forwarder, parallelizable execution across all network nodes and provable convergence with explicit rate bounds.This analytical tractability is what enables real-time distributed optimization—a capability absent in prior ICN streaming systems.

\section{Conclusion}
\label{Conclusion}
In this work, we introduced \textbf{COMETS}, the first coordinated multi-destination video transmission framework that integrates \emph{in-network aggregation} with \emph{distributed optimization} to address the scalability and fairness limitations of existing streaming architectures. Unlike client-driven DASH or server-driven MoQ, COMETS transforms adaptation into a collaborative, hop-by-hop negotiation, fundamentally shifting streaming from isolated endpoint control to coordinated in-network intelligence.  

\textbf{Achievements.} Our results demonstrate that COMETS delivers consistently high QoE under hundreds of concurrent users, achieving superior bandwidth efficiency, stable playback, and near-perfect fairness compared to state-of-the-art baselines. This is enabled by its architectural innovations: (i) grouping receivers and aligning resolutions to eliminate redundant flows, (ii) delegating adaptation logic to forwarders to remove central bottlenecks, and (iii) providing a deployable overlay that integrates seamlessly with today’s CDN and MoQ infrastructures.  

\textbf{Optimization highlights.} A key novelty of COMETS lies in its formalization of coordinated adaptation as a unified ILP problem. By leveraging dual decomposition with closed-form updates and a lightweight greedy feasibility reconstruction, COMETS achieves near-optimal allocation decisions with sub-50\,ms convergence time at scale---up to 3.7$\times$ faster than centralized solvers. This combination of \emph{theoretical rigor} and \emph{practical scalability} sets COMETS apart from prior ICN-based approaches that relied only on local heuristics or reactive congestion signals.  

\textbf{Properties.} Together, these contributions give COMETS three defining properties: (1) \emph{scalable coordination}, by distributing computation proportionally to network size; (2) \emph{fair and efficient resource use}, by harmonizing quality across groups; and (3) \emph{practical deployability}, by requiring no changes to the underlying Internet architecture.  

\textbf{Broader impact.} By aligning with emerging transport initiatives such as MoQ and the evolution of CDN overlays, COMETS provides a realistic migration path toward next-generation media systems. We believe the combination of ICN-inspired primitives with scalable optimization offers a promising direction for future large-scale, low-latency streaming and real-time applications.  

Future work will extend COMETS to heterogeneous edge environments, cross-layer optimization across network and transport layers, and integration with real deployments to further validate its robustness and applicability.

\section*{ACKNOWLEDGMENTS}
This work has partly been supported by the Guangdong provincial project under Grant 2023QN10X048, Guangzhou Municipal Key Laboratory on Future Networked Systems under Grant 2024A03J0623, the Guangdong Provincial Key Lab of Integrated Communication, Sensing and Computation for Ubiquitous Internet of Things under Grant 2023B1212010007, the Guangzhou Municipal Science and Technology Project under Grant 2023A03J0011, the Guangdong provincial project 2023ZT10X009, the Natural Science Foundation of China under Grant U23A20339, Business Finland 6GSoft project under Grant 8541/31/2022, the National Key Research and Development Program of China under Grant 2024YFE0200603, the National Science and Technology Major Project of China on Mobile Information Networks under Grant 2024ZD1300400, the National Natural Science Foundation of China under Grant 62371412, and Guangdong Basic and Applied Basic Research Natural Science Funding Scheme under Grant 2024A1515011184. 
The authors would also like to thank Dr. Yiqing Zhai, Dr. Yu Zhang and Dr. Bowen Tan from The Hong Kong University of Science and Technology (Guangzhou) for their valuable assistance in verifying and refining the optimization designs.
\bibliographystyle{IEEEtran}
% \balance
\bibliography{ref}

@article{zhang2024networked,
  title={Networked metaverse systems: Foundations, gaps, research directions},
  author={Zhang, Yulong and Kutscher, Dirk and Cui, Ying},
  journal={IEEE Open Journal of the Communications Society},
  year={2024},
  publisher={IEEE}
}

@inproceedings{geng2023sok,
  title={SoK: Distributed computing in ICN},
  author={Geng, Wei and Zhang, Yulong and Kutscher, Dirk and Kumar, Abhishek and Tarkoma, Sasu and Hui, Pan},
  booktitle={Proceedings of the 10th ACM Conference on Information-Centric Networking},
  pages={88--100},
  year={2023}
}

@article{li2014probe,
  title={Probe and adapt: Rate adaptation for HTTP video streaming at scale},
  author={Li, Zhi and Zhu, Xiaoqing},
  journal={IEEE journal on selected areas in communications},
  volume={32},
  number={4},
  pages={719--733},
  year={2014},
  publisher={IEEE}
}

@article{zhou2016mdash,
  title={mDASH: A Markov decision-based rate adaptation approach for dynamic HTTP streaming},
  author={Zhou, Chao and Lin, Chia-Wen and Guo, Zongming},
  journal={IEEE Transactions on Multimedia},
  volume={18},
  number={4},
  pages={738--751},
  year={2016},
  publisher={IEEE}
}

@article{spiteri2019theory,
  title={From theory to practice: Improving bitrate adaptation in the DASH reference player},
  author={Spiteri, Kevin and Sitaraman, Ramesh and Sparacio, Daniel},
  journal={ACM Transactions on Multimedia Computing, Communications, and Applications (TOMM)},
  volume={15},
  number={2s},
  pages={1--29},
  year={2019},
  publisher={ACM New York, NY, USA}
}

@article{tan2021game,
  title={Game theory based dynamic adaptive video streaming for multi-client over NDN},
  author={Tan, Xiaobin and Xu, Lei},
  journal={IEEE Transactions on Multimedia},
  volume={24},
  pages={3491--3505},
  year={2021},
  publisher={IEEE}
}

@article{altamimi2020qoe,
  title={QoE-fair DASH video streaming using server-side reinforcement learning},
  author={Altamimi, Sa’di and Shirmohammadi, Shervin},
  journal={ACM Transactions on Multimedia Computing, Communications, and Applications (TOMM)},
  volume={16},
  number={2s},
  pages={1--21},
  year={2020},
  publisher={ACM New York, NY, USA}
}

@article{bentaleb2022bob,
  title={BoB: Bandwidth prediction for real-time communications using heuristic and reinforcement learning},
  author={Bentaleb, Abdelhak and Akcay},
  journal={IEEE Transactions on Multimedia},
  volume={25},
  pages={6930--6945},
  year={2022},
  publisher={IEEE}
}

@inproceedings{gurel2023media,
  title={Media over QUIC: Initial testing, findings and results},
  author={Gurel, Zafer and Erkilic Civelek, Tugce and Bodur, Atakan and Bilgin, Senem and Yeniceri, Deniz and Begen, Ali C},
  booktitle={Proceedings of the 14th Conference on ACM Multimedia Systems},
  pages={301--306},
  year={2023}
}

@article{ravuri2023adaptive,
  title={Adaptive partially reliable delivery of immersive media over QUIC-HTTP/3},
  author={Ravuri, Hemanth Kumar and Vega, Maria Torres and Der Van Hooft, Jeroen and Wauters, Tim and De Turck, Filip},
  journal={Ieee Access},
  volume={11},
  pages={38094--38111},
  year={2023},
  publisher={IEEE}
}

@ARTICLE{10689417,
  author={Dwijaksara, Made Harta and Hilman, Muhammad Hafizhuddin },
  journal={IEEE Access}, 
  title={Rate Adaptation Technique for Media Streaming Over QUIC With Limited Backhaul}, 
  year={2024},
  volume={12},
  number={},
  pages={139028-139041},
  doi={10.1109/ACCESS.2024.3466468}}

@inproceedings{liu2016hop,
  title={Hop-by-hop adaptive video streaming in content centric network},
  author={Liu, Zhengyang and Wei, Yiran},
  booktitle={2016 IEEE International Conference on Communications (ICC)},
  pages={1--7},
  year={2016},
  organization={IEEE}
}

@inproceedings{Alt,
  title={CBA: Contextual quality adaptation for adaptive bitrate video streaming},
  author={Alt, Bastian and Ballard, Trevor},
  booktitle={IEEE INFOCOM 2019-IEEE Conference on Computer Communications},
  pages={1000--1008},
  year={2019},
  organization={IEEE}
}

@article{spiteri2020bola,
  title={BOLA: Near-optimal bitrate adaptation for online videos},
  author={Spiteri, Kevin and Urgaonkar, Rahul and Sitaraman, Ramesh K},
  journal={IEEE/ACM transactions on networking},
  volume={28},
  number={4},
  pages={1698--1711},
  year={2020},
  publisher={IEEE}
}

@article{garcia2019practical,
  title={Practical evaluation of VMAF perceptual video quality for WebRTC applications},
  author={Garc{\'\i}a, Boni and L{\'o}pez-Fern{\'a}ndez, Luis and Gort{\'a}zar, Francisco and Gallego, Micael},
  journal={Electronics},
  volume={8},
  number={8},
  pages={854},
  year={2019},
  publisher={MDPI}
}

@misc{schulzrinne2003rfc3550,
  title={RFC3550: RTP: A transport protocol for real-time applications},
  author={Schulzrinne, Henning and Casner, Steven and Frederick, R and Jacobson, Van},
  year={2003},
  publisher={RFC Editor}
}

@inproceedings{ware2019beyond,
  title={Beyond jain's fairness index: Setting the bar for the deployment of congestion control algorithms},
  author={Ware, Ranysha and Mukerjee},
  booktitle={Proceedings of the 18th ACM Workshop on Hot Topics in Networks},
  pages={17--24},
  year={2019}
}

@inproceedings{yin2015control,
  title={A control-theoretic approach for dynamic adaptive video streaming over HTTP},
  author={Yin, Xiaoqi and Jindal, Abhishek and Sekar, Vyas and Sinopoli, Bruno},
  booktitle={Proceedings of the 2015 ACM Conference on Special Interest Group on Data Communication},
  pages={325--338},
  year={2015}
}

@misc{pwc2024,
   author = {{PwC}},
   title = {Streaming the game: How the rise of digital platforms is changing sports consumption},
   howpublished = {\url{https://www.pwc.com/us/en/industries/tmt/library/sports-streaming-platforms.html}},
   year = {2024},
   note = {Accessed: April 2025}
}

@misc{rossvideo2025,
   author = {{Ross Video}},
   title = {Sports Audience Statistics: The Data Behind Changing Fan Behavior},
   howpublished = {\url{https://www.rossvideo.com/blog/sports-audience-statistics-the-data-behind-changing-fan-behavior/}},
   year = {2025},
   month = {February},
   note = {Accessed: April 2025}
}

@inproceedings{yu2024secure,
  title={Secure web objects: Building blocks for Metaverse interoperability and decentralization},
  author={Yu, Tianyuan and Ma, Xinyu and Patil, Varun and Kocaogullar, Yekta and Zhang, Yulong and Burke, Jeff and Kutscher, Dirk and Zhang, Lixia},
  booktitle={2024 IEEE International Conference on Metaverse Computing, Networking, and Applications (MetaCom)},
  pages={25--33},
  year={2024},
  organization={IEEE}
}

@inproceedings{bagheri2010bandwidth,
  title={Bandwidth Adapted Hierarchical Multicast Overlay},
  author={Bagheri, Maryam and Movaghar, Ali },
  booktitle={2010 Fifth International Multi-conference on Computing in the Global Information Technology},
  pages={262--267},
  year={2010},
  organization={IEEE}
}

@inproceedings{kwan2005overlay,
  title={On overlay multicast tree construction and maintenance},
  author={Kwan, TT-M and Yeung},
  booktitle={2005 International Conference on Collaborative Computing: Networking, Applications and Worksharing},
  pages={6},
  year={2005},
  organization={IEEE}
}

@book{boyd2004convex,
  title={Convex optimization},
  author={Boyd, Stephen P and Vandenberghe, Lieven},
  year={2004},
  publisher={Cambridge university press}
}

@article{su2009drafting,
  title={Drafting behind Akamai: Inferring network conditions based on CDN redirections},
  author={Su, Ao-Jan and Choffnes},
  journal={IEEE/ACM transactions on networking},
  volume={17},
  number={6},
  pages={1752--1765},
  year={2009},
  publisher={IEEE}
}

@article{shobiri2023cdns,
  title={CDNs’ dark side: Security problems in CDN-to-origin connections},
  author={Shobiri, Behnam and Mannan, Mohammad and Youssef, Amr},
  journal={Digital Threats: Research and Practice},
  volume={4},
  number={1},
  pages={1--22},
  year={2023},
  publisher={ACM New York, NY}
}

@article{pathan2007taxonomy,
  title={A taxonomy and survey of content delivery networks},
  author={Pathan, Al-Mukaddim Khan and Buyya, Rajkumar and others},
  journal={Grid computing and distributed systems laboratory, University of Melbourne, Technical Report},
  volume={4},
  number={2007},
  pages={70},
  year={2007}
}

@ARTICLE{7995137,
  author={Samain, Jacques and Carofiglio, Giovanna},
  journal={IEEE Transactions on Multimedia}, 
  title={Dynamic Adaptive Video Streaming: Towards a Systematic Comparison of ICN and TCP/IP}, 
  year={2017},
  volume={19},
  number={10},
  pages={2166-2181},
  doi={10.1109/TMM.2017.2733340}}

@ARTICLE{8007216,
  author={Saltarin, Jonnahtan and Bourtsoulatze, Eirina and Thomos, Nikolaos and Braun, Torsten},
  journal={IEEE Transactions on Multimedia}, 
  title={Adaptive Video Streaming With Network Coding Enabled Named Data Networking}, 
  year={2017},
  volume={19},
  number={10},
  pages={2182-2196},
  doi={10.1109/TMM.2017.2737950}}

@ARTICLE{9195016,
  author={Wu, Fan and Yang, Wang and Ren, Ju and Lyu, Feng and Yang, Peng and Zhang, Yaoxue and Shen, Xuemin},
  journal={IEEE Transactions on Multimedia}, 
  title={NDN-MMRA: Multi-Stage Multicast Rate Adaptation in Named Data Networking WLAN}, 
  year={2021},
  volume={23},
  number={},
  pages={3250-3263},
  doi={10.1109/TMM.2020.3023282}}

@inproceedings{thomas2016applications,
  title={Applications and deployments of server and network assisted DASH (SAND)},
  author={Thomas, Emmanuel and van Deventer, MO and Stockhammer, Thomas and Begen, M-L and Oyman, Ozgur},
  booktitle={IBC 2016 Conference},
  pages={22},
  year={2016},
  organization={IET}
}

@book{bertsekas2003convex,
  title={Convex analysis and optimization},
  author={Bertsekas, Dimitri and Nedic, Angelia and Ozdaglar, Asuman},
  volume={1},
  year={2003},
  publisher={Athena Scientific}
}

@INPROCEEDINGS{8712644,
  author={Awiphan, Suphakit and Poobai, Kanin},
  booktitle={22nd International Computer Science and Engineering Conference}, 
  title={Adaptive Video Streaming on Named Data Networking with IoT-Assisted Content Delivery}, 
  year={2018},
  volume={},
  number={},
  pages={1-4},
  doi={10.1109/ICSEC.2018.8712644}}

@inproceedings{yang2021high,
  title={High performance adaptive video streaming using NDN WLAN multicast},
  author={Yang, Wang and Wu, Fan and Tian, Kaijin},
  booktitle={Proceedings of the 8th ACM Conference on Information-Centric Networking},
  pages={42--51},
  year={2021}
}

@article{boyd2007notes,
  title={Notes on decomposition methods},
  author={Boyd, Stephen and Xiao, Jacob},
  journal={Notes for EE364B, Stanford University},
  volume={635},
  pages={1--36},
  year={2007}
}

@inproceedings{raake2017bitstream,
  title={A bitstream-based, scalable video-quality model for HTTP adaptive streaming: ITU-T P. 1203.1},
  author={Raake, Alexander and Garcia, Marie-Neige and Robitza, Werner and List, Peter and G{\"o}ring, Steve and Feiten, Bernhard},
  booktitle={2017 Ninth international conference on quality of multimedia experience (QoMEX)},
  pages={1--6},
  year={2017},
  organization={IEEE}
}

@book{cormen2022introduction,
  title={Introduction to algorithms},
  author={Cormen, Thomas H and Leiserson, Charles E and Rivest, Ronald L and Stein, Clifford},
  year={2022},
  publisher={MIT press}
}

@techreport{irtf-icnrg-flic-06,
    number =    {draft-irtf-icnrg-flic-06},
    type =      {Internet-Draft},
    institution =   {Internet Engineering Task Force},
    publisher = {Internet Engineering Task Force},
    note =      {Work in Progress},
    url =       {https://datatracker.ietf.org/doc/draft-irtf-icnrg-flic/06/},
    author =    {Christian Tschudin and Christopher A. Wood and Marc Mosko and David R. Oran},
    title =     {{File-Like ICN Collections (FLIC)}},
    pagetotal = 41,
}

@misc{cisco2024qos,
  title={{Quality of Service Design Overview}},
  author={{Cisco Systems}},
  year={2024},
  howpublished={\url{https://www.cisco.com/c/en/us/td/docs/solutions/Enterprise/WAN_and_MAN/QoS_SRND_40/QoSIntro_40.html}}
}

@inproceedings{wang2025vifusion,
  title={ViFusion: In-Network Tensor Fusion for Scalable Video Feature Indexing},
  author={Wang, Yisu and Zhu, Yixiang and Li, Xinjiao and Zhang, Yulong and Wu, Ruilong and Kutscher, Dirk},
  booktitle={Proceedings of the 2025 International Conference on Multimedia Retrieval},
  pages={1452--1460},
  year={2025}
}

@inproceedings{chai2025inds,
  title={INDS: Incremental Named Data Streaming for Real-Time Point Cloud Video},
  author={Chai, Ruonan and Zhu, Yixiang and Li, Xinjiao and Li, Jiawei and Meng, Zili and Kutscher, Dirk},
  booktitle={Proceedings of the 33rd ACM International Conference on Multimedia},
  pages={12102--12110},
  year={2025}
}

\begin{IEEEbiography}[{\includegraphics[width=1in,height=1.25in,clip,keepaspectratio]{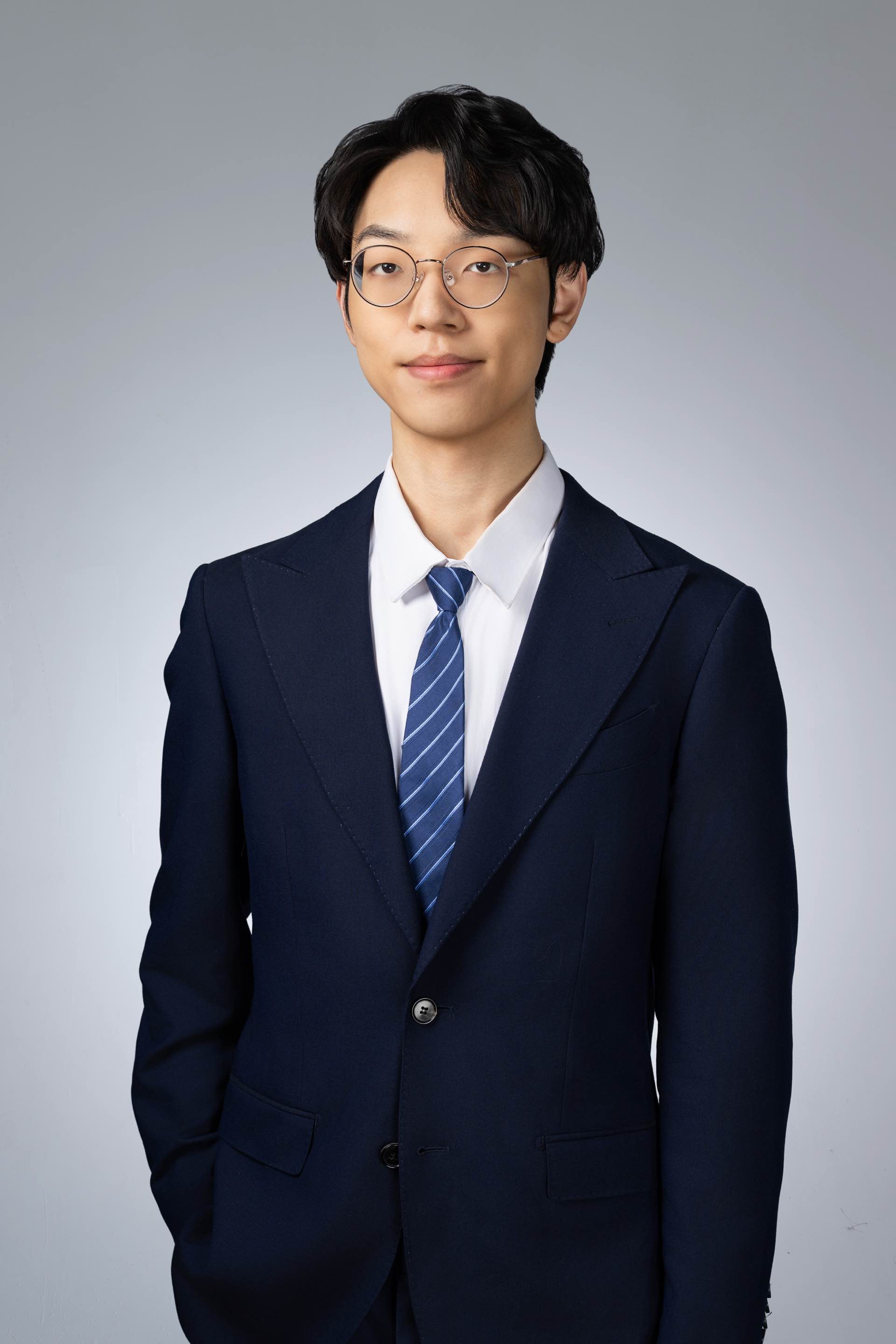}}]{Yulong Zhang}
received the B.Eng. degree from the Wuhan University of Technology, China, in 2018. He is currently pursuing the Ph.D. degree with The Hong Kong University of Science and Technology (Guangzhou). His main research interests include designing and optimizing emerging networked systems, distributed computing and networking (compute-first networking, computing in the network), and Internet architecture and decentralized communication.
\end{IEEEbiography}

\begin{IEEEbiography}[{\includegraphics[width=1in,height=1.25in,clip,keepaspectratio]{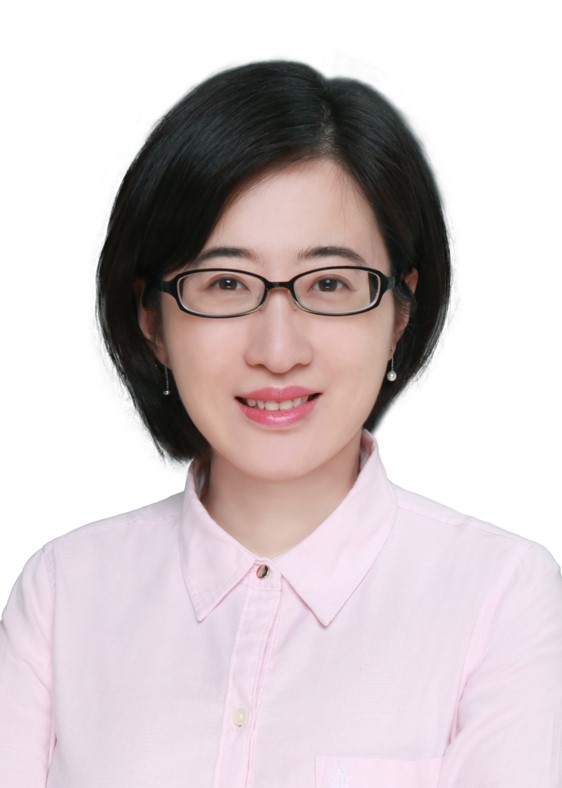}}]{Ying Cui}
received her B.Eng degree in Electronic and Information Engineering from Xi'an Jiao Tong University, China, in 2007 and her Ph.D. degree from the Hong Kong University of Science and Technology, Hong Kong, in 2012. She held visiting positions at Yale University, US, in 2011 and Macquarie University, Australia, in 2012. From June 2012 to June 2013, she was a postdoctoral research associate at Northeastern University, US. From July 2013 to December 2014, she was a postdoctoral research associate at the Massachusetts Institute of Technology, US. From January 2015 to July 2022, she was an Associate Professor at Shanghai Jiao Tong University, China. Since August 2022, she has been an Associate Professor with the IoT Thrust at The Hong Kong University of Science and Technology (Guangzhou). Her current research interests include optimization, learning, IoT communications, integrated sensing and communications, and edge intelligence. She was selected to the National Young Talent Program in 2014. She received Best Paper Awards from IEEE ICC 2015, IEEE GLOBECOM 2021, and IEEE GLOBECOM 2025. She serves as an Editor for IEEE Transactions on Communications and an Executive Editor for IEEE Transactions on Wireless Communications. She served as an Editor for the IEEE Transactions on Wireless Communications from 2018 to 2024.
\end{IEEEbiography}

\begin{IEEEbiography}[{\includegraphics[width=1in,height=1.25in,clip,keepaspectratio]{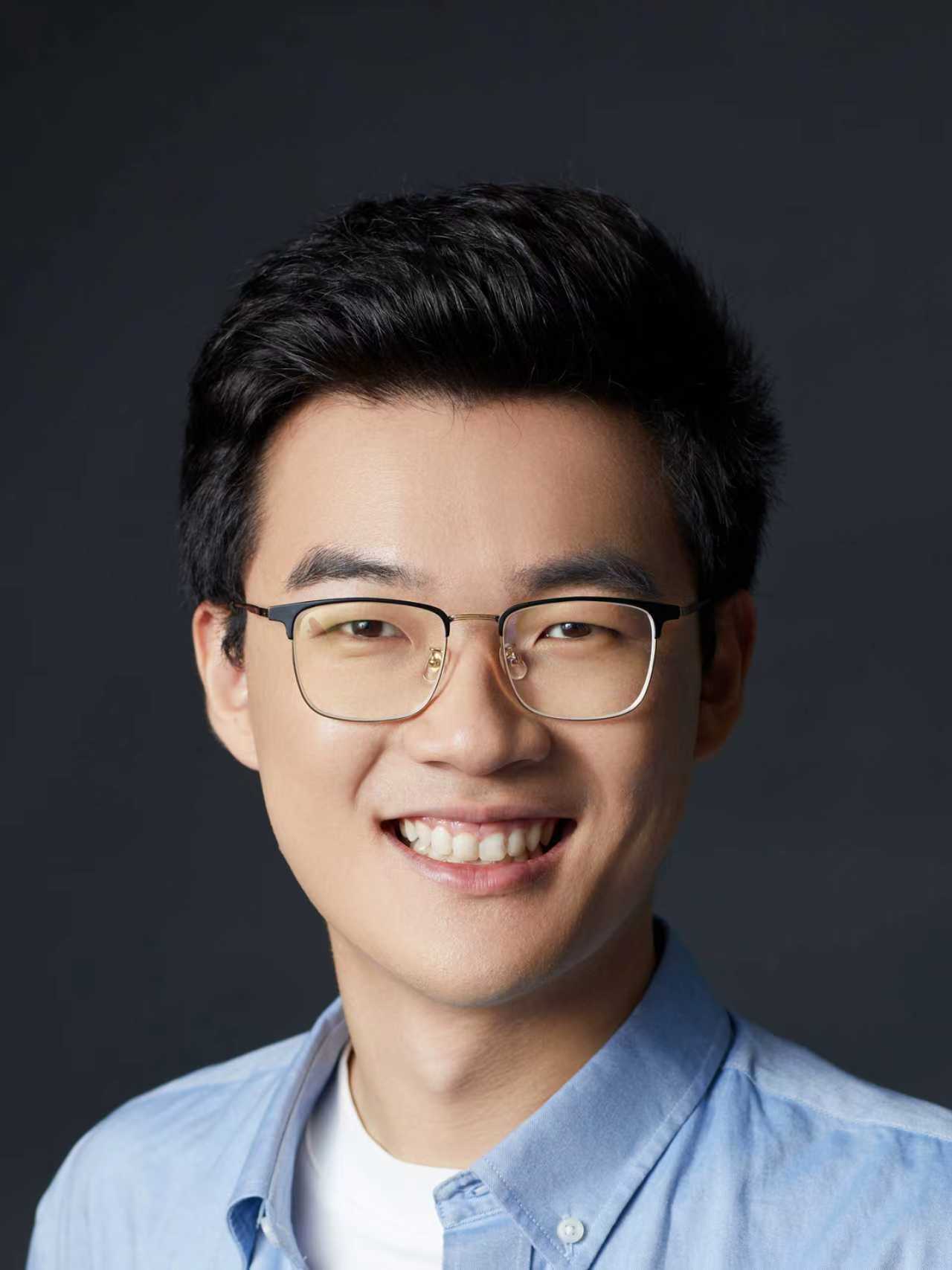}}]{Zili Meng}
(S'18-M'23) received the B.Eng. and Ph.D. degrees from Tsinghua University. He is currently an Assistant Professor with the Department of Electronic and Computer Engineering, The Hong Kong University of Science and Technology. His research interests include multimedia streaming and networking.
\end{IEEEbiography}

\begin{IEEEbiography}[{\includegraphics[width=1in,height=1.25in,clip,keepaspectratio]{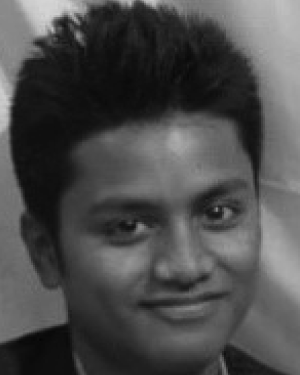}}]{Abhishek Kumar}
(Member, IEEE and IEEE Signal Processing Society) is a Senior Lecturer with the Faculty of Information Technology, University of Jyväskylä, Finland. He received his B.Tech., M.S., and Ph.D. degrees from National Institute of Technology Patna, Korea Advanced Institute of Science and Technology, and University of Helsinki in 2014, 2016, and 2023, respectively. He also serves as the secretary to the IEEE Finland Chapter for Signal Processing, and Circuits \& Systems (Jan 2025--Dec 2026). His current research spans edge intelligence, intent-driven neuro-symbolic AI, and cyber-physical systems with a special focus on privacy and security angles.
\end{IEEEbiography}

\begin{IEEEbiography}[{\includegraphics[width=1in,height=1.25in,clip,keepaspectratio]{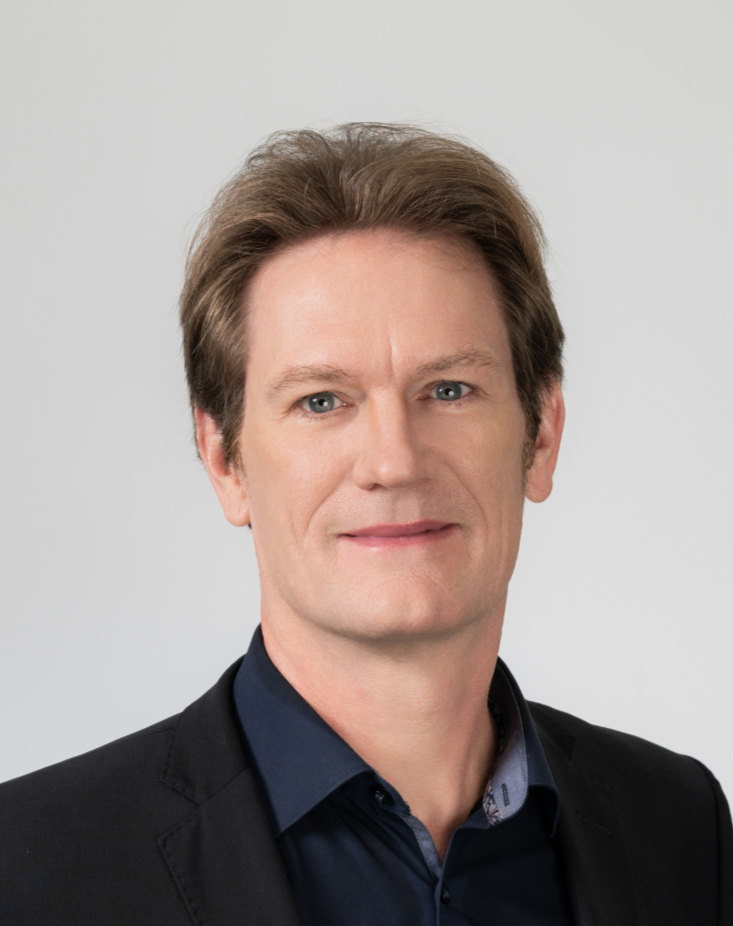}}]{Dirk Kutscher}
received the Ph.D. degree from Universit{\"a}t Bremen, Germany. He is a Professor with The Hong Kong University of Science and Technology (Guangzhou), where he is the Director of the Future Networked Systems Laboratory. Previously, he was the CTO of Virtual Networking with Huawei Research Germany, a Chief Networking Researcher with NEC Europe, and a Visiting Researcher with KDDI R\&D, Japan. He is currently the Chair of the Internet Research Task Force (IRTF). He has authored several international patents, IETF RFCs, books, and highly cited and awarded research papers. His main interests lie in the intersection of distributed computing and networking for compute-first networking and in Internet architecture. 
\end{IEEEbiography}

\clearpage

% \appendix
\appendices

\section{Rationale for the Distributed Optimization Approach}
\label{sec:rationale}

The problem formulated in (1) is an ILP, which is NP-hard. To address its inherent challenges, particularly in large-scale, dynamic network environments, we analyze several potential solution paradigms before motivating our choice.

\subsection{Centralized ILP Solvers}
Commercial solvers (e.g., Gurobi, CPLEX) can find the global optimum. However, they face two critical limitations:
\begin{enumerate}
\item \textbf{Exponential Complexity:} Solving time grows exponentially with the number of integer variables. For a network with $V$ nodes and $L$ resolution levels, we have $\mathcal{O}(V L)$ node selection variables and $\mathcal{O}(EL)$ transmission variables, making the problem intractable for networks with hundreds of nodes.

\item \textbf{Global Information Requirement:} They require a central controller with complete and instantaneous knowledge of the entire network topology, link capacities, and all user states, which violates the distributed nature of NDN and introduces significant signaling overhead.
\end{enumerate}

\subsection{AI-based Methods}
Recent advancements have applied machine learning, particularly Deep Neural Networks (DNNs), to solve combinatorial optimization problems. These methods, however, are not well-suited for our problem due to: 
\begin{enumerate}
\item \textbf{Generalization and Robustness Issues:} A trained model performs well only on problem instances with similar statistical distributions (e.g., network topology, user density) as the training data. Significant changes in the network environment, which are common in real-world scenarios, would necessitate costly retraining. 
\item \textbf{Lack of Interpretability and Optimality Guarantees:} DNNs act as "black boxes," providing no guarantee on the quality of the obtained solution, which is a significant drawback for network resource allocation where performance bounds are critical.
\end{enumerate}

\subsection{Lagrangian Dual Decomposition}
In contrast, we adopt Lagrangian dual decomposition. This choice is motivated by its inherent advantages for our problem structure: 
\begin{enumerate}
\item \textbf{Decomposability:} By dualizing the coupling constraints, the original intractable problem naturally decomposes into independent, smaller-scale subproblems for each node. This aligns perfectly with the distributed architecture of NDN. 
\item \textbf{Theoretical Soundness:} The framework provides a systematic, iterative approach with convergence properties and yields a duality gap that bounds the suboptimality of any feasible solution. 
\item \textbf{Efficiency:} The resulting subproblems admit either closed-form or highly efficient algorithmic solutions, leading to low computational complexity at each node.
\end{enumerate}

Therefore, to balance optimality, computational complexity, and practical deployability in a distributed system, Lagrangian dual decomposition presents the most promising path forward.

\section{Proof of Lemma~\ref{lemma_decomposition}}
\label{app:proof_decomposition}

\begin{IEEEproof}
The Lagrangian $\mathcal{L}$ in \eqref{eq:lagrangian_function} can be rewritten by grouping terms:

\begin{align}
\mathcal{L} = &\sum_{s \in \mathcal{S}} \sum_{i \in \mathcal{O}_s} \sum_{l \in \mathcal{L}} \lambda_{1,i,l} y_{s,i,l} + \sum_{u \in \mathcal{U}} \sum_{l \in \mathcal{L}} (w_u Q_l - \lambda_{1,u,l}) x_{u,l} \notag\\
&+ \sum_{f \in \mathcal{F}} \sum_{l \in \mathcal{L}} \left(\sum_{k \in \mathcal{O}_f} \lambda_{2,f,k,l} - \lambda_{1,f,l}\right) x_{f,l} \notag\\
&+ \sum_{f \in \mathcal{F}} \sum_{k \in \mathcal{O}_f} \sum_{l \in \mathcal{L}} (\lambda_{1,k,l} - \lambda_{2,f,k,l}) y_{f,k,l}.
\end{align}

Each group is independent: 
\begin{itemize}
    \item The first term depends only on $\mathbf{y}_{s,i}$ with constraints \eqref{c6}, \eqref{c9}
    \item The second term depends only on $\mathbf{x}_u$ with constraints \eqref{c2}, \eqref{c3}, \eqref{c9}
    \item The third term depends only on $\mathbf{x}_f$ with constraint $x_{f,l} \in [0,1]$
    \item The fourth term depends only on $\mathbf{y}_{f,k}$ with constraints \eqref{c6}, \eqref{c9}
\end{itemize}

No cross terms or shared constraints remain after dualizing \eqref{c4} and \eqref{c5}. Thus, 
\begin{equation}
\begin{split}
g(\boldsymbol{\lambda}_1, \boldsymbol{\lambda}_2) &= \sum_{s \in \mathcal{S}} \sum_{i \in \mathcal{O}_s} g_{s,i}(\boldsymbol{\lambda}_1) 
+ \sum_{u \in \mathcal{U}} g_u(\boldsymbol{\lambda}_1) \\
&+ \sum_{f \in \mathcal{F}} g_{f,x}(\boldsymbol{\lambda}_1, \boldsymbol{\lambda}_2) 
+ \sum_{f \in \mathcal{F}} \sum_{k \in \mathcal{O}_f} g_{f,k}(\boldsymbol{\lambda}_1, \boldsymbol{\lambda}_2)
\end{split}
\end{equation}
where each $g_*$ is the optimal value of the corresponding subproblem. The master problem minimizes this sum, which is equivalent to the dual problem by definition.
\end{IEEEproof}

\end{document}